\documentclass[opre,nonblindrev]{informs3arxiv}

\DoubleSpacedXI 

\usepackage{endnotes}
\let\footnote=\endnote
\let\enotesize=\normalsize
\def\notesname{Endnotes}%
\def\makeenmark{\hbox to1.275em{\theenmark.\enskip\hss}}
\def\enoteformat{\rightskip0pt\leftskip0pt\parindent=1.275em
  \leavevmode\llap{\makeenmark}}

\usepackage[utf8]{inputenc}
\usepackage{graphicx,tikz}
\usepackage[tbtags]{mathtools}
\usepackage{amsmath,amsfonts,amssymb,bm}
\usepackage{color,dsfont}
\usepackage{url}
\usepackage{algorithm,algorithmic,setspace}
\usepackage{hhline}
\usepackage{enumerate}
\usepackage{array}
\usepackage{booktabs}
\usepackage{mathrsfs,bbm,bm}
\usepackage{tikz}
\usepackage{cleveref}
\usepackage{accents}
\usepackage[round]{natbib}
\usepackage{pgfplots}
\pgfplotsset{compat=1.9}

\usepackage{float}

\usetikzlibrary {positioning}
\usepackage{pgf}
\usepackage{pgfplots}
\usetikzlibrary{shapes.geometric, arrows}
\tikzstyle{node} = [ellipse, minimum width=1cm, minimum height=.5cm, text centered, text width=1cm, draw=black, fill=blue!40]
\tikzstyle{arrow} = [thick, ->, >=stealth]
\tikzstyle{arrow1} = [red, thin, ->, >=stealth]
\usepackage{rotating}
\usetikzlibrary{fillbetween}
\usetikzlibrary{intersections}
\pgfdeclarelayer{bg}
\pgfsetlayers{bg,main}
\usepgfplotslibrary{fillbetween}

\usepackage[caption=false]{subfig}

\usepackage{etoolbox}
\let\bbordermatrix\bordermatrix
\patchcmd{\bbordermatrix}{8.75}{4.75}{}{}
\patchcmd{\bbordermatrix}{\left(}{\left[}{}{}
\patchcmd{\bbordermatrix}{\right)}{\right]}{}{}

\newcounter{algo}
\AtBeginEnvironment{algorithm}{\OneAndAHalfSpacedXI}
\AtBeginEnvironment{tabular}{\OneAndAHalfSpacedXII}

\newcommand{\AAA}{\mathcal{A}}

\newcommand{\FFF}{\mathcal{F}}
\newcommand{\PPP}{\mathcal{P}}
\renewcommand{\NNN}{\mathcal{N}}

\newcommand{\LLL}{\mathcal{L}}
\newcommand{\UUU}{\mathcal{U}}
\newcommand{\GGG}{\mathcal{G}}
\newcommand{\YYY}{\mathcal{Y}}
\newcommand{\ZZZ}{\mathcal{Z}}

\newcommand{\EE}{\mathbb{E}}
\newcommand{\RR}{\mathbb{R}}
\newcommand{\ZZ}{\mathbb{Z}}
\newcommand{\PP}{\mathbb{P}}
\newcommand{\NN}{\mathbb{N}}

\newcommand{\one}{\bm{\mathbf{1}}}
\newcommand{\zero}{\mathbf{0}}

\newcommand{\Rsen}{R}

\newcommand{\RsenRVx}{R^{\text{RV}}}

\DeclareMathOperator{\Int}{int}
\DeclareMathOperator{\vertex}{vert}

\DeclareMathOperator{\gr}{gr}

\newcommand{\of}[1]{\ensuremath{\left( #1 \right)}}

\newcommand{\cb}[1]{\ensuremath{ \left\{ #1 \right\} }}

\newcommand{\sqb}[1]{\ensuremath{ \left[ #1 \right] }}

\newcommand{\ignore}[1]{}

\newcommand{\norminfX}{\ensuremath{ \left\Vert \bm{X} \right\Vert }_\infty}

\newcommand{\norminfpbar}{\ensuremath{ \left\Vert \bm{{\bar{p}}} \right\Vert }_\infty}

\newcommand{\norm}[1]{\ensuremath{ \left\Vert #1 \right\Vert }}
\newcommand{\abs}[1]{\ensuremath{ \left| #1 \right| }}

\newcommand{\pnew}{p'}
\newcommand{\snew}{s'}
\newcommand{\tnew}{t'}

\newcommand{\transposeT}{^\mathsf{T}}
\newcommand{\transpose}[1]{{#1}^\mathsf{T}}

\newcommand{\elll}{\ell,\hat{\ell}}

\newcommand{\systemRV}{\of{\bm{{x}}, \bm{{\bar{p}}}, \bm{{\pi}}, \alpha, \beta}}

\newcommand{\XsystemRV}{\of{\bm{{X}}, \bm{{\bar{p}}}, \bm{{\pi}}, \alpha, \beta}}

\usepackage{natbib}
 \bibpunct[, ]{(}{)}{,}{a}{}{,}%
 \def\bibfont{\small}%
 \def\BIBand{and}%

\TheoremsNumberedThrough     
\ECRepeatTheorems

\EquationsNumberedThrough    

\begin{document}



\RUNAUTHOR{Ararat and Meimanjan}

\RUNTITLE{Computation of Systemic Risk Measures}

\TITLE{Computation of Systemic Risk Measures:\\ A Mixed-Integer Programming Approach}

\ARTICLEAUTHORS{%
\AUTHOR{\c{C}a{\u{g}}{\i}n Ararat}
\AFF{Department of Industrial Engineering, Bilkent University, Ankara, Turkey,\\ \EMAIL{cararat@bilkent.edu.tr}} 
\AUTHOR{Nurtai Meimanjan}
\AFF{Institute for Statistics and Mathematics, Vienna University of Economics and Business, Vienna, Austria,\\ \EMAIL{nurtai.meimanjan@wu.ac.at}}
} 

\ABSTRACT{%
Systemic risk is concerned with the instability of a financial system whose members are interdependent in the sense that the failure of a few institutions may trigger a chain of defaults throughout the system. Recently, several systemic risk measures have been proposed in the literature that are used to determine capital requirements for the members subject to joint risk considerations. We address the problem of computing systemic risk measures for systems with sophisticated clearing mechanisms. In particular, we consider an extension of the Rogers-Veraart network model where the operating cash flows are unrestricted in sign. We propose a mixed-integer programming problem that can be used to compute clearing vectors in this model. Due to the binary variables in this problem, the corresponding (set-valued) systemic risk measure fails to have convex values in general. We associate nonconvex vector optimization problems with the systemic risk measure and provide theoretical results related to the weighted-sum and Pascoletti-Serafini scalarizations of this problem. Finally, we test the proposed formulations on computational examples and perform sensitivity analyses with respect to some model-specific and structural parameters.
}%


 \HISTORY{August 1, 2023.}

\maketitle

%


\section{Introduction}

\subsection{Motivation and Literature Review}\label{lit}

Financial contagion is usually associated with a chain of failures in a financial system triggered by external correlated shocks as well as direct or indirect interdependencies among the members of the system. From an economic point of view, it leads to undesirable consequences such as financial crisis, necessity for bailout loans, economic regression, rise in national debt, and so on. A good example is a bank run, when a large number of holders withdraw their money from a bank due to panic or decrease in confidence in the bank, causing insolvency of the bank. In turn, the bank may call its claims from the other banks, decreasing confidence in them and causing new bank runs. Being unable to meet their liabilities, some of the banks may become bankrupt and, thus, aggravate the contagion even further.  Unlike the more traditional institutional risk, systemic risk is related to the strength of an entire financial system against financial contagions.

In this paper, we consider financial systems in which members have direct links to each other through contractual liabilities. Besides these endogenous links, each member has its economic activities with entities that are exogenous to the system, resulting in some cash, referred to as \emph{operating cash flow}, to be used in meeting the endogenous liabilities. When the members realize their operating cash flows, the actual interbank payments are determined through a clearing procedure. As an example of such systems, \cite{eisenberg-noe} models a financial system as a static directed network of banks where interbank liabilities are attached to the arcs. Assuming a positive operating cash flow for each bank, the paper develops two approaches to calculate a clearing vector, that is, a vector of payments to meet interbank liabilities. The first is a simple algorithm, called the \emph{fictitious default algorithm}, which gradually calculates a clearing vector by finitely many updates. The second is a laconic mathematical programming problem with linear constraints determined by the liabilities, the operating cash flows, and an arbitrary strictly increasing objective function. In particular, one can choose a linear objective function so that a clearing vector is calculated as an optimal solution of a \emph{linear programming problem}.

As an important extension, \cite{rogers-veraart} introduces default costs to the model in \cite{eisenberg-noe}. In addition, one of the main focuses in \cite{rogers-veraart} is the investigation of the necessity of bailing out procedures for the defaulting institutions. It is shown that under strictly positive default costs, it might be beneficial for some of the solvent institutions to take over insolvent institutions. For a detailed review of clearing systems, the reader is referred to the survey \cite{kabanov}, which focuses on the existence and uniqueness of clearing vectors as well as their calculations by certain variations of the fictitious default algorithm in \cite{eisenberg-noe}. However, to the best of our knowledge, none of these works builds on the mathematical programming approach of \cite{eisenberg-noe}.

The operating cash flows of the members of a network are typically subject to uncertainty due to correlated risk factors. Hence, these cash flows can be modeled as one possible realization of a random vector with possibly correlated components. Then, the resulting clearing vector is a deterministic function of the operating cash flow random vector, where the deterministic function is defined through the underlying clearing mechanism. Based on the random clearing vector, one can define various \emph{systemic risk measures} to calculate the necessary capital allocations for the members of the network in order to control some (nonlinear) averages over different scenarios. This is the main focus of a recent stream of research that started with \cite{chen}. Using the clearing mechanism, one defines a random aggregate quantity associated with the clearing vector, such as the total debt paid in the system or the total equity made by all members as a result of clearing. This aggregate quantity can be seen as a deterministic and scalar function, called the \emph{aggregation function}, of the operating cash flow vector. In \cite{chen}, a systemic risk measure is defined as a scalar functional of the operating cash flow vector that measures the risk of the random aggregate quantity through a convex risk measure \citep[Chapter~4]{stoch.finance} such as negative expected value, average value-at-risk or entropic risk measure.

The value of the systemic risk measure in \cite{chen} can be seen as the total capital requirement for the system to keep the risk of the aggregate quantity at an acceptable level. However, since the total capital is used only after the shock is aggregated, the allocation of this total back into the members of the system remains a question to be addressed by an additional procedure. To that end, set-valued and scalar systemic risk measures that are considered ``sensitive" to capital levels are proposed in \cite{feinstein} and \cite{biagini}, respectively. These systemic risk measures look for deterministic capital allocation vectors that are directly used to augment the random operating cash flow vector. Thus, the new augmented cash flow vector is aggregated and the risk of the resulting random aggregate quantity is controlled by a convex risk measure as in \cite{chen}. In particular, the value of the set-valued systemic risk measure in \cite{feinstein} is the set of all ``feasible" capital allocation vectors, which addresses the measurement and allocation of systemic risk as a joint problem.

The sensitive systemic risk measures studied in \cite{feinstein} and \cite{biagini} have convenient theoretical properties when the underlying aggregation function is simple enough. In \cite{ararat}, assuming a monotone and concave aggregation function, it has been shown that the set-valued sensitive systemic risk measure is a convex set-valued risk measure in the sense of \citet{hamel-heyde-rudloff} and dual representations are obtained in terms of the conjugate function of the aggregation function. In particular, the aggregation function for the Eisenberg-Noe model, assuming positive operating cash flows as in the original formulation in \cite{eisenberg-noe}, is monotone and concave, and an explicit dual representation is obtained for the corresponding systemic risk measure of this model.

\subsection{Contribution of the Paper}

In this paper, we are concerned with the \emph{computation} of a sensitive systemic risk measure. We relate the value of this systemic risk measure to a vector (multiobjective) optimization problem whose ``efficient frontier" corresponds to the boundary of the systemic risk measure. The vector optimization problem has a risk constraint written in terms of the aggregation function. The main challenge in solving this problem is that the aggregation function needs to be evaluated for every scenario of the underlying probability space as well as for every choice of the capital allocation vector, which is the decision variable of the optimization problem. For the standard Eisenberg-Noe model, thanks to the linear programming characterization of the clearing vectors, one can formulate the aggregation function in terms of a linear programming problem parametrized by the scenario and the capital allocation vector. Hence, the ultimate vector optimization problem can be seen as a nested optimization problem.

We focus particularly on models beyond the Eisenberg-Noe framework with positive operating cash flows. More precisely, when computing systemic risk measures, we add the following two features to the Eisenberg-Noe model: 1) signed operating cash flows, 2) default costs as in the Rogers-Veraart model. It turns out that both features have a similar type of singularity that can be formulated in terms of \emph{binary variables}, a key idea exploited in this paper. By combining these two features, we introduce a signed version of the Rogers-Veraart model. One of our main contributions is to develop a mixed-integer programming (MIP) problem whose optimal solution yields a clearing vector in this model. For the computations, we choose the objective function of this optimization problem in such a way that the optimal value gives the total debt paid at clearing; in this case, we calculate the aggregation function as the optimal value of a mixed-integer linear programming (MILP) problem. As special cases, we consider the simplified settings in which only one of the features 1) and 2) arise, and provide a reduced formulation of the MIP problem in each case.

The existence of binary variables in the associated optimization problems results in \emph{lack of concavity} for the corresponding aggregation functions. Consequently, in contrast to the existing literature on set-valued and scalar systemic risk measures, the sensitive systemic risk measures for the two models do not possess the nice theoretical feature of being convex. In particular, the dual representations studied for systemic risk measures in \cite{ararat} are not applicable in our setting. Indeed, we even have that the values of these systemic risk measures fail to be convex sets, in general. Therefore, one of our fundamental observations is that \emph{binary variables} and the accompanying \emph{lack of concavity/convexity} show up naturally at the cost of using more sophisticated aggregation mechanisms beyond the standard Eisenberg-Noe framework.

Although our main interest is on the signed Rogers-Veraart model for computations, we follow a unified approach by using a general aggregation function defined in terms of a mixed-integer optimization problem for the theoretical development. To be able to approximate the nonconvex values of the corresponding systemic risk measure, we associate a (generally nonconvex) vector optimization problem with it. As a general paradigm, algorithms for solving vector optimization problems iterate by solving certain \emph{scalarization problems} (e.g., weighted-sum scalarizations) along with additional computational procedures (e.g., vertex enumeration subroutines). For instance, Benson's algorithm \citep{benson} for linear vector optimization, the Benson-type algorithm in \cite{lohne-rudloff-ulus} for convex vector optimization, and the Benson-type algorithm in \cite{non-conv.benson} for nonconvex vector optimization follow this pattern. Consequently, the validity of such algorithms is conditional on the ability to solve the scalarization problems. 

We develop extensive mixed-integer formulations for two scalarization problems that are commonly used in vector optimization: the weighted-sum scalarization and the Pascoletti-Serafini scalarization, which consists of calculating the minimum step-length to enter a set with a fixed direction. We prove that both scalarization problems for the signed Rogers-Veraart model can be formulated as MIP problems. We also prove some results related to the feasibility and boundedness of these problems. We solve these scalarization problems as subroutines of the nonconvex algorithm in \cite{non-conv.benson}. It should be noted that the choice of this algorithm is not arbitrary at all; indeed, due to the existence of binary variables in the formulations, the corresponding vector optimization problems for both models are nonconvex, hence the use of an algorithm that works without convexity assumptions is essential.

We perform a detailed computational study for both models as well as sensitivity analyses with respect to some model parameters such as the default cost parameters in the Rogers-Veraart model, the threshold level used in the risk constraint, and also some parameters determining the interconnectedness of the network.

The rest of the main part of the paper is organized as follows. In \Cref{systemic_risk_measures}, we present a framework for systemic risk measures via acceptance sets based on expected values on a general probability space and aggregation functions based on optimization problems, without focusing on a particular network model. In \Cref{systemic_risk_models}, we propose an extension of the Rogers-Veraart network model in which the operating cash flows of the financial institutions can have positive and negative values. Our focus is on the calculation of clearing vectors through novel mixed-integer programming formulations. In \Cref{RVsystrisk}, we study systemic risk measures in the context of the signed Rogers-Veraart model and provide mixed-integer programming formulations for the scalarizations of these risk measures. We present the computational results in \Cref{computational_results}.

\section{Optimization Problems for Systemic Risk Measures} \label{systemic_risk_measures}

In this section, we consider the computation of \emph{sensitive systemic risk measures}. These are set-valued functionals whose input is the random operating cash flow vector and the output yields a set of capital allocation vectors that make the network acceptable. A systemic risk measure has two main ingredients: an \emph{aggregation function}, which is defined in terms of the underlying network model, and an \emph{acceptance set}, which evaluates the risk of the corresponding random output of the aggregation function. The literature on systemic risk measures summarized in \Cref{lit} focuses mainly on the case where the aggregation function is concave and the acceptance set is convex, which results in a convex-valued systemic risk measure. In this paper, we handle aggregation functions that are non-concave and the corresponding systemic risk measures fail to have convex values, in general. To do so, we assume that the aggregation function is defined through a certain mixed-integer programming problem. In \Cref{systemic_risk_models}, we will study a concrete network model whose aggregation function fits into this abstract form and is non-concave; the Rogers-Veraart model will be included as a special case.

The proofs of all results in this section are given in \Cref{appendixB}.

Let us introduce the notation of the paper. The power set of a set $A$ is denoted by $2^A$. We denote by $\NN,\ZZ,\RR$ the sets of all natural numbers (starting from $1$), integers, real numbers, respectively. Given $a, b\in\RR$, we write $a\wedge b = \min\{a,b\}$, $a\vee b = \max\{a,b\}$, $a^+ = 0\vee a$, and $a^- = 0 \vee (-a)$. Let $n\in\NN$. Given $\bm{a}=\transpose{(a_1,\ldots,a_n)}, \bm{b}=\transpose{(b_1,\ldots,b_n)}\in\RR^n$, we write $\bm{a}\wedge\bm{b} = \transpose{(a_1\wedge b_1, \ldots, a_n\wedge b_n)}$, $\bm{a}\vee\bm{b} = \transpose{(a_1\vee b_1, \ldots, a_n\vee b_n)}$, $\bm{a}^+ = \zero_n \vee \bm{a}$, and $\bm{a}^- =\zero_n \vee (-\bm{a})$, where $\zero_n = \transpose{(0,\ldots,0)}\in\RR^n$. We sometimes use $\one_n = \transpose{(1,\ldots,1)}\in\RR^n$ and the $n\times n$ identity matrix $\bm{I}_n$ as well. We write $\bm{a}\le\bm{b}$ if and only if $a_i\le b_i$ for each $i\in\{1,\ldots,n\}$. In this case, we also define the hyperrectangle $[\bm{a},\bm{b}] = [a_1,b_1]\times\ldots\times[a_n,b_n]\subseteq \RR^n$. Using $\leq$ on $\RR^n$, we define $\RR^n_+\coloneqq \{\bm{x}\in\RR^n\mid \zero_n\leq \bm{x}\}$, whose elements are said to be \emph{positive}. We also define $\RR^n_{++}\coloneqq\{\bm{x}\in\RR^n\mid \forall i\in \{1,\ldots,n\}\colon  0<x_i\}$. For $A,B\subseteq \RR^n$ and $\bm{x}\in\RR^n$, we write $A+B=\{\bm{a}+\bm{b}\mid \bm{a}\in A,\ \bm{b}\in B\}$ and $\bm{x}+A= \{\bm{x}\}+A$; we denote by $\Int(A)$ the interior of $A$. Given a function $f\colon\RR^n\to\RR$, we say that $f$ is strictly increasing if $\bm{a}\le\bm{b}$ and $\bm{a}\neq\bm{b}$ imply $f\of{\bm{a}}<f\of{\bm{b}}$ for every $\bm{a},\bm{b}\in\RR^n$.

Let $(\Omega,\FFF,\mathbb{P})$ be a probability space. In particular, $\Omega$ can be either countable or uncountable. We denote by $L^\infty(\RR^n)$ the Banach space of all random vectors $\bm{X}=\transpose{(X_1,\ldots,X_n)}\colon\Omega\to\RR^n$ that are identified up to $\mathbb{P}$-almost sure equality and with finite $L^\infty$-norm $\norminfX$, where
	\begin{equation*}
		\norminfX \coloneqq \inf\{c>0\mid \mathbb{P}\{\abs{X_1}\leq c,\ldots, \abs{X_n}\leq c\}=1\}.
	\end{equation*}
	When $\bm{X}$ is deterministic, $\norminfX$ coincides with the $\ell^\infty$-norm of $\bm{X}$. Throughout, (in)equalities between random variables are understood in the $\mathbb{P}$-almost sure sense. We also recall some elements of measurability for set-valued functions, the reader is referred to \citet[Section 1.3]{molchanov} for further details. Let $m\in\NN$ and consider a set-valued function $\mathcal{D}\colon \Omega\to 2^{\RR^m}$ with closed values, i.e., $\mathcal{D}(\omega)$ is a closed set for every $\omega\in\Omega$. We say that $\mathcal{D}$ is \emph{measurable} if $\mathcal{D}^{-1}(A)\coloneqq\{\omega\in\Omega\mid \mathcal{D}(\omega)\cap A\neq\emptyset\}\in\FFF$ for every closed set $A\subseteq \RR^m$. In this case, a random vector $\bm{Y}\colon\Omega\to \RR^m$ is called a \emph{measurable selection} of $\mathcal{D}$ if $\bm{Y}(\omega)\in \mathcal{D}(\omega)$ for $\mathbb{P}$-almost every $\omega\in\Omega$. Let $L^\infty(\mathcal{D})$ denote the set of all measurable selections of $\mathcal{D}$ in $L^\infty(\RR^m)$, i.e.,
	\[
	L^\infty(\mathcal{D})\coloneqq\{\bm{Y}\in L^\infty(\RR^m)\mid \mathbb{P}\{\omega\in\Omega\mid \bm{Y}(\omega)\in \mathcal{D}(\omega)\}=1\}.
	\]

Without specifying a particular model, we consider a financial network with $n$ institutions. The indices of the nodes are collected in the set $\NNN = \cb{1,\ldots,n}$. 
We fix a random vector $\bm{X}\in L^\infty(\RR^n)$ that is used as the \emph{random operating cash flow vector} of the network, i.e., $X_i(\omega)$ is the value of the assets of node $i\in\NNN$ once scenario $\omega\in\Omega$ is observed as a consequence of a \emph{random shock}. Accordingly, each node may or may not be able to meet its obligations within the network.

We use the notion of grouping, also discussed in \cite{feinstein}, to keep the dimension of the systemic risk measure at a reasonable level for computational purposes. This notion allows one to categorize the members of the network into groups and assign the same capital level for all the members of a group. To that end, let $G\in\NN$ and denote by $\GGG=\cb{1,\ldots,G}$ the set of all groups in the network. For the computations in \Cref{computational_results}, we will use $G=2$ or $G=3$ groups. Let $\of{\NNN_\ell}_{\ell\in\GGG}$ be a partition of $\NNN$, where $\NNN_\ell$ denotes the set of all institutions that belong to group $\ell\in\GGG$. Without loss of generality, we assume that node indices respect the order of group indices, i.e., for every $\ell_1,\ell_2\in\GGG$ and $i_1\in\NNN_{\ell_1}, i_2\in\NNN_{\ell_2}$, we have $i_1<i_2$ if $\ell_1<\ell_2$. For each $\ell\in\GGG$, let $n_\ell\coloneqq|\NNN_\ell|$ and denote by $\bm{B}_\ell\in\RR^{G\times n_\ell}$ the matrix having 1's in the $\ell$\textsuperscript{th} row and 0's elsewhere.
Let $\bm{B}\in\RR^{G\times n}$ be the \emph{grouping matrix} defined by
\begin{equation}\label{grouping_matrix_B}
	\bm{B} \coloneqq 
	\sqb{\bm{B}_1  \ \ldots \ \bm{B}_G}.
\end{equation}
Accordingly, a capital allocation vector $\bm{z}\in\RR^G$ for the groups corresponds to a capital allocation vector $\bm{B}^{\mathsf{T}}\bm{z}\in\RR^n$ for the institutions. In some practical situations, it might be more natural or desirable to search for capital allocation vectors already within a restricted domain in $\RR^G$. For this purpose, we fix a nonempty closed convex set $\ZZZ\subseteq\RR^G$ such that $\ZZZ+\RR^G_+\subseteq \ZZZ$. An important special case is $\ZZZ=\bm{z}^{\text{LB}}+\RR^G_+$, where $\bm{z}^{\text{LB}}\in\RR^G$ is a predefined lower bound on the capital allocation vectors.

The central object of this paper is the \emph{sensitive systemic risk measure} defined by
\begin{equation}\label{sensitive_set-valued}
	\Rsen\of{\bm{X}}\coloneqq\big\{\bm{z}\in\ZZZ\mid \Lambda(\bm{X}+\transpose{\bm{B}}\bm{z})\in\AAA\big\},
\end{equation}
where $\Lambda\colon \RR^n\to\RR\cup\cb{-\infty}$ is an aggregation function and $\AAA\subseteq L^\infty(\RR)$ is an \emph{acceptance set}, that is, the set of all random aggregate outputs that are at an acceptable level of risk.

In \eqref{sensitive_set-valued}, we assume that $\Lambda$ is a general  aggregation function defined through a parametric family of optimization problems as follows:
\begin{equation}\label{aggregation_OPT}
	\Lambda\of{\bm{x}} \coloneqq \sup\cb{f\of{\bm{p}}\mid \of{\bm{p}, \bm{s}}\in\YYY\of{\bm{x}}, \bm{p}\in\RR^{n}, \bm{s}\in\ZZ^{d}}.
\end{equation}
Here, $f\colon\RR^{n}\to\RR$ is a strictly increasing and continuous function, $d\in\NN\cup\{0\}$, and $\YYY\colon \RR^n\to 2^{\RR^{n}\times\ZZ^{d}}$ is a set-valued constraint function. We work under the following assumption.

\begin{assumption}\label{assumption}
(i) The set $\gr \YYY\coloneqq \{(\bm{x},\bm{p},\bm{s})\in \RR^n\times\RR^n\times\ZZ^d \mid (\bm{p},\bm{s})\in\YYY(\bm{x})\}$, called the graph of $\YYY$, is closed. (ii) There exists a nonempty compact set $\bar{\YYY}\subseteq \RR^n$ such that $\YYY(\bm{x})\subseteq \bar{\YYY}$ for every $\bm{x}\in\RR^n$.
\end{assumption}

We comment on the implications of \Cref{assumption} in the next remark.

\begin{remark}\label{rem:graph}
	By \citet[Corollary 14.14]{rockafellar-wets}, \Cref{assumption}(i) ensures that the set-valued function $\omega \mapsto \YYY(\bm{Y})(\omega)\coloneqq \YYY(\bm{Y}(\omega))$ is measurable whenever $\bm{Y}\in L^\infty(\RR^n)$. Moreover, \Cref{assumption}(ii) implies that every measurable selection of $\YYY(\bm{Y})$ is in $L^\infty(\RR^n\times\ZZ^d)$. In particular, $L^\infty(\YYY(\bm{Y}))\neq\emptyset$. Finally, (i) and (ii) together imply that $\YYY\of{\bm{x}}$ is either the empty set or a nonempty compact set for every $\bm{x}\in\RR^n$. Hence, when $\YYY(\bm{x})=\emptyset$, we have $\Lambda(\bm{x})=-\infty$ by the usual convention for supremum; when $\YYY(\bm{x})\neq\emptyset$, we have $\Lambda(\bm{x})\in\RR$ and the supremum in \eqref{aggregation_OPT} is attained.
\end{remark}

Moreover, in \eqref{sensitive_set-valued}, we assume that $\AAA$ is a halfspace-type acceptance set defined by
\begin{equation}\label{acceptance_set}
	\AAA = \cb{Y \in L^\infty(\RR) \mid \EE\sqb{Y}\ge\gamma},
\end{equation}
where $\EE$ denotes the expectation operator with respect to $\PP$ and $\gamma\in\RR$ is a suitable threshold for aggregate values. From a risk measurement point of view, this choice of $\AAA$ is quite restrictive as it corresponds to a linear (risk-neutral) evaluation of risk. To limit the technical exposition, we work under this structure in the main text and consider a more general structure in \Cref{sec:polyrisk}, as explained in the next remark.

\begin{remark}\label{rem:extension}
	More generally, one can choose $\AAA$ as the acceptance set of a \emph{monetary risk measure} $\rho\colon L^\infty(\RR)\to\RR$ in the sense of \citet[Section 4.1]{stoch.finance}: $\AAA=\{Y\in L^\infty(\RR) \mid \rho(Y)\leq 0\}$. More precisely, such $\rho$ is \emph{monotone}, i.e., $Y\geq Y^\prime$ implies $\rho(Y)\leq \rho(Y^\prime)$ for every $Y,Y^\prime\in L^\infty(\RR)$, and it is \emph{translative}, i.e., $\rho(Y+r)=\rho(Y)-r$ for every $Y\in L^\infty(\RR)$ and $r\in\RR$. As a special case, taking $\rho(Y):=\gamma - \EE[Y]$, $Y\in L^\infty(\RR)$, reproduces \eqref{acceptance_set}. In \Cref{sec:polyrisk}, we extend our analysis to the case where $\rho$ is a polyhedral risk measure, which also covers average value-at-risk as an example.
\end{remark}

With \eqref{acceptance_set}, the systemic risk measure in \eqref{sensitive_set-valued} becomes
\begin{equation}\label{senSystemicRiskMeasureOPT}
	\Rsen\of{\bm{X}} = \big\{\bm{z}\in \ZZZ\mid\EE[\Lambda(\bm{X}+\transpose{\bm{B}}\bm{z})]\ge\gamma\big\}.
\end{equation} 
Let us consider the vector optimization problem
\begin{equation}\label{vector_optimization_OPT}
		\text{minimize} \quad \bm{z}\in\ZZZ\quad \text{with respect to}\quad \le\quad  
		\text{subject to}\quad \EE[\Lambda(\bm{X}+\transpose{\bm{B}}\bm{z})]\ge\gamma,
\end{equation}
where we recall that $\leq$ denotes the componentwise ordering on $\RR^G$. Note that $\Rsen(\bm{X})$ coincides with the so-called \emph{upper image} of this vector optimization problem in the sense that
\begin{equation}\label{upperset}
	\Rsen(\bm{X}) = \big\{\bm{z}\in\ZZZ\mid \EE[\Lambda(\bm{X}+\transpose{\bm{B}}\bm{z})]\ge\gamma\big\}+\RR^G_+.
\end{equation}
Next, we provide a reformulation of the systemic risk measure by introducing binary variables.

\begin{theorem}\label{theorem_OPT}
	Suppose that \Cref{assumption} holds. Let $\bm{X}\in L^\infty(\RR^n)$ with $R(\bm{X})\neq \emptyset$, i.e., there exists $\bm{z}\in\ZZZ$ such that $\EE[\Lambda(\bm{X}+\transpose{\bm{B}}\bm{z})]\geq \gamma$. Then,
		\[
		R(\bm{X}) = \cb{\bm{z}\in\ZZZ\mid \EE[f(\bm{P})]  \ge \gamma,\ (\bm{P}, \bm{S})\in L^\infty(\YYY(\bm{X}+\transpose{\bm{B}}\bm{z})),\ \bm{P}\in L^\infty(\RR^{n}),\ \bm{S}\in L^\infty(\ZZ^{d})}.
		\]
\end{theorem}

The representation of $R(\bm{X})$ in \Cref{theorem_OPT} can be seen as a \emph{two-stage mixed-integer stochastic programming formulation} of the vector optimization problem in \eqref{vector_optimization_OPT}: the capital allocation vector $\bm{z}$ is determined in the first-stage; the realizations of $\bm{P}, \bm{S}$ are determined in the second-stage once the uncertainty is resolved.

In the next two subsections, we establish mathematical programming characterizations of certain scalarization problems associated with $\Rsen(\bm{X})$ by exploiting the structure of the optimization-based aggregation function $\Lambda$.

\subsection{Weighted-Sum Scalarizations}\label{secP1}

For each $\bm{w}\in\RR^G_+\backslash\cb{\zero_G}$, we consider the weighted-sum scalarization problem
\begin{equation}\label{P1_OPT}
	\PPP_1\of{\bm{w}} = \inf_{z\in\Rsen(\bm{X})}\transpose{\bm{w}}\bm{z}=\inf_{\bm{z}\in\ZZZ} \big\{\transpose{\bm{w}}\bm{z} \mid \EE[\Lambda(\bm{X}+\transpose{\bm{B}}\bm{z})]\ge\gamma\big\}.
\end{equation}

The following corollary provides an alternative formulation for $\PPP_1\of{\bm{w}}$.

\begin{corollary}\label{theorem_P1_OPT}
	Suppose that \Cref{assumption} holds. Let $\bm{X}\in L^\infty(\RR^n)$ with $R(\bm{X})\neq \emptyset$. Let $\bm{w}\in\RR^G_+\backslash\cb{\zero_G}$ and consider the problem
		\begin{align}\tag{$\text{P}_1(\bm{w})$}\label{z_OPT}
			\emph{minimize}\quad &\transpose{\bm{w}}\bm{z} && \\ 
			\emph{subject to}\quad&\EE[f(\bm{P})]  \ge \gamma, && \notag \\
			&(\bm{P}, \bm{S})\in L^\infty(\YYY(\bm{X}+\transpose{\bm{B}}\bm{z})),&& \notag \\ 
			& \bm{z}\in \ZZZ, \quad \bm{P}\in L^\infty(\RR^{n}),\quad \bm{S}\in L^\infty(\ZZ^{d}).&&\notag 
		\end{align}
		Then, the optimal value of \eqref{z_OPT} equals $\PPP_1\of{\bm{w}}$. Moreover, if $\ZZZ$ is a lower bounded set in the sense that $\ZZZ\subseteq\bm{z}^{\textnormal{LB}}+\RR^G_+$ for some $\bm{z}^{\textnormal{LB}}\in\RR^G$, then $\PPP_1(\bm{w})\in\RR$.
\end{corollary}

The problem \eqref{z_OPT} can be infinite-dimensional in general. The next remark discusses an important special case which we will focus on in the computational experiments of \Cref{computational_results}.

\begin{remark}\label{rem:P1finite}
	When $\Omega$ is a finite sample space with $\FFF=2^{\Omega}$, \eqref{z_OPT} reduces to a finite-dimensional optimization problem.
\end{remark}

\begin{remark}\label{ideal}
	For each group $\ell\in\GGG$, let $\bm{e}^\ell\in\RR^G$ be the corresponding standard unit vector and assume that $\PPP_1(\bm{e}^\ell)\in\RR$. Then, we may define the ideal point $\bm{z}^\text{ideal}\in\RR^G$ of the vector optimization problem in \eqref{vector_optimization_OPT} as the point obtained by minimizing each component of the objective function, i.e.,
		\begin{equation}
			\bm{z}^\text{ideal} \coloneqq \transpose{\of{\PPP_1(\bm{e}^1),\ldots,\PPP_1(\bm{e}^G)}}\in\RR^G.
		\end{equation}
		By \Cref{theorem_P1_OPT}, $\bm{z}^\text{ideal}$ is well-defined when $\ZZZ$ is a lower bounded set, and one can solve $G$ optimization problems,
		namely, the problems $(\text{P}_1(\bm{e}^\ell))$, $\ell\in\GGG$, to calculate $\bm{z}^\text{ideal}$. It should be noted that, when $\bm{z}^{\text{ideal}}$ is well-defined, it may not be an element of $R(\bm{X})$, in general. When $\ZZZ=\RR^G$ and $\bm{z}^{\text{ideal}}$ is well-defined, working with $\ZZZ=\RR^G$ is equivalent to working with $\ZZZ=\bm{z}^{\text{LB}}+\RR^G_+$, where $\bm{z}^{\text{LB}}\in\RR^G$ is a lower bound on $\bm{z}^\text{ideal}$, i.e., $\bm{z}^{\text{LB}}\leq \bm{z}^\text{ideal}$. The existence of the ideal point (or lower boundedness) is usually assumed in the literature to ensure that an approximation algorithm converges in finite time, which is also the case in \cite{non-conv.benson} (see \Cref{algorithm}). However, the ideal point may not be well-defined in general when $\ZZZ=\RR^G$. In such cases, $R(\bm{X})$, the upper image of the vector optimization problem, is ``too large" to be approximated by these algorithms.
\end{remark}

\begin{remark}\label{rem:capitalP1}
	Weighted-sum scalarizations can be used to construct simple capital allocation rules. In some practical situations, the decision-maker (e.g., a regulator) may wish to have a capital allocation rule whose output is a single vector rather than a set. In this case, we may choose $w_\ell=|\NNN_\ell|$ for each $\ell\in\GGG$ so that $\transpose{\bm{w}}\bm{z}$ corresponds to the total capital allocated to the network. With this choice of $\bm{w}$, let $\text{CAR}_1^{\bm{w}}(\bm{X})$ denote an arbitrarily fixed minimizer of \eqref{z_OPT}. Then, the mapping $\bm{X}\mapsto \text{CAR}_1^{\bm{w}}(\bm{X})$ defines an \emph{efficient capital allocation rule} in the sense of \citet[Definition 3.3]{feinstein}. When the systemic risk measures are convex-valued, under certain regularity conditions on the recession cone of $R(\bm{X})$, \citet[Lemma 3.9]{feinstein} shows that every efficient capital allocation rule is in the form of a weighted-sum scalarization problem ($\bm{w}$ may depend on $\bm{X}$ though), which is not the case in our nonconvex framework.
	\end{remark}

\subsection{Pascoletti-Serafini Scalarizations}\label{secP2}

Weighted-sum scalarizations are used to calculate supporting hyperplanes for the value of a systemic risk measure and they can be sufficient to characterize the entire risk set when the set is convex. In our nonconvex case, we make use of additional scalarizations that are used to calculate the minimum step-lengths to enter the set from possibly outside points. Such scalarizations are well-known in vector optimization; see \citet{pascolettiserafini}, \citet{p2_reference_1}, \citet{p2_reference_2}, for instance.

Let us fix a direction vector $\bm{c}\in\RR^G_+\setminus\{\bm{0}_G\}$. For each $\bm{v}\in\ZZZ$, we consider the Pascoletti-Serafini scalarization problem
\begin{align}\label{P2_OPT}
	\PPP^{\bm{c}}_2\of{\bm{v}} &\coloneqq\inf\big\{\mu\in\RR\mid \bm{v}+\mu\bm{c}\in \Rsen\of{\bm{X}}\big\}\\
	&= \inf \big\{\mu\in\RR \mid\EE[\Lambda(\bm{X}+\transpose{\bm{B}}(\bm{v}+\mu\bm{c}))]\ge\gamma,\ \bm{v}+\mu\bm{c}\in \ZZZ\big\},\notag 
\end{align}
which can be interpreted as the minimum step-length in the direction $\bm{c}$ from the point $\bm{v}$ to hit the boundary of the set $\Rsen\of{\bm{X}}$. When computing systemic risk measures, we will consider the cases $\bm{c}=\one_G$, the vector of ones, and $\bm{c}=\bm{e}^{\ell}$, the $\ell^{\text{th}}$ standard unit vector in $\RR^G$ with $\ell\in \GGG$.

The following corollary provides an alternative formulation for $\PPP^{\bm{c}}_2\of{\bm{v}}$.

\begin{corollary}\label{theorem_P2_OPT}
	Suppose that \Cref{assumption} holds. Let $\bm{X}\in L^\infty(\RR^n)$ with $R(\bm{X})\neq \emptyset$. Let $\bm{c}\in\RR^G_+\setminus\{\zero_G\}$ and $\bm{v}\in\ZZZ$. Consider the problem
		\begin{align}\tag{$\text{P}^{\bm{c}}_2(\bm{v})$}\label{P2_OPT_lin}
			\emph{minimize}\quad &\mu\in\RR && \\ 
			\emph{subject to}\quad& \EE[f(\bm{P})] \ge \gamma, && \notag \\
			&(\bm{P}, \bm{S})\in L^\infty(\YYY(\bm{X}+\transpose{\bm{B}}(\bm{v}+\mu\bm{c}))),&&\notag \\ 
			& \bm{v}+\mu\bm{c}\in \ZZZ, \quad \bm{P}\in L^\infty(\RR^{n}),\quad \bm{S}\in L^\infty(\ZZ^{d}).&&\notag 
		\end{align}
		Then, the optimal value of \eqref{P2_OPT_lin} equals $\PPP^{\bm{c}}_2\of{\bm{v}}$. Moreover, if $\ZZZ$ is a lower bounded set, then $\PPP^{\bm{c}}_2(\bm{v})>-\infty$ and the following statements hold: (i) Suppose further that there exists $\bm{z}\in R(\bm{X})$ such that $z_\ell\leq v_\ell$ for every $\ell\in\GGG$ with $c_\ell=0$. Then, $\PPP^{\bm{c}}_2(\bm{v})\in\RR$. (ii) If $\bm{c}\in\RR^G_{++}$, then $\PPP_2^{\bm{c}}(\bm{v})\in\RR$.
\end{corollary}

\begin{remark}\label{rem:P2finite}
	Similar to \Cref{rem:P1finite}, \eqref{P2_OPT_lin} reduces to a finite-dimensional problem when $\Omega$ is finite.
\end{remark}

\begin{remark}\label{rem:capitalP2}
	Similar to weighted-sum scalarizations (see \Cref{rem:capitalP1}), Pascoletti-Serafini scalarizations can also be used to construct simple capital allocation rules. Suppose that the decision-maker has a pre-determined plan of using $\bm{v}\in\ZZZ$ as a capital allocation vector. This could be calculated based on the historical data of the institutions. In case no such plan exists, one may also start with $\bm{v}=\zero_G$. The decision-maker may choose to allocate additional budget only for the institutions in group 1, say, the group of big banks. Then, the minimal additional capital required for each big bank is given by $\PPP_2^{\bm{c}}(\bm{v})$, where $\bm{c}=\bm{e}^1$. Moreover, the mapping $\bm{X}\mapsto \text{CAR}_2^{\bm{c},\bm{v}}(\bm{X})\coloneqq \bm{v}+\PPP_2^{\bm{c}}(\bm{v})\bm{c}$ gives a \emph{weakly efficient capital allocation rule}, i.e., it satisfies \citet[Definition 3.3]{feinstein} except that the capital allocation vector is only a \emph{weakly} minimal (i.e., boundary) point of $R(\bm{X})$. For a similar purpose, functionals that are in the form of Pascoletti-Serafini scalarizations have also been considered in the literature on multi-asset markets under the name \emph{liquidation value}; see \cite{lepinette-nonconvex} for a recent discussion in nonconvex financial markets.
\end{remark}

\subsection{The Nonconvex Benson-Type Algorithm} \label{algorithm}

The mixed-integer programming formulations for the weighted-sum and Pascoletti-Serafini scalarizations developed in Sections~\ref{secP1}, \ref{secP2} can be used to solve these problems via commercial optimization software. In general, methods for solving the scalarizations can be embedded into any vector optimization meta-algorithm that makes use of these scalarizations. In this section, assuming that such methods are available, we formulate a procedure based on the Benson-type algorithm for nonconvex multi-objective programming problems developed in \citet{non-conv.benson}, which we describe briefly here. For a linear choice of the function $f$, we will argue in \Cref{sec:RVsc} that the scalarizations can indeed be solved as MILP problems when the signed Rogers-Veraart model is considered; we also comment on more efficient implementations for this model in \Cref{rem:bisection}. In particular, the procedure is guaranteed to work correctly in that setting.

Let $\LLL\subseteq\RR^G$. A point $\bm{v}\in \LLL$ is called a \emph{vertex} of $\LLL$ if there exists a neighborhood $N$ of $\bm{v}$ for which $\bm{v}$ cannot be expressed as a strict convex combination of two distinct points in $\LLL\cap N$. The set of all vertices of $\LLL$ is denoted by $\vertex(\LLL)$. Given a point $\bm{z}\in\RR^G$, we define $\LLL|_{\bm{z}} \coloneqq \cb{\bm{v}\in \LLL \mid  \bm{v}\leq\bm{z}}$.

We consider the approximation of the systemic risk measure $R(\bm{X})$ defined in \eqref{senSystemicRiskMeasureOPT}, where $\bm{X}\in L^\infty(\RR^n)$. We assume that \Cref{assumption} holds, $R(\bm{X})\neq\emptyset$, $\ZZZ=\bm{z}^{\text{LB}}+\RR^G_+$ for some $\bm{z}^{\text{LB}}\in \RR^G_+$, and the ideal point $\bm{z}^\text{ideal}$ is well-defined with $\bm{z}^{\text{LB}}\leq \bm{z}^{\text{ideal}}$. Hence, our results in Sections \ref{secP1}, \ref{secP2} are applicable. The approximation is performed with respect to a user-defined approximation error $\epsilon>0$ and an upper bound $\bm{z}^\text{UB}\in R(\bm{X})$ that limits the approximated region of $R(\bm{X})$. Let $\LLL,\UUU\subseteq\RR^G$ be given. We say that $\LLL$ is an \emph{outer approximation} for $\UUU$ if $\UUU \subseteq \LLL$ and $\LLL|_{\bm{z}^\text{UB}} \subseteq \UUU + B(\zero_G,\epsilon)$, where $B\of{\zero_G,\epsilon}$ is the closed Euclidean ball in $\RR^G$ centered at $\zero_G$ with radius $\epsilon$. We say that $\UUU$ is an \emph{inner approximation} for $\LLL$ if $\LLL$ is an outer approximation for $\UUU$. The aim is to find an inner and outer approximation for $R(\bm{X})$.

The pseudocode of the algorithm is provided in \Cref{alg1}. At initialization, the algorithm finds the ideal point $\bm{z}^\text{ideal}$ (see \Cref{ideal}), which can be computed by solving $G$ weighted-sum scalarizations (see \Cref{theorem_P1_OPT}). The upper bound vector is set as $\bm{z}^{\text{UB}}=\bm{z}^{\text{ideal}}+(\PPP_2^{\bm{e}^1}(\bm{z}^\text{ideal}),\ldots, \PPP_2^{\bm{e}^G}(\bm{z}^{\text{ideal}}))^{\mathsf{T}}$. The inner approximation is set as $\UUU^0 \coloneqq \bm{z}^\text{UB} + \RR^G_+$ and the outer approximation is set as $\LLL^0 \coloneqq \bm{z}^\text{ideal} + \RR^G_+$. Then, $\UUU^0, \LLL^0$ satisfy $\UUU^0  \subseteq R(\bm{X}) \subseteq \LLL^0$.

At an arbitrary iteration $t\in\NN$, for a vertex $\bm{v}^t \in \vertex( \mathcal{L}^t|_{\bm{z}^{\text{UB}}})$ such that $\bm{v}^t + \epsilon\one_G \notin \Int(\UUU^t)$, the algorithm solves a Pascoletti-Serafini scalarization to find $\PPP^{\bm{c}}_2(\bm{v}^t)$ with direction vector $\bm{c}=\one_G$ (see \Cref{theorem_P2_OPT}). With $\bm{y}^t \coloneqq \bm{v}^t + \PPP_2^{\one_G}(\bm{v}^t)\one_G$, which is a boundary point of $R(\bm{X})$, the algorithm updates the outer approximation by excluding the cone $\bm{y}^t - \RR^G_{++}$ from $\LLL^t$, and it updates the inner approximation by adding the cone $\bm{y}^t + \RR^G_+$ to $\UUU^t$ (line 8). Therefore, $\UUU^t \subseteq \UUU^{t+1} \subseteq R(\bm{X}) \subseteq \LLL^{t+1} \subseteq \LLL^t$. At the end of the iteration, $\vertex( \mathcal{L}^{t+1})$ is computed by a subroutine described in \citet{gourion-luc}. The algorithm stops at the first iteration $T\in\NN$ at which $\vertex( \mathcal{L}^T|_{\bm{z}^{\text{UB}}})+ \epsilon\one_G \subseteq \Int(\UUU^T)$ is achieved. The sets $\UUU^T$ and $\LLL^T$ are the returned inner and outer approximations of $R(\bm{X})$, respectively. Thanks to the existence of the ideal point, \citet[Theorem 4.1]{non-conv.benson} guarantees that the algorithm stops after finitely many iterations.

A graphical illustration of the algorithm is given in \Cref{sec:illustration} for the case $G=2$.

\begin{algorithm}[th]
	\caption{Inner and outer approximation algorithm for $R(\bm{X})$}	
	\algsetup{
		linenosize=\small,
		linenodelimiter=.
	}
	\label{alg1}
	\begin{algorithmic}[1]
		\STATE Compute $\bm{z}^{\text{ideal}}=\transpose{(\PPP_1(\bm{e}^1), \ldots, \PPP_1(\bm{e}^G))}$; $\bm{z}^{\text{UB}}=\bm{z}^{\text{ideal}}+(\PPP_2^{\bm{e}^1}(\bm{z}^\text{ideal}),\ldots, \PPP_2^{\bm{e}^G}(\bm{z}^{\text{ideal}}))^{\mathsf{T}}$.
		\STATE Fix $\epsilon>0$; set $\LLL^0 = \bm{z}^\text{ideal} + \RR^G_+$, $\UUU^0 = \bm{z}^\text{UB} + \RR^G_+$, $t = 0$, $S=\emptyset$;
		\REPEAT
		\STATE Choose $\bm{v}^t \in \vertex(\LLL^t|_{\bm{z}^\text{UB}})\setminus S$;
		\IF{$\bm{v}^t + \epsilon\one_G \in \Int(\UUU^t)$}
		\STATE $S\leftarrow S\cup\cb{\bm{v}^t}$;
		\ELSE 
		\STATE Compute $\PPP^{\one_G}_2(\bm{v}^t)$; set $\bm{y}^t =\bm{v}^t + \PPP^{\one_G}_2(\bm{v}^t)\one_G$, $\LLL^{t+1} \coloneqq \LLL^t \setminus \of{\bm{y}^t - \RR^G_{++}}$, $\UUU^{t+1} \coloneqq \UUU^t \cup \of{\bm{y}^t+\RR^G_+}$; compute $\vertex(\LLL^{t+1})$; $t \leftarrow t + 1$;
		\ENDIF
		\UNTIL{$\vertex(\LLL^t|_{\bm{z}^{\text{UB}}}) \subseteq S$}
		\RETURN		 
		$T=t$, $\LLL^T$, $\UUU^T$.
	\end{algorithmic}
\end{algorithm}

\section{Optimization Problems for the Signed Rogers-Veraart Model} \label{systemic_risk_models}

In this section, we propose a seniority-based extension of the Rogers-Veraart model by allowing signed operating cash flows and provide an MIP formulation of clearing vectors. Then, we consider the calculation of clearing vectors in three special cases where the MIP formulation simplifies: the original Rogers-Veraart model, a signed version of the Eisenberg-Noe model, and the original Eisenberg-Noe model.

The proofs of all results in this section are given in \Cref{appendixC}.

\subsection{The Signed Rogers-Veraart Model}\label{signed_rv_model}

We consider a network that represents a financial system with $n\in\NN$ institutions. The next definition extends the model in \cite{rogers-veraart} by considering signed operating cash flows.

\begin{definition}\label{signed_RVsystem}
	A quintuple $\systemRV$ is called a \emph{signed Rogers-Veraart network} if $\bm{x}=\transpose{\of{x_1,\ldots,x_n}}\in\RR^n$, $\bm{{\bar{p}}}=\transpose{\of{\bar{p}_1,\ldots,\bar{p}_n}}\in\RR^n_{++}$, $\bm{\pi} = \of{\pi_{ij}}_{i,j\in\NNN}\in\RR^{n\times n}$ is a right stochastic matrix with $\pi_{ii} = 0$ for each $i\in\NNN$, and $\alpha,\beta\in(0,1]$. A signed Rogers-Veraart network $\systemRV$ is called a \emph{Rogers-Veraart network} if $\bm{x}\in\RR^n_+$, it is called a \emph{signed Eisenberg-Noe network} if $\alpha=\beta=1$, and it is called an \emph{Eisenberg-Noe network} if $\bm{x}\in\RR^n_+$ and $\alpha=\beta=1$.
\end{definition}

In \Cref{signed_RVsystem}, for every $i\in\NNN$, $\bar{p}_i > 0$ denotes the total amount of liabilities to the other nodes in the system and $x_i\in\RR$ denotes the operating cash flow of node $i$. More precisely, $x_i$ can be considered as the difference between the total value of the assets of node $i$ that are external to the network and the total cost associated with the business activities of node $i$. Hence, $x_i>0$ corresponds to the case where the external assets dominate the business costs and $x_i<0$ corresponds to the opposite case. We call $\bm{x}$ the \emph{operating cash flow vector} and $\bm{{\bar{p}}}$ the \emph{total obligation vector}.

For every $i,j\in\NNN$ with $i\neq j$, $\pi_{ij}$ denotes the fraction of the total liability of node $i$ owed to node $j$. In order for this to make sense, $\bm{\pi}$ is naturally assumed to be a right stochastic matrix; we call it the \emph{relative liabilities matrix}. The assumption $\pi_{ii} = 0$ means that node $i$ cannot have liabilities to itself. From these two assumptions, it is immediate that $\sum_{j=1}^{n}\pi_{ji} < n$, i.e., no node owns all the claims in the network. Note that, given $\bm{{\bar{p}}}$ and $\bm{\pi}$, the nominal liability $l_{ij}$ of node $i$ to node $j$ can be calculated as $l_{ij} = \pi_{ij}\bar{p}_i$.

It is assumed that a defaulting node may not be able to use all of its liquid assets to meet its obligations. For this purpose, we use $\alpha$ as the fraction of the positive operating cash flow and $\beta$ as the fraction of the cash inflow from other nodes that can be used by a defaulting node to meet its obligations. Logically, no default costs are incurred when the operating cash flow is negative. For convenience, we introduce the piecewise linear function $\bm{\varphi}^\alpha=\transpose{(\varphi^\alpha_1,\ldots,\varphi^\alpha_n)}\colon\RR^n\to\RR^n$ defined by
\begin{equation}\label{phialpha}
	\bm{\varphi}^\alpha(\bm{x})\coloneqq (\alpha \bm{x} )\wedge \bm{x}.
\end{equation}

Let $\systemRV$ be a signed Rogers-Veraart network. For each $i\in\NNN$, let $p_i\ge0$ be the sum of all payments made by node $i$ to the other nodes in the network. Then, $\bm{p} = \transpose{\of{p_1, \ldots, p_n}}\in\RR^n_+$ is called a \emph{payment vector}. It is assumed that all internal liabilities of node $i$ are of equal seniority so that the payment to every other node $j\in\NNN$ will be proportional to the relative liability $\pi_{ij}$. We define a clearing (payment) vector next.

\begin{definition}\label{clearing_vector_defn_signed_RV}
	A vector $\bm{p}\in\sqb{\zero_n,\bm{{\bar{p}}}}$ is called a \emph{clearing vector} for $\systemRV$ if it satisfies the following properties for every $i\in\NNN$:
	\begin{itemize}
		\item \emph{Immediate default:} If $(\bm{x}+\bm{\pi}\transposeT\bm{p})_i < 0$, then $p_i = 0$.
		\item \emph{Partial liquidity:} If $0 \le (\bm{x}+\bm{\pi}\transposeT\bm{p} )_i < \bar{p}_i$, then $p_i = \of{\bm{\varphi}^\alpha(\bm{x}) + \beta\bm{\pi}\transposeT\bm{p}}_i^+$.
		\item \emph{Full commitment:} If $(\bm{x}+\bm{\pi}\transposeT\bm{p})_i \ge \bar{p}_i$, then $p_i = \bar{p}_i$.
	\end{itemize}
\end{definition}

In \Cref{clearing_vector_defn_signed_RV}, \emph{immediate default} asserts that a node with negative total cash inflow defaults immediately without meeting any of its obligations. This property amounts to saying that, although all internal liabilities of a node $i\in\NNN$ are of equal seniority, a negative operating cash flow $x_i<0$ has seniority over the internal liabilities of node $i$ and no internal payment is made before the net operating cost $-x_i$ is paid in full. For this reason, we refer to our approach as a \emph{seniority-based approach}; see \Cref{rem:naiveapproach} for an alternative approach. In the case of positive total cash inflow, we further compare it with the total obligation. If the total cash inflow is not sufficient to cover all obligations, then \emph{partial liquidity} states that the node may not be able to utilize all of its assets for payments due to default costs. More precisely, these costs are incurred for a positive operating cash flow and for the payments that are received from the other nodes. After default cost reductions, the remaining value of the total cash inflow is used to meet some of the obligations provided that this value is positive. Finally, if the total cash inflow is sufficient to cover all obligations, then there is \emph{full commitment} to do so.

\begin{remark}\label{rem:lim_absRV}
	It is easy to see that \emph{partial liquidity} and \emph{full commitment} imply the following:
	\begin{itemize}
		\item \emph{Limited liability:} If $(\bm{x}+\transpose{\bm{\pi}}\bm{p})_i \ge 0$, then $p_i \le (\bm{x}+\transpose{\bm{\pi}}\bm{p} )_i$. 
	\end{itemize}
	Indeed, if $0 \le (\bm{x}+\bm{\pi}\transposeT\bm{p} )_i < \bar{p}_i$, then \emph{partial liquidity} implies $p_i = \of{\bm{\varphi}^\alpha(\bm{x})+ \beta\bm{\pi}\transposeT\bm{p}}_i^+\leq \of{\bm{x} + \bm{\pi}\transposeT\bm{p}}_i^+=\of{\bm{x} + \bm{\pi}\transposeT\bm{p}}_i$; if $(\bm{x}+\bm{\pi}\transposeT\bm{p})_i \ge \bar{p}_i$, then \emph{full commitment} implies $p_i=\bar{p}_i\leq \of{\bm{x} + \bm{\pi}\transposeT\bm{p}}_i$.
	\emph{Limited liability} ensures that a node cannot pay more than it has. Suppose further that $\bm{x}\in\RR^n_+$, which corresponds to the setting in \cite{rogers-veraart}. In this case, \emph{partial liquidity} and \emph{full commitment} also imply the following:
	\begin{itemize}
		\item \emph{Absolute priority:} If $(\bm{x}+\transpose{\bm{\pi}}\bm{p})_i \ge 0$, then either $p_i = \bar{p}_i$ or $p_i = (\alpha\bm{x} + \beta\bm{\pi}\transposeT\bm{p})_i$.
	\end{itemize}
	Indeed, if the hypothesis of full commitment holds, then $p_i=\bar{p}_i$; if the hypothesis of partial liquidity holds, then $p_i = \of{\bm{\varphi}^\alpha(\bm{x}) + \beta\bm{\pi}\transposeT\bm{p}}_i^+=\of{\alpha\bm{x}+ \beta\bm{\pi}\transposeT\bm{p}}_i^+=\of{\alpha\bm{x}+ \beta\bm{\pi}\transposeT\bm{p}}_i$ since $x_i\geq 0$. \emph{Absolute priority} asserts that a node either meets its obligations in full or else it pays as much as it has after default cost reductions. Conversely, \emph{absolute priority} and \emph{limited liability} imply \emph{partial liquidity}.
	Indeed, if $0 \le (\bm{x}+\bm{\pi}\transposeT\bm{p} )_i < \bar{p}_i$, then limited liability yields $p_i\neq \bar{p}_i$ and absolute priority leaves us with $p_i = (\alpha\bm{x} + \beta\bm{\pi}\transposeT\bm{p})_i=(\bm{\varphi}^\alpha(\bm{x})+ \beta\bm{\pi}\transposeT\bm{p})^+_i$ since $x_i\geq 0$. However, \emph{absolute priority} and \emph{limited liability} do not imply \emph{full commitment} in general. Here is a counterexample. Consider a two-node network $\systemRV$, where $\bm{x}=(10,10)^{\mathsf{T}}$, $\pi_{11}=\pi_{22}=0$, $\pi_{12}=\pi_{21}=1$, $\bm{p}=(20,15)^{\mathsf{T}}$, $\alpha=\beta=0.5$ and let $\bm{{\bar{p}}}=(20,25)^{\mathsf{T}}$. Then, $\bm{p}$ satisfies \textit{absolute priority} and \textit{limited liability}. However, we have $(\bm{x}+\transpose{\bm{\pi}}\bm{p})_2=30\geq 25=\bar{p}_2$ but $p_2=15\neq 25 = \bar{p}_2$. Hence, \emph{full commitment} is violated.
\end{remark}

\begin{remark}\label{rem:lim_absEN}
	As a continuation of \Cref{rem:lim_absRV}, when $\bm{x}\in\RR^n_+$ and $\alpha=\beta=1$, the signed Rogers-Veraart model reduces to the Eisenberg-Noe model, where a clearing vector is defined as any vector that satisfies \emph{limited liability} and \emph{absolute priority} \citep[Definition 1]{eisenberg-noe}. In our setting, \Cref{clearing_vector_defn_signed_RV} reduces to this definition when the Eisenberg-Noe model is considered.
\end{remark}

\begin{remark}\label{rem:naiveapproach}
	In \citet[Section 2.2]{eisenberg-noe}, it is argued that the operating cash flow vector $\bm{x}$ is assumed to be positive \emph{without loss of generality} since a strictly negative operating cash flow of any node can be regarded as its liability to an artificial ``sink node." The sink node itself does not have any liabilities or operating cash flow. We call this approach the \emph{sink node approach}. In \Cref{sec:naiveapproach}, we include a detailed discussion of the sink node approach in the setting of the Eisenberg-Noe model and argue that the seniority-based approach proposed in this paper has significant advantages over the sink node approach from modeling and computational perspectives.
\end{remark}

Let $\bm{\Phi}^{\text{RV}}=\transpose{(\Phi^{\text{RV}}_1,\ldots,\Phi^{\text{RV}}_n)}\colon[\zero_n,\bm{{\bar{p}}}]\to[\zero_n,\bm{{\bar{p}}}]$ be defined by
\begin{equation}\label{fixed_point_signed_RV}
	\Phi^{\text{RV}}_i\of{\bm{p}} \coloneqq
	\begin{cases}
		0 & \quad\text{if }(\bm{x}+\bm{\pi}\transposeT\bm{p})_i < 0, \\
		\of{\bm{\varphi}^\alpha(\bm{x}) + \beta\bm{\pi}\transposeT\bm{p}}_i^+ & \quad\text{if }0 \le (\bm{x}+\bm{\pi}\transposeT\bm{p})_i < \bar{p}_i, \\
		\bar{p}_i & \quad\text{if }(\bm{x}+\bm{\pi}\transposeT\bm{p} )_i \ge \bar{p}_i,
	\end{cases}
\end{equation}
for each $i\in\NNN$. By \Cref{clearing_vector_defn_signed_RV}, it is immediate that a vector $\bm{p}\in [\zero_n,\bm{{\bar{p}}}]$ is a clearing vector for $\systemRV$ if and only if $\bm{p}$ is a fixed point of $\bm{\Phi}^{\text{RV}}$, that is, $\bm{\Phi}^{\text{RV}}(\bm{p})=\bm{p}$.

\begin{remark}\label{rem:RVdefn}
	When $\bm{x}\in\RR^n_+$, the above fixed point characterization of a clearing vector reduces to the definition of a clearing vector in \citet[Definition 2.6]{rogers-veraart}.
	\end{remark}

\begin{remark}\label{rem:gamma1}
	All nodes meet their liabilities in full if and only if $\bm{{\bar{p}}}$ is a clearing vector, or equivalently, $\bm{\Phi}^{\text{RV}}(\bm{{\bar{p}}})=\bm{{\bar{p}}}$, i.e., $\bm{x}\geq (\bm{I}_n-\bm{\pi})^{\mathsf{T}}\bm{{\bar{p}}}$. In other words, $(\bm{I}_n-\bm{\pi})^{\mathsf{T}}\bm{{\bar{p}}}$ is the (componentwise) minimum operating cash flow vector which ensures that all nodes are in full commitment.
	\end{remark}

The next theorem is the main result of this section and it provides a mathematical programming formulation for calculating clearing vectors.

\begin{theorem}\label{signed_RV_theorem}
	(Signed Rogers-Veraart model) Let $\bm{x}\in\RR^n$ and denote by $\Lambda^{\textnormal{RV}}(\bm{x})$ the optimal value of the problem
	\begin{align}\tag{RV$(\bm{x})$}\label{signed_RV_MILP_explicit}
		\emph{maximize} \quad 			& f(\bm{p}) && \\ 
		\emph{subject to} \quad 	& \bar{p}_is_i - m_i t_i \le x_i + (\bm{\pi}\transposeT\bm{p})_i, &&i\in\NNN,  \label{signed_RV_MILP_constraint_1} \\ 
		& \bar{p}_is_i\le p_i \le \bar{p}_i(1 - t_i), &&i\in\NNN,  \label{signed_RV_MILP_constraint_2}\\ 
		& p_i \le \varphi^\alpha_i(\bm{x}) + \beta(\bm{\pi}\transposeT\bm{p})_i + (m_i+\bar{p}_i)(s_i + t_i), &&i\in\NNN,  \label{signed_RV_MILP_constraint_4} \\ 
		& 0 \le p_i \le \bar{p}_i,\quad s_i, t_i \in \cb{0,1}, && i\in\NNN,\label{signed_RV_MILP_constraint_7}
	\end{align}
	where $f\colon [\bm{0}_n,\bm{{\bar{p}}}]\to\RR$ is a strictly increasing continuous function, and $m_i\in[ x_i^-,+\infty)$ is a constant for each $i\in\NNN$. Then, the feasible region of \eqref{signed_RV_MILP_explicit} is free of the choice of $(m_i)_{i\in\NNN}$, \eqref{signed_RV_MILP_explicit} has an optimal solution, and $\Lambda^{\textnormal{RV}}(\bm{x})\in\RR$. Moreover, if $(\bm{p}, \bm{s}, \bm{t})$ is an optimal solution of \eqref{signed_RV_MILP_explicit}, then $\bm{p}$ is a clearing vector for $\systemRV$. In particular, $\systemRV$ has at least one clearing vector.
\end{theorem}

\begin{remark}\label{rem:binary}
	In \eqref{signed_RV_MILP_explicit}, for each $i\in\NNN$, the variable $t_i$ is associated with the distinction between immediate default and partial liquidity (first two branches of $\Phi^{\text{RV}}_i$ in \eqref{fixed_point_signed_RV}) whereas $s_i$ is related to the distinction between partial liquidity and full commitment (last two branches of $\Phi^{\text{RV}}_i$). This point will be made clearer by Corollaries~\ref{pos_RV_cor},~\ref{signed_EN_cor}; see also \Cref{signed_RV_lemma_1}.
\end{remark}

\begin{remark}\label{rem:f}
	It should be noted that the optimization problem in \Cref{signed_RV_theorem} finds a clearing vector regardless of the choice of $f$, although the corresponding optimal value depends on $f$. The choice of this function should be in accordance with the desired interpretation of $\Lambda^{\text{RV}}(\bm{x})$ in terms of the network. For the computations in \Cref{computational_results}, we will assume that $f$ is linear, that is, $f(\bm{p})=\transpose{\bm{c}}\bm{p}$, $\bm{p}\in [\bm{0}_n,\bm{{\bar{p}}}]$, for some $\bm{c}\in\RR^n_{++}$. Then, the optimization problem \eqref{signed_RV_MILP_explicit} becomes an MILP problem. In this case, choosing $\bm{c}=\bm{1}_n$ yields that $\Lambda^{\text{RV}}(\bm{x})$ is the total payment made by all institutions at clearing, which is free of the particular clearing vector that solves the MILP optimally. In the setting of the Eisenberg-Noe model, a different choice of $\bm{c}$ is proposed in \citet[Section 4.4]{ararat}. There, it is assumed that, in addition to the existing liabilities within the network, every institution has a positive liability to an external entity (node $0$), which can be interpreted as society. Then, by passing to an updated relative liabilities matrix $\tilde{\bm{\pi}}\in\RR^{(n+1)\times(n+1)}$, the vector $\bm{c}=\transpose{(\tilde{\pi}_{10},\ldots,\tilde{\pi}_{n0})}$ is used, which yields that $\Lambda^{\text{RV}}(\bm{x})$ is the total payment made to society by all institutions at clearing. In addition to these linear choices of $f$, one may also take $f$ as a strictly increasing concave function, in which case $f(\bm{p})$ can be interpreted as the overall \emph{utility} of the network under a clearing vector $\bm{p}$. For instance, letting $f(\bm{p})=\sum_{i=1}^n p_i(2\bar{p}_i-p_i)=f_{\max}-\sum_{i=1}^n (\bar{p}_i-p_i)^2$, $\bm{p}\in [\bm{0}_n,\bm{{\bar{p}}}]$, gives a quadratic utility function, where $f_{\max}\coloneqq\sum_{i=1}^n \bar{p}_i^2$ is the maximum utility achieved when all obligations are met. Then, the optimization problem becomes a \emph{mixed integer quadratic programming} problem. Beyond the linear and quadratic cases, solving \eqref{signed_RV_MILP_explicit} for an arbitrary $f$ is a computationally demanding task in general.
\end{remark}

\subsection{Special Cases of the Signed Rogers-Veraart Model}\label{RVspecialcases}

In the next corollary, we focus on the standard Rogers-Veraart model. In this case, a simpler version of the optimization problem in \Cref{signed_RV_theorem} can be used for calculating a clearing vector. Remarkably, this version does not use any big-$M$ constants.

\begin{corollary}\label{pos_RV_cor}
	(Rogers-Veraart model) Let $\bm{x}\in\RR^n_{+}$ and denote by $\Lambda^{\textnormal{RV}_+}(\bm{x})$ the optimal value of the problem
	\begin{align}\tag{$\text{RV}_+(\bm{x})$}\label{pos_RV_MILP_explicit}
		\emph{maximize} \quad 			& f(\bm{p}) && \\ 
		\emph{subject to} \quad 	& \bar{p}_is_i \le x_i + (\bm{\pi}\transposeT\bm{p})_i, &&i\in\NNN,  \label{pos_RV_MILP_constraint_1} \\ 
		& p_i \le \alpha x_i + \beta(\bm{\pi}\transposeT\bm{p})_i + \bar{p}_i s_i , &&i\in\NNN,  \label{pos_RV_MILP_constraint_4} \\ 
		& 0 \le p_i \le \bar{p}_i,\quad s_i \in \cb{0,1}, && i\in\NNN,\label{pos_RV_MILP_constraint_7}
	\end{align}
	where $f\colon [\bm{0}_n,\bm{{\bar{p}}}]\to\RR$ is a strictly increasing continuous function. Then, \eqref{pos_RV_MILP_explicit} has an optimal solution and $\Lambda^{\textnormal{RV}_+}(\bm{x})=\Lambda^{\textnormal{RV}}(\bm{x})\in\RR$. Moreover, if $(\bm{p}, \bm{s})$ is an optimal solution of \eqref{pos_RV_MILP_explicit}, then $\bm{p}$ is a clearing vector for $\systemRV$. In particular, $\systemRV$ has at least one clearing vector.
\end{corollary}

Next, we consider the signed version of the Eisenberg-Noe model. Since default costs do not incur in this case, the binary vector $\bm{s}$ in \eqref{signed_RV_MILP_explicit} can be eliminated and a simpler optimization problem can be solved to calculate a clearing vector.

\begin{corollary}\label{signed_EN_cor}
	(Signed Eisenberg-Noe model) Suppose that $\alpha=\beta=1$. Let $\bm{x}\in\RR^n$ and denote by $\Lambda^{\textnormal{EN}}(\bm{x})$ the optimal value of the problem
	\begin{align}\tag{EN$(\bm{x})$}\label{signed_EN_MILP_explicit}
		\emph{maximize} \quad 			& f(\bm{p}) && \\ 
		\emph{subject to} \quad & p_i \le \bar{p}_i(1 - t_i), &&i\in\NNN,  \label{signed_EN_MILP_constraint_2}\\ 
		& p_i \le  x_i + (\bm{\pi}\transposeT\bm{p})_i + (m_i+\bar{p}_i)t_i, &&i\in\NNN,  \label{signed_EN_MILP_constraint_4} \\ 
		& 0 \le p_i \le \bar{p}_i,\quad t_i \in \cb{0,1}, && i\in\NNN,\label{signed_EN_MILP_constraint_7}
	\end{align}
	where $f\colon [\bm{0}_n,\bm{{\bar{p}}}]\to\RR$ is a strictly increasing continuous function and $m_i\in[x_i^-,+\infty)$ is a constant for each $i\in\NNN$. Then, the feasible region of \eqref{signed_EN_MILP_explicit} is free of the choice of $(m_i)_{i\in\NNN}$, \eqref{signed_EN_MILP_explicit} has an optimal solution, and $\Lambda^{\textnormal{EN}}(\bm{x})=\Lambda^{\textnormal{RV}}(\bm{x})\in\RR$. Moreover, if $(\bm{p}, \bm{t})$ is an optimal solution of \eqref{signed_EN_MILP_explicit}, then $\bm{p}$ is a clearing vector for $(\bm{x},\bm{{\bar{p}}},\bm{\pi},1,1)$. In particular, $(\bm{x},\bm{{\bar{p}}},\bm{\pi},1,1)$ has at least one clearing vector.
\end{corollary}

\begin{remark}\label{noconcave_signed_EN}
	The functions $\Lambda^{\text{RV}_+}$, $\Lambda^\text{EN}$ fail to be concave in general, even when $f$ is a linear function; we illustrate this point by providing two counterexamples in \Cref{sec:counter}.
\end{remark}

Finally, we consider the standard Eisenberg-Noe model, where the two features modeled by binary variables do not show up. Hence, the optimization problem in \Cref{signed_RV_theorem} reduces to the continuous optimization problem in \citet[Lemma 4]{eisenberg-noe} as we show next.

\begin{corollary}\label{pos_EN_cor}
	(Eisenberg-Noe model) Suppose that $\alpha=\beta=1$. Let $\bm{x}\in\RR^n_+$ and denote by $\Lambda^{\textnormal{EN}_+}(\bm{x})$ the optimal value of the problem
	\begin{align}\tag{EN$_+(\bm{x})$}\label{pos_EN_LP_explicit}
		\emph{maximize} \quad 			& f(\bm{p}) && \\ 
		\emph{subject to} \quad & p_i \le  x_i + (\bm{\pi}\transposeT\bm{p})_i , &&i\in\NNN,  \label{pos_EN_LP_constraint_4} \\ 
		& 0 \le p_i \le \bar{p}_i,&& i\in\NNN,\label{pos_EN_LP_constraint_7}
	\end{align}
	where $f\colon [\bm{0}_n,\bm{{\bar{p}}}]\to\RR$ is a strictly increasing continuous function. Then, \eqref{pos_EN_LP_explicit} has an optimal solution and $\Lambda^{\textnormal{EN}_+}(\bm{x})=\Lambda^{\textnormal{RV}}(\bm{x})\in\RR$. Moreover, if $\bm{p}$ is an optimal solution of \eqref{pos_EN_LP_explicit}, then it is a clearing vector for $(\bm{x},\bm{{\bar{p}}},\bm{\pi},1,1)$. In particular, $(\bm{x},\bm{{\bar{p}}},\bm{\pi},1,1)$ has at least one clearing vector.
\end{corollary}

\section{Systemic Risk Measure for the Signed Rogers-Veraart Model}\label{RVsystrisk}

In this section, we add randomness to the operating cash flow of the signed Rogers-Veraart model. Let $\bm{X}\in L^\infty(\RR^n)$; $\bm{{\bar{p}}}\in\RR^n_{++}$; $\alpha,\beta\in(0,1]$. Let $\bm{\pi}\in\RR^{n\times n}$ be a right stochastic matrix with $\pi_{ii}=0$ for each $i\in\NN$. We consider the system $\XsystemRV$, which has the property that $(\bm{X}(\omega), \bm{{\bar{p}}}, \bm{{\pi}}, \alpha, \beta)$ is a signed Rogers-Veraart network in the sense of \Cref{signed_RVsystem} for each $\omega\in\Omega$. Then, the corresponding systemic risk measure is defined by
	\begin{equation}\label{sensitive_set-valuedRV}
		\RsenRVx(\bm{X})\coloneqq\big\{\bm{z}\in\ZZZ\mid \EE[\Lambda^{\text{RV}}(\bm{X}+\transpose{\bm{B}}\bm{z})]\geq \gamma\big\},
	\end{equation}
	where $\Lambda^{\text{RV}}$ is the aggregation function defined in \Cref{signed_RV_theorem} and $\gamma \in \RR$. We choose the function $f\colon[\bm{0}_n,\bm{{\bar{p}}}]\to\RR$ in the formulation of $\Lambda^{\text{RV}}$ as
	\begin{equation}\label{fchoice}
		f(\bm{p}) \coloneqq \transpose{\one}_n\bm{p},\quad \bm{p}\in[\bm{0}_n,\bm{{\bar{p}}}].
	\end{equation}
	With this choice of $f$, as noted in \Cref{rem:f}, $\Lambda^{\text{RV}}(\bm{X}(\omega))$ yields the total payment made by all institutions at clearing under scenario $\omega\in\Omega$. To ensure that $\RsenRVx(\bm{X})$ is a lower bounded set, as discussed in \Cref{ideal}, we assume that
	\begin{equation}\label{zLB}
	\ZZZ\subseteq \bm{z}^{\text{LB}}+\RR^G_+
	\end{equation}
	for some lower bound vector $\bm{z}^{\text{LB}}\in\RR^G$. Accordingly, for each $i\in\NNN$, we define
	\[
	M_i\coloneqq \|(X_i+(\transpose{\bm{B}}\bm{z}^{\text{LB}})_i)^-\|_\infty
	\]
	and take $m_i=M_i$ when calculating $\Lambda^\text{RV}(\bm{X}(\omega)+\transpose{\bm{B}}\bm{z})$ by \Cref{signed_RV_theorem} for each $\omega\in\Omega$ and $\bm{z}\in\ZZZ$.
	
The main results of this section are on the scalarizations of $R^{\text{RV}}(\bm{X})$, which are given in \Cref{sec:RVsc}. The proofs are deferred to \Cref{app:RVsystrisk}. In \Cref{rem:bisection}, we comment on the implementation of \Cref{alg1} to approximate $R^{\text{RV}}(\bm{X})$.

\subsection{Results on Scalarizations}\label{sec:RVsc}

We start with a lemma that characterizes the nonemptiness of $R^{\text{RV}}(\bm{X})$ in terms of an inequality between the parameters $\bm{{\bar{p}}}$ and $\gamma$.

\begin{lemma}\label{lem:RV_RM}
	Let $\bm{{\hat{z}}}\coloneqq (\|\bm{X}^-\|_\infty+\frac{1}{\alpha} \norminfpbar)\one_G$, suppose that $\bm{{\hat{z}}}\in\ZZZ$. (When $\ZZZ=\bm{z}^{\textnormal{LB}}+\RR^G_+$, this condition reads as $\bm{z}^{\textnormal{LB}} \leq \bm{{\hat{z}}}$.) The following are equivalent: (a) $\bm{{\hat{z}}}\in R^{\textnormal{RV}}(\bm{X})$, (b) $R^{\textnormal{RV}}(\bm{X})\neq \emptyset$,  (c) $\gamma \leq \transpose{\one}_n\bm{{\bar{p}}}$.
\end{lemma}

\begin{remark}\label{rem:gamma}
	The choice of $f$ in \eqref{fchoice} and the definition of the systemic risk measure in \eqref{sensitive_set-valuedRV} imply that $\gamma$ is the target mean amount of total liabilities met in the system at clearing. In line with \Cref{lem:RV_RM}, for the computational experiments in \Cref{computational_results}, we will express this threshold as $\gamma=\gamma^{\text{p}}\transpose{\one}_n\bm{{\bar{p}}}$, where $\gamma^{\text{p}}\in [0,1]$ is the target mean fraction of total liabilities met at clearing.
\end{remark}

Let $\bm{w}\in\RR^G_+\setminus\{\bm{0}_G\}$. The corresponding weighted-sum scalarization of $R^{\text{RV}}(\bm{X})$ is given by
\begin{equation}\label{P1_signed_RV}
	\PPP_1^{\text{RV}}(\bm{w}) \coloneqq \inf_{\bm{z}\in\ZZZ} \big\{\transpose{\bm{w}}\bm{z} \mid \EE[\Lambda^{\text{RV}}(\bm{X}+\transpose{\bm{B}}\bm{z})]\ge\gamma\big\}.
\end{equation}
The next result provides an MILP formulation for this problem as a consequence of \Cref{theorem_P1_OPT}.

\begin{corollary}\label{P1_signed_RV_proposition}
	Suppose that $\gamma \le \transpose{\one}_n\bm{{\bar{p}}}$ and $\bm{{\hat{z}}}\in\ZZZ$, where $\bm{{\hat{z}}}$ is defined by \Cref{lem:RV_RM}. Consider the MILP problem
		\begin{align}\tag{$\text{P}^\text{RV}_1(\bm{w})$}\label{P1_MILP_signed_RV_explicit}
			\emph{minimize}\quad &\transpose{\bm{w}}\bm{z} \\
			\emph{subject to}\quad& \EE[\transpose{\one}_n\bm{P}]\ge \gamma, \label{P1_MILP_signed_RV_constraint_1}\\
			& \bar{p}_i S_i - M_i T_i \le X_i + (\transpose{\bm{B}}\bm{z})_i + (\bm{\pi}\transposeT\bm{P})_i, &i\in\NNN, \label{P1_MILP_signed_RV_constraint_2} \\ 
			&\bar{p}_iS_i\leq  P_i \le \bar{p}_i(1 - T_i), &i\in\NNN,  \label{P1_MILP_signed_RV_constraint_3}\\
			& P_i \le \varphi^\alpha_i(\bm{X} + \transpose{\bm{B}}\bm{z})+ \beta(\bm{\pi}\transposeT\bm{P})_i + (M_i+\bar{p}_i)(S_i + T_i), &i\in\NNN,  \label{P1_MILP_signed_RV_constraint_5} \\ 
			& \bm{z}\in \ZZZ, \quad \bm{P}\in L^\infty([\zero_n,\bm{{\bar{p}}}]), \quad \bm{S}, \bm{T}\in L^\infty(\{0,1\}^n),& i\in\NNN.\label{P1_MILP_signed_RV_constraint_8}
		\end{align}
	 (Here, the nonlinear constraint \eqref{P1_MILP_signed_RV_constraint_5} is written for brevity and it stands for two linear constraints obtained by the two branches of the function $\varphi_i^\alpha$; see \eqref{phialpha}.) Then, the optimal value of \eqref{P1_MILP_signed_RV_explicit} equals $\PPP_1^{\textnormal{RV}}(\bm{w})$. Moreover, the problem has an optimal solution and
		\begin{equation}\label{eq:P1RV}
			\PPP_1^{\textnormal{RV}}(\bm{w})\leq  \transpose{\bm{w}}\bm{{\hat{z}}}=\of{\|\bm{X}^-\|_\infty + \frac{1}{\alpha}\norminfpbar}\transpose{\bm{w}}\one_G.
		\end{equation}
\end{corollary}

Let $\bm{c}\in\RR^G_+\setminus\{\zero_G\}$ and $\bm{v}\in\mathcal{Z}$. The corresponding Pascoletti-Serafini scalarization problem for $R^{\text{RV}}(\bm{X})$ is given by
\begin{equation}\label{P2_signed_RV}
	\PPP_2^{\text{RV},\bm{c}}(\bm{v}) =\inf \big\{\mu\in\RR \mid\EE[\Lambda^{\text{RV}}(\bm{X}+\transpose{\bm{B}}(\bm{v}+\mu\bm{c}))]\ge\gamma,\ \bm{v}+\mu\bm{c}\in \ZZZ\big\}.
\end{equation}
The next corollary shows that the above problem can be reformulated as an MILP problem.

\begin{corollary}\label{corollary_P2_signed_RV}
	Suppose that $\gamma \le \transpose{\one}_n\bm{{\bar{p}}}$ and $\bm{{\hat{z}}}\in\ZZZ$, where $\bm{{\hat{z}}}$ is defined by \Cref{lem:RV_RM}. Consider the MILP problem
		\begin{align}\tag{$\text{P}^{\text{RV},\bm{c}}_2(\bm{v})$} \label{P2_MILP_signed_RV_explicit}
			\emph{minimize}\quad &\mu\in\RR\\
			\emph{subject to}\quad& \EE[\transpose{\one}_n\bm{P}] \ge \gamma, \label{P2_MILP_signed_RV_constraint_1}\\
			& 
			\bar{p}_iS_i - M_i T_i \le X_i +\of{\transpose{\bm{B}}(\bm{v}+\mu\bm{c})}_i + (\bm{\pi}\transposeT\bm{P})_i, &i\in\NNN,   \label{P2_MILP_signed_RV_constraint_2} \\ 
			& \bar{p}_iS_i\leq  P_i \le \bar{p}_i(1 - T_i),  &i\in\NNN,  \label{P2_MILP_signed_RV_constraint_3}\\ 
			& 
			P_i \le \varphi^\alpha_i\of{\bm{X}+ \transpose{\bm{B}}(\bm{v}+ \mu\bm{c})}+ \beta(\bm{\pi}\transposeT\bm{P})_i + (M_i+\bar{p}_i)(S_i + T_i), &i\in\NNN, \label{P2_MILP_signed_RV_constraint_5} \\ 
			& \bm{v}+\mu\bm{c} \in \ZZZ, \quad \bm{P}\in L^\infty([\bm{\zero}_n,\bm{{\bar{p}}}]), \quad \bm{S}, \bm{T} \in L^\infty(\cb{0,1}^n).&\label{P2_MILP_signed_RV_constraint_8}
		\end{align}
		Suppose further that there exists $\bm{z}\in R^{\textnormal{RV}}(\bm{X})$ such that $z_\ell\leq v_\ell$ for every $\ell\in\GGG$ with $c_\ell=0$. Then, the optimal value of \eqref{P2_MILP_signed_RV_explicit} equals $\PPP_2^{\textnormal{RV},\bm{c}}(\bm{v})$, the problem has an optimal solution, and
		\begin{equation}\label{eq:P2RV}
			\PPP_2^{\textnormal{RV}}(\bm{v})\leq  \max_{\ell\in\GGG\colon c_\ell>0}\frac{z_\ell -v_\ell}{c_\ell}.
		\end{equation}
\end{corollary}

We conclude this subsection by commenting on the standard Rogers-Veraart model.

\begin{remark}\label{rem_pos_RV}
Similar to the treatment in this section, we may consider the standard Rogers-Veraart model, where the random operating cash flow vector is positive, i.e., $\bm{X}(\omega)\in \RR^n_+$ for every $\omega\in\Omega$. In this case, as shown in \Cref{pos_RV_cor}, the mixed-integer optimization problem in \Cref{signed_RV_theorem} simplifies as one type of binary variables and the big-$M$ constants can be eliminated. We define $\Lambda^{\text{RV}_+}(\bm{x})$ by \Cref{pos_RV_cor} for $\bm{x}\in\RR^n_+$ and by $\Lambda^{\text{RV}_+}(\bm{x})=-\infty$ for $\bm{x}\in\RR^n\setminus\RR^n_+$. The corresponding systemic risk measure is given by $R^{\text{RV}_+}(\bm{X})=\{\bm{z}\in\ZZZ\mid \EE[\Lambda^{\text{RV}_+}(\bm{X}+\transpose{\bm{B}}\bm{z})]\geq \gamma \}$, where $\gamma\in\RR$ and $f$ is as given in \eqref{fchoice}. In this formulation, we implicitly have the domain constraint $\bm{X}+\transpose{\bm{B}}\bm{z}\in L^\infty(\RR_+^n)$, which can be incorporated by taking $\ZZZ=\{\bm{z}\in\RR^G\mid \bm{X}+\transpose{\bm{B}}\bm{z}\in L^\infty(\RR_+^n)\}$. Clearly, $\ZZZ$ is a lower bounded set in this case and, by \Cref{theorem_P1_OPT} and \Cref{ideal}, it follows that the ideal point $\bm{z}^\text{ideal}$ is well-defined. Then, similar to \Cref{lem:RV_RM}, it can be checked that $R^{\text{RV}_+}(\bm{X})\neq\emptyset$ if and only if $\gamma \leq \transpose{\one}_n\bm{{\bar{p}}}$. Finally, by combining \Cref{pos_RV_cor} with Corollaries~\ref{theorem_P1_OPT},~\ref{theorem_P2_OPT}, it is possible to provide MILP formulations for the weighted-sum and Pascoletti-Serafini scalarizations of $R^{\text{RV}_+}(\bm{X})$ analogous to Corollaries~\ref{P1_signed_RV_proposition},~\ref{corollary_P2_signed_RV}. We omit these formulations for brevity.
\end{remark}

\subsection{A Note on the Implementation of \Cref{alg1}}\label{rem:bisection}

	The weighted-sum and Pascoletti-Serafini scalarization problems can be considered as two-stage mixed-integer stochastic programming problems with an expectation constraint, which are special types of large-scale mixed-integer programming problems. The formulations \eqref{P1_MILP_signed_RV_explicit} and \eqref{P2_MILP_signed_RV_explicit} are the so-called \emph{deterministic equivalent formulations} of these problems. As is well-known in the literature on such problems (see \cite{tutorial} and the references therein), using commercial solvers directly on the deterministic equivalent formulations suffers from excessive computation times. In general, more efficient solution techniques are developed based on decompositions, which we leave for future research in the setting of systemic risk measures. In our implementation, weighted-sum scalarizations are solved only at the initialization stage for finding the ideal point, hence relying on the deterministic equivalent formulation is still reasonable. However, the main loop of \Cref{alg1} solves a Pascoletti-Serafini scalarization for many iterations (line 8). For this purpose, we use a simple bisection search on the one-dimensional first-stage variable $\mu$. We set $\mu^{\text{LB}}\leftarrow 0$ and $\mu^{\text{UB}}\leftarrow \mu^{\max}\coloneqq \max\{\mu\geq 0\mid \bm{v}+\mu\bm{c}\leq \bm{z}^{\text{UB}}\}$. For $\mu=\frac{\mu^{\text{LB}}+\mu^{\text{UB}}}{2}$, we check if $\bm{v}+\mu\bm{c}\in R^{\text{RV}}(\bm{X})$, which can be done by solving the small-scale MILP problem in \Cref{signed_RV_theorem} for each scenario. If this is the case, we set $\mu^{\text{UB}}\leftarrow \mu$; otherwise we set $\mu^{\text{LB}}\leftarrow \mu$. We repeat this process until $\mu^{\text{UB}}-\mu^{\text{LB}}<\delta$ for a tolerance level $\delta\in(0,\mu^{\max})$; hence, we detect $\PPP_2^{\text{RV},\bm{c}}(\bm{X})$ up to an approximation error of $\delta$ in at most $\lceil \log_2(\frac{\mu^{\max}}{\delta})\rceil$ iterations.

\section{Computational Results and Analysis} \label{computational_results}

In this section, we implement \Cref{alg1} for the standard Rogers-Veraart model and perform a detailed sensitivity analysis. In \Cref{app:comp}, we perform a similar analysis for the signed Rogers-Veraart and the signed Eisenberg-Noe models. We run the algorithm on Java Hotspot(TM) (Release 18.0.1.1) by calling Gurobi Interactive Shell (Version 9.5.2) \citep{gurobi}. We use a computer with an Apple M1 chip and a 16 GB RAM.

In order to have finite-dimensional MILP problems for the scalarizations, we assume that $\Omega=\{\omega^1,\ldots,\omega^K\}$ is a finite sample space with $K\in\NN$ outcomes (with $\FFF=2^\Omega$) as in Remarks~\ref{rem:P1finite},~\ref{rem:P2finite}. In line with \Cref{rem:gamma}, we write the threshold parameter as $\gamma = \gamma^\text{p}(\transpose{\one}\bm{{\bar{p}}})$, where $\gamma^\text{p}\in\sqb{0,1}$.

\subsection{Data Generation}\label{sec:datagen}

We consider a network with $n$ banks that are decomposed into $G=2$ or $G=3$ groups. Recall that $\GGG = \{1,\ldots,G\}$, $\NNN = \bigcup_{\ell\in\GGG}{\NNN_\ell} = \cb{1,\ldots,n}$, and $n_\ell = |\NNN_\ell|$. When $G=2$, the groups $\ell = 1$ and $\ell = 2$ correspond to big and small banks, respectively. When $G=3$, the groups $\ell = 1$, $\ell = 2$ and $\ell = 3$ correspond to big, medium and small banks, respectively.

In order to construct a network $\XsystemRV$, the interbank liabilities matrix $\bm{l}\coloneqq(l_{ij})_{i,j\in\NNN}\in\RR^{n\times n}_+$ and the random operating cash flow vector $\bm{X}$ are generated in the following fashion. For $\bm{l}$, we use an Erd{\"o}s-R{\'e}nyi random graph model \citep{ersod-renyi,gilbert}. First, we fix a \emph{connectivity probabilities matrix} $\bm{q}^\text{con}\coloneqq(q^\text{con}_{\elll})_{\elll\in\GGG}\in\RR^{G\times G}$ and an \emph{intergroup liabilities matrix} $\bm{l}^\text{gr}\coloneqq(l^\text{gr}_{\elll})_{\elll\in\GGG}\in\RR^{G\times G}$. For any two banks $i,j\in\NNN$ with $i\in\NNN_\ell$,  $j\in\NNN_{\hat{\ell}}$ and $\ell,\hat{\ell}\in\GGG$, $q^\text{con}_{\elll}$ is interpreted as the probability that bank $i$ owes amount $l^\text{gr}_{\elll}$ to bank $j$. Then, the liability $l_{ij}$ is generated by the Bernoulli trial given by $l_{ij} =l^\text{gr}_{\elll}$ if $U_{ij} < q^\text{con}_{\elll}$, and $l_{ij}=0$ otherwise, where $U_{ij}$ is the realization of a random variable with the standard uniform distribution on a separate probability space. The relative liabilities matrix $\bm{\pi}$ and the total obligation vector $\bm{{\bar{p}}}$ are calculated accordingly.

We assume that all $K$ scenarios in $\Omega$ are equally likely to happen. We generate the random operating cash flow vector $\bm{X}$ as $K$ independently drawn instances of a random vector $\bm{{\tilde{X}}}$ (on a separate probability space) whose cumulative distribution function is stated in terms of a Gaussian copula with gamma marginal distributions. Specifically, the components of $\bm{{\tilde{X}}}$ have a common standard deviation $\sigma$, and there is a common correlation $\varrho$ between $\tilde{X}_i$ and $\tilde{X}_j$ whenever $i,j\in\NNN$ with $i\neq j$. For the marginal distributions of $\bm{{\tilde{X}}}$, shape parameters $\bm{\kappa} =\transpose{(\kappa_1,\dots,\kappa_G)}$ and scale parameters $\bm{\theta}=\transpose{(\theta_1,\ldots,\theta_G)}$ are fixed in accordance with the choices of $\sigma,\varrho$. In particular, the mean value is $\nu_\ell \coloneqq \kappa_\ell \theta_\ell$ and the standard deviation is $\sigma = \sqrt{\kappa_\ell} \theta_\ell$ for each group $\ell\in\GGG$.

\subsection{A Two-Group Rogers-Veraart Network with $45$ Nodes}

In this subsection, we consider a network with $n = 45$ banks distributed into two groups with $n_1 = 15$, $n_2 = 30$. We take the parameters as $K = 50$, $\varrho = 0.05$, $\alpha=0.7$, $\beta=0.9$, $\gamma^{\text{p}}=0.9$, and
\[
\bm{q}^\text{con} = 
\begin{bmatrix}
	0.5 & 0.1 \\
	0.3 & 0.5
\end{bmatrix},\quad
\bm{l}^\text{gr} = 
\begin{bmatrix}
	200 & 100 \\
	50 & 50
\end{bmatrix}.
\]
We choose the shape and scale parameters of gamma distributions as $\bm{\kappa} = \transpose{(100, 64)}$,	$\bm{\theta} = \transpose{(1, \frac{5}{4})}$. Then, the mean values of the random operating cash flows in the corresponding groups are $\bm{\nu} = \transpose{(100,	80)}$, and the common standard deviation is $\sigma = 10$. We take the approximation error as $\epsilon = 1$.

In the tables, we report the following quantities: number of iterations ($T$), number of Pascoletti-Serafini scalarizations solved ($\#(\PPP_2)$), average time spent per Pascoletti-Serafini scalarization ($\overline{\text{time}}(\PPP_2)$, in seconds), total algorithm time ($\text{time}(\text{total})$, in 
seconds).

\subsubsection{Rogers-Veraart $\alpha$ Parameter}

\begin{table}[!tbp]
	\centering
	\resizebox{0.4\textwidth}{!}{
	\begin{tabular}{|c|c|c|c|c|}
		\hline
		 $\alpha$ & 
		 $T$ & 
		  $\#(\PPP_2)$& 
		 $\overline{\text{time}}(\PPP_2)$ (sec.)& 
		 $\text{time}(\text{total})$ (sec.)\\ \hline\hline
		0.1	& 590		&  295	&  1.80	&  531	\\ \hline
		0.3	& 866   	&  433	&  1.76	&  762	\\ \hline
		0.5	& 1050		&  526	&  1.86	&  979		\\ \hline
		0.7	& 1082		&  541	&  1.92	&  1037	\\ \hline
		0.9	& 1110		&  555	&  2.02	&  1119		\\ \hline	
	\end{tabular}
}
\quad 
	\resizebox{0.4\textwidth}{!}{
	\begin{tabular}{|c|c|c|c|c|}
	\hline
		$\beta$ & 
		$T$ & 
		$\#(\PPP_2)$& 
		$\overline{\text{time}}(\PPP_2)$ (sec.)& 
		$\text{time}(\text{total})$ (sec.)\\ \hline\hline
		0.1	&  407	&  205	&  1.50	&  307	\\ \hline
		0.3	&  514	&  258	&  1.61	&  416	\\ \hline
		0.5	&  634	&  318	&  1.62	&  515	\\ \hline
		0.7	&  762	&  382	&  1.77	&  676	\\ \hline
		0.9	& 1082		&  541	&  1.92	&  1037	\\ \hline
	\end{tabular}
}
\vspace{5pt}
\caption{Computational performance of \Cref{alg1} relative to $\alpha$ (left) and $\beta$ (right).}
\label{table_alpha_2D}
\end{table}

\begin{figure}[!tbp]
	\centering
		\subfloat[$\alpha$]{\includegraphics[width=0.52\textwidth]{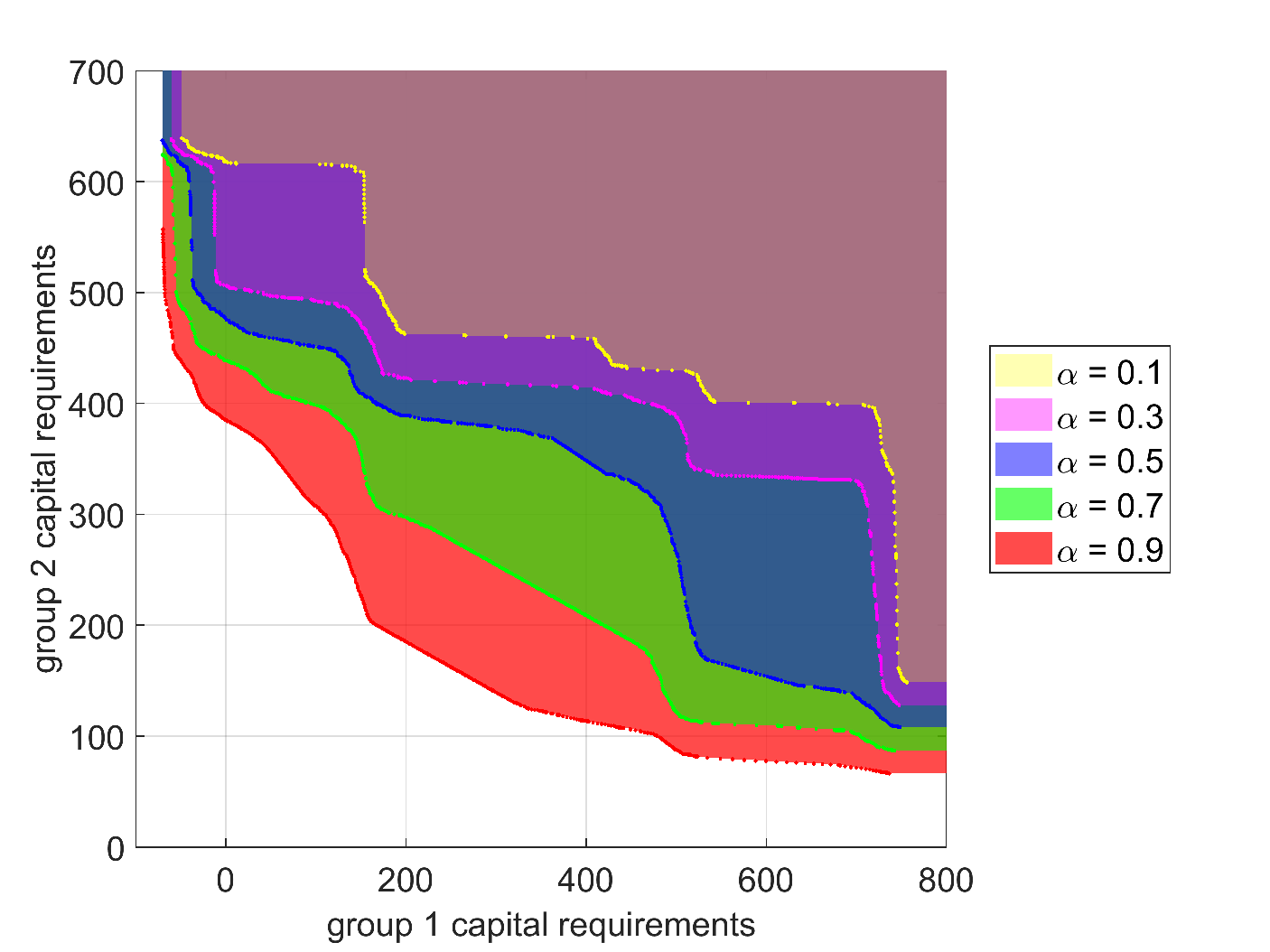}}
	\negthinspace
		\subfloat[$\beta$]{\includegraphics[width=0.52\textwidth]{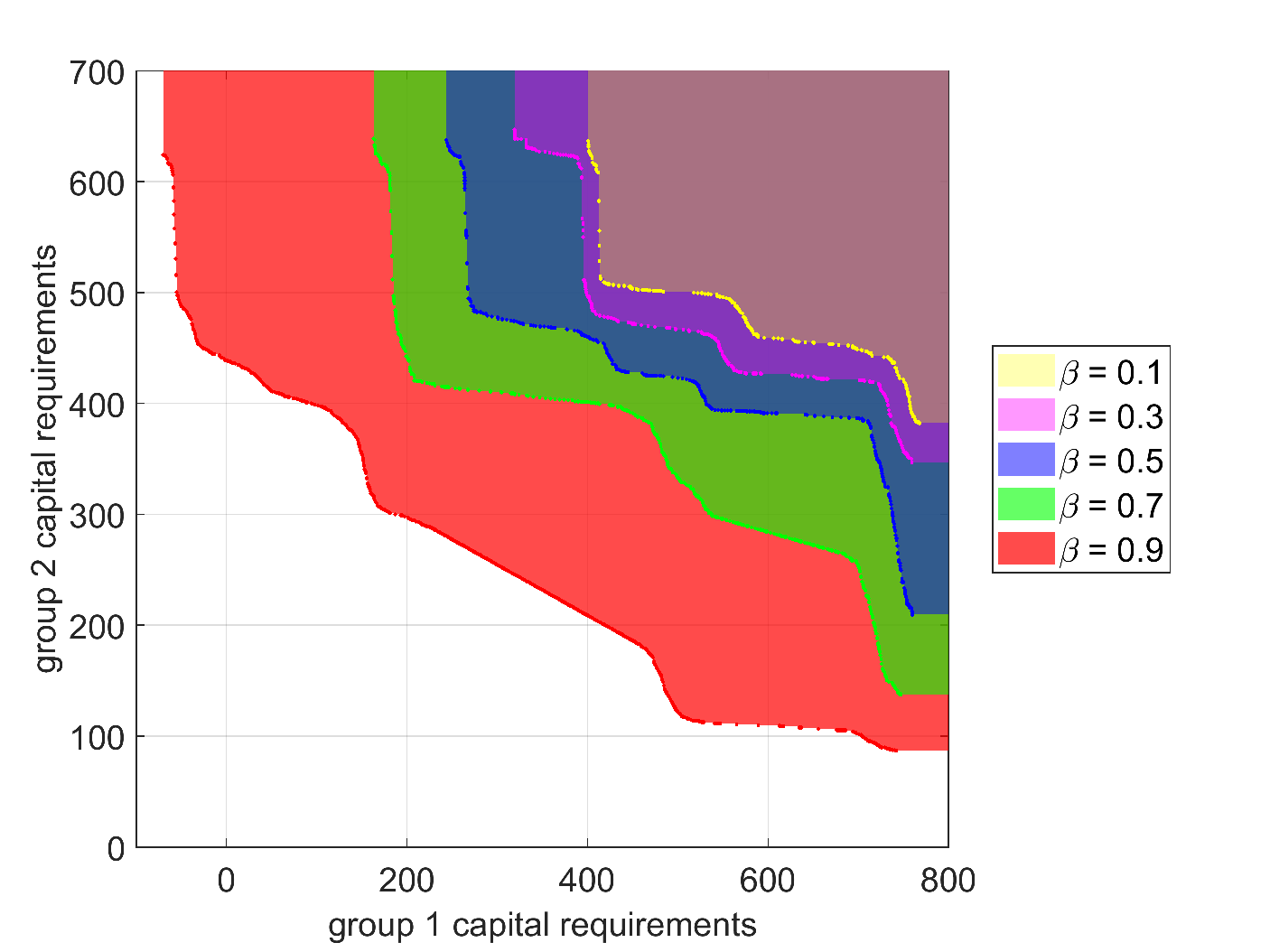}}
	\caption{Inner approximations of the Rogers-Veraart systemic risk measures for different $\alpha$ and $\beta$ values.}
	\label{figure_alpha_2D_inner}
\end{figure}

In this part, we perform a sensitivity analysis with respect to $\alpha$, the liquid fraction of the operating cash flow that can be used by a defaulting node to meet its obligations. The generated network remains the same in all cases. \Cref{table_alpha_2D} (left) illustrates the computational performance of the algorithm for $\alpha\in\{0.1, 0.3, 0.5, 0.7, 0.9\}$ and \Cref{figure_alpha_2D_inner}(a) consists of the inner approximations of the corresponding Rogers-Veraart systemic risk measure.

In \Cref{figure_alpha_2D_inner}(a), we observe that the Rogers-Veraart systemic risk measure expands significantly as $\alpha$ increases. Hence, both big and small banks get less strict capital requirements as default costs decrease. One can also observe that, in each case, allocating zero capital to all groups is not an available option. In \Cref{table_alpha_2D} (left), we notice that the average time spent per Pascoletti-Serafini scalarization shows only a slight increase with $\alpha$ while the number of scalarization problems has a more dramatic increase with $\alpha$.

\subsubsection{Rogers-Veraart $\beta$ Parameter}

In this part, we perform a sensitivity analysis with respect to $\beta$, the liquid fraction of the realized claims from the other nodes that can be used by a defaulting node to meet its obligations. \Cref{table_alpha_2D} (right) shows the computational performance of the algorithm for $\beta\in\{0.1, 0.3, 0.5, 0.7, 0.9\}$ and \Cref{figure_alpha_2D_inner}(b) provides the inner approximations of the systemic risk measure.

In \Cref{figure_alpha_2D_inner}(b), we note that the Rogers-Veraart systemic risk measure expands significantly as $\beta$ increases. In addition, as in the sensitivity analysis for $\alpha$, we observe that the number of scalarization problems increases with $\beta$. In both cases, this might be related to the arc length of the \emph{efficient frontier} of the systemic risk measure, i.e., the boundary excluding the vertical and horizontal line segments, in the sense that the longer it gets, the more scalarization problems are required to approximate it.

\begin{table}[!hb]
	\centering
	\resizebox{0.4\textwidth}{!}{
	\begin{tabular}{|c|c|c|c|c|}
		\hline
		$\gamma^{\text{p}}$ & 
		$T$ & 
		$\#(\PPP_2)$& 
		$\overline{\text{time}}(\PPP_2)$ (sec.)& 
		$\text{time}(\text{total})$ (sec.)\\ \hline\hline
		0.1 	&  2		&  0		&  0	&  0		\\ \hline
		0.2 	&  28		&  14		&  0.47	&  7			\\ \hline
		0.3 	&  100		&  50		&  0.93 &  46	\\ \hline
		0.4 	&  186		&  93		&  1.07 &  100		\\ \hline
		0.5 	&  300		&  150		&  1.40	&  210		\\ \hline
		0.6 	&  426		&  214		&  1.65	&  352		\\ \hline
		0.7 	&  720		&  361		&  1.87	&  676	\\ \hline
		0.8 	&  724		&  362		&  2.22	&  803		\\ \hline
		0.9	& 1082		&  541	&  1.92	&  1037	\\ \hline
		0.95 	&  800		&  400		&  1.57	&  627		\\ \hline
		0.99 	&  130 		&  65		&  0.44	&  28			\\ \hline
		1.00 	&  2		&  0		&  0	&  0	\\ \hline
	\end{tabular}
}
\quad
	\resizebox{0.4\textwidth}{!}{
	\begin{tabular}{|c|c|c|c|c|}
		\hline
		$n_1$ & 
		$T$ & 
		$\#(\PPP_2)$& 
		$\overline{\text{time}}(\PPP_2)$ (sec.)& 
		$\text{time}(\text{total})$ (sec.)\\ \hline\hline
		5	&  10		&  6	& 0.32 	&  2			\\ \hline
		10	&  826	&  412	&  1.79	&  738		\\ \hline
		15	&  1082		&  541	&  1.92	&  1037 	\\ \hline
		20	&  1024		&  512	&  1.92	&  981		\\ \hline
		25	&  1088		&  544	&  1.88	&  1023		\\ \hline
		30	&  718		&  358	&  1.93	&  692		\\ \hline
		35	&  368	&  185	&  1.71	&  317		\\ \hline
		40	&  212		& 107	&  1.94	&  207		\\ \hline
	\end{tabular}
}
\vspace{5pt}
	\caption{Computational performance of \Cref{alg1} relative to $\gamma^\text{p}$ (left) and $n_1$ (right).}
	\label{table_gamma_2D_RV}
\end{table}

\subsubsection{Threshold Level}

\begin{figure}[!tbp]
	\centering
		\subfloat[$\gamma^\text{p}$]{\includegraphics[width=0.52\textwidth]{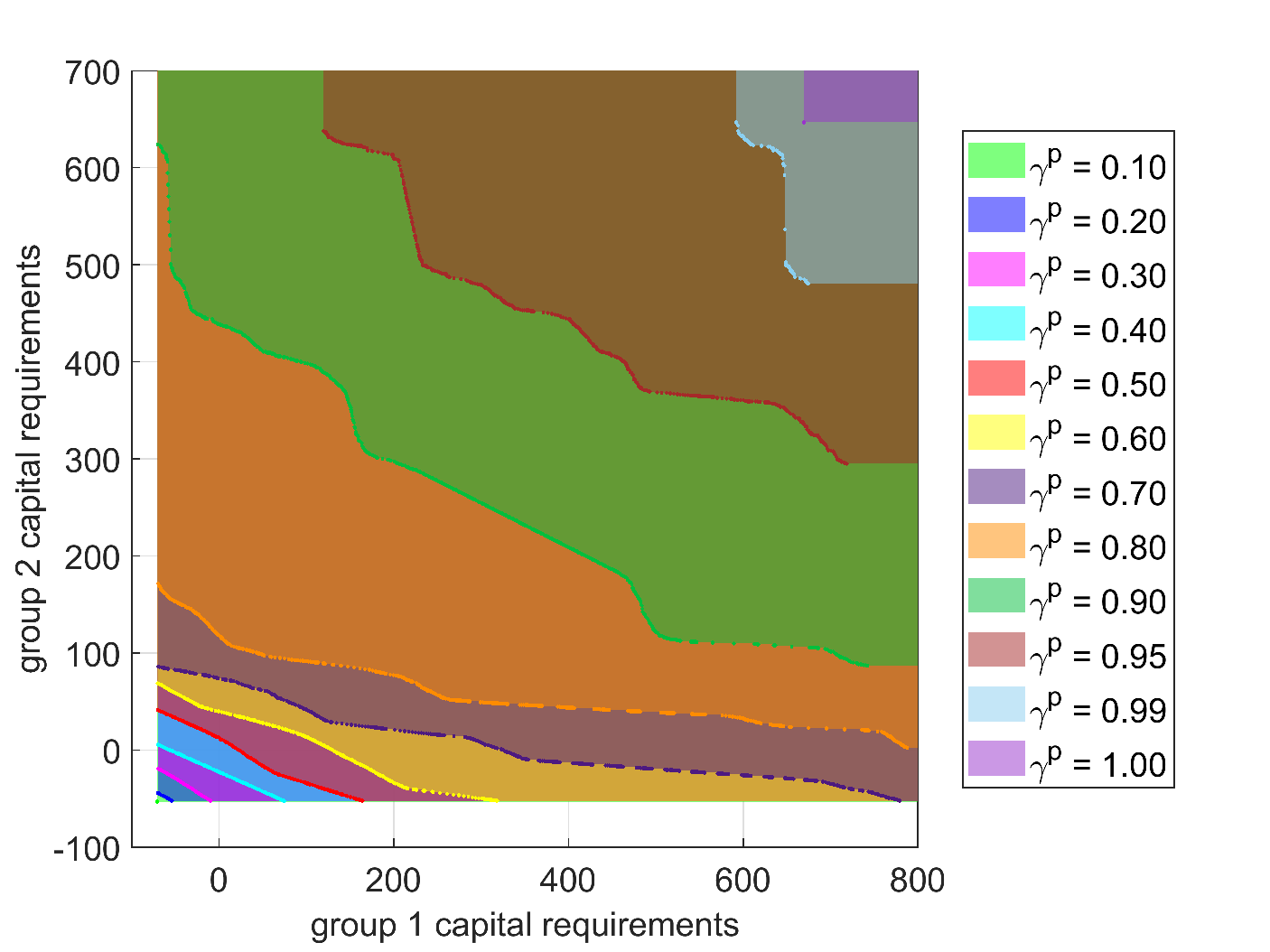}}
	\negthinspace\negthinspace\negthinspace 
		\subfloat[$n_1$]{\includegraphics[width=0.52\textwidth]{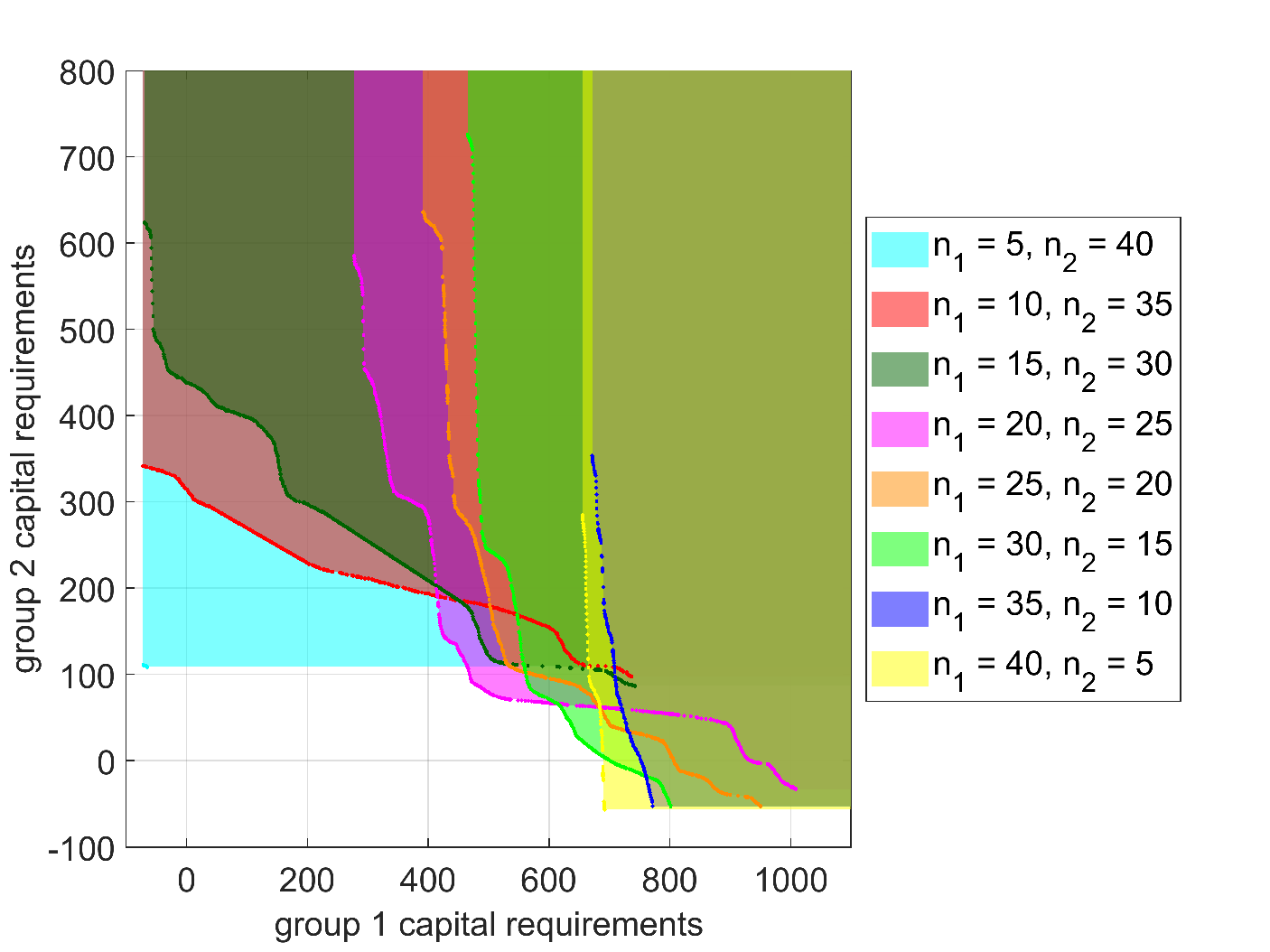}}
	\caption{Inner approximations of the Rogers-Veraart systemic risk measures for different $\gamma^{\text{p}}$ and $(n_1,n_2)$ values.}
	\label{figure_gamma_2D_RV_inner}
\end{figure}

Next, we compare different $\gamma^\text{p}$ levels. \Cref{table_gamma_2D_RV} (left) shows the computational performance of the algorithm for $\gamma^\text{p}\in\cb{0.1, 0.2, \ldots, 0.9, 0.95, 0.99, 1}$ and \Cref{figure_gamma_2D_RV_inner}(a) consists of the inner approximations of the corresponding systemic risk measure.

In \Cref{table_gamma_2D_RV} (left), we see that the average time spent per Pascoletti-Serafini scalarization and the total algorithm time are high around $\gamma^\text{p}=0.8$ and $\gamma^\text{p}=0.9$. In addition, the number of these scalarizations increases with $\gamma^{\text{p}}$ up to $\gamma^\text{p} = 0.9$ and then it decreases. We observe in \Cref{figure_gamma_2D_RV_inner}(a) that the systemic risk measure becomes a smaller set as $\gamma^\text{p}$ increases, which is consistent with the definition in \Cref{rem_pos_RV}. The case $\gamma^\text{p}=1.00$ corresponds to the extreme requirement that all interbank liabilities be met in full under all scenarios. In view of \Cref{rem:gamma1}, we have $R^{\text{RV}_+}(\bm{X})=\bm{{\bar{z}}}+\RR^G_+$, where $\bm{{\bar{z}}}$ is the componentwise minimum vector that satisfies $\bm{X}+\bm{B}^{\mathsf{T}}\bm{{\bar{z}}}\geq (\bm{I}_n-\bm{\pi})^{\mathsf{T}}\bm{{\bar{p}}}$. Hence, the computation of $R^{\text{RV}_+}(\bm{X})$ becomes trivial, explaining the values in the last line of \Cref{table_gamma_2D_RV} (left).

\subsubsection{Distribution of Nodes among Groups}

Here, we perform a sensitivity analysis by changing the distribution of nodes among the groups for a fixed total number of nodes, $n=45$. The number of big banks takes values $n_1\in \cb{5, 10, 15, 20, 25, 30, 35, 40}$. Then, the number of small banks is $n_2 = n - n_1$. \Cref{table_gamma_2D_RV} (right) shows the computational performance of the algorithm and \Cref{figure_gamma_2D_RV_inner}(b) provides the inner approximations of the corresponding systemic risk measure.

Note that the average time spent per Pascoletti-Serafini scalarization in \Cref{table_gamma_2D_RV} (right) is relatively low for $n_1=5$ but does not fluctuate much otherwise. In addition, the number of scalarizations and the total algorithm time are larger around $n_1 = 20$. From \Cref{figure_gamma_2D_RV_inner}(b), we observe that, as the number of big banks increases, small banks get a wider range of capital allocation options, as opposed to big banks. This happens because the total number of banks is fixed and the group with less number of banks has more claims to the other group's banks in this setup.

\subsection{A Three-Group Rogers-Veraart Network with $60$ Nodes}\label{RV3}

In this subsection, we consider a Rogers-Veraart network with $n = 60$ nodes distributed into three groups with $n_1 = 10$, $n_2 = 20$, $n_3 = 30$. There are $K = 50$ scenarios; we take $\alpha = \beta = 0.9$, $\gamma^\text{p}= 0.99$, $\varrho = 0.05$, $\bm{\kappa} =\transpose{(100, 81, 64)}$, $\bm{\theta} = \transpose{(1, \frac{10}{9}, \frac{5}{4})}$, and
\[
\bm{q}^\text{con} = 
\begin{bmatrix}
	0.4 & 0.2 & 0.1 \\
	0.2 & 0.3 & 0.2 \\
	0.1 & 0.2 & 0.2
\end{bmatrix},\quad
\bm{l}^\text{gr} = 
\begin{bmatrix}
	200 & 190 & 180 \\
	190 & 190 & 180 \\
	180 & 180 & 170
\end{bmatrix}.
\]

\begin{figure}[!tp]
	\centering
		\includegraphics[width=0.5\textwidth]{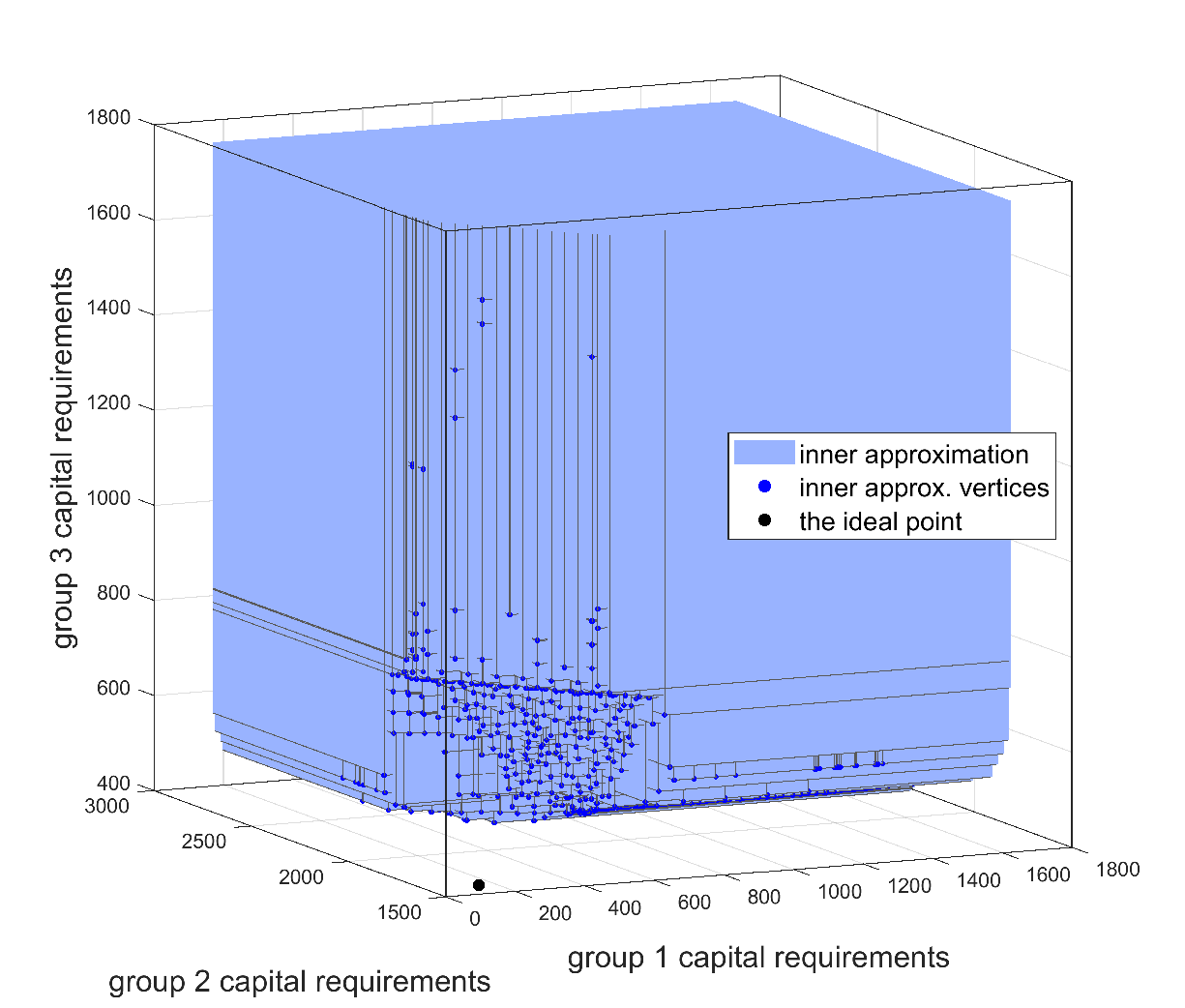}
	\caption{Inner approximation of the three-group Rogers-Veraart systemic risk measure with 60 nodes, 50 scenarios, and approximation error $\epsilon = 40$.}
	\label{figure_3D_RV}
\end{figure}

For $\epsilon = 40$, we observe the following values for the performance of \Cref{alg1}: $|\vertex(\LLL^T)|=975$, $T=53917$, $\#(\PPP_2)=19382$, $\overline{\text{time}}(\PPP_2)=0.4$ seconds, 
$\text{time}(\text{total})=138.1$ minutes. (For this implementation, we use the deterministic equivalent formulations of Pascoletti-Serafini scalarizations rather than the bisection search described in \Cref{rem:bisection}.)  \Cref{figure_3D_RV} provides the inner approximation of the corresponding three-group Rogers-Veraart systemic risk measure, which clearly illustrates that the systemic risk measure does not have convex values in general.

%
%
%


\ACKNOWLEDGMENT{We thank the anonymous referees and the editors whose comments guided us in preparing the current version of the manuscript with major improvements compared to earlier versions. Nurtai Meimanjan acknowledges support from the OeNB anniversary fund, project number 17793, and from the Vienna Graduate School on Computational Optimization, Austrian Science Fund (FWF), project number W1260-N35.}




\ECSwitch

\ECHead{Additional Results, Proofs, Counterexamples, and Discussions}

\section{Systemic Risk Measures Based on Polyhedral Risk Measures}\label{sec:polyrisk}

In this section, we extend the framework of \Cref{systemic_risk_measures} to the case where $\AAA$ is a lower level set of a polyhedral risk measure. The theory of polyhedral risk measures is studied in \cite{polyrisk} in detail, we recall some definitions and results before introducing their use in our setting.

Given $p\in [1,+\infty)$, we denote by $L^p(\RR)$ the set of all random variables $Y\colon\Omega\to\RR$ that are identified up to $\mathbb{P}$-almost sure equality and with $\norm{Y}_p\coloneqq (\EE[\abs{Y}^p])^{\frac{1}{p}}<+\infty$; the space $L^\infty(\RR)$ is already defined in \Cref{systemic_risk_measures}. Given $p\in[1,+\infty]$, recall that a functional $\rho\colon L^p(\RR)\to [-\infty,+\infty]$ is called a \emph{(monetary) risk measure} if the following properties hold for every $Y,Y^\prime\in L^p(\RR)$, $r\in\RR$:
\begin{itemize}
	\item \emph{Monotonicity:} If $Y\geq Y^\prime$, then $\rho(Y)\leq \rho(Y^\prime)$.
	\item \emph{Translativity:} We have $\rho(Y+r)=\rho(Y)-r$.
\end{itemize}

\begin{definition}\label{defn:poly}
	\cite[Definition 2.1]{polyrisk} Let $p\in[1,+\infty]$. A functional $\rho\colon L^p(\RR)\to [-\infty,+\infty]$ is called \emph{polyhedral} if there exist $n_1,n_2\in\NN$; $\bm{c},\bm{{\tilde{c}}}\in\RR^{n_1}$; $\bm{b}, \bm{{\tilde{b}}}\in\RR^{n_2}$; a nonempty polyhedral set $\mathcal{U}_1\subseteq \RR^{n_1}$; and a nonempty polyhedral cone $\mathcal{U}_2\subseteq\RR^{n_2}$ such that
	\begin{align*}
		\rho(Y)=\inf\cb{\transpose{\bm{c}}\bm{u}+\EE[\transpose{\bm{{\tilde{c}}}}\bm{U}]\mid \bm{u}\in \mathcal{U}_1,\ \bm{U}\in L^p(\mathcal{U}_2),\ \transpose{\bm{b}}\bm{u}+\transpose{\bm{{\tilde{b}}}}\bm{U}=Y}
	\end{align*}
	for every $Y\in L^p(\RR)$.
\end{definition}

When $\rho$ is a polyhedral functional as in \Cref{defn:poly}, $\rho(Y)$ can be seen as the optimal value of a two-stage stochastic programming problem where $\bm{u}$ is the first-stage decision variable and $\bm{U}$ is the second-stage decision variable. The quantities $n_1,n_2, \bm{c},\bm{{\tilde{c}}}, \bm{b}, \bm{{\tilde{b}}}, \mathcal{U}_1, \mathcal{U}_2$ are the fixed parameters of the problem; $Y\in L^p(\RR)$ is also a parameter but we treat the optimal value of the problem as a function of this parameter.

Given a nonempty set $\mathcal{U}\subseteq \RR^q$ with $q\in\NN$, the \emph{polar cone} of $\mathcal{U}$ is defined as 
\[
\mathcal{U}^+ \coloneqq \{\bm{z}\in\RR^q\mid \forall \bm{u}\in \mathcal{U}\colon \transpose{\bm{z}}\bm{u}\geq 0\}.
\]

\begin{proposition}\label{prop:poly}
	\cite[Theorem 2.4, Proposition 2.9]{polyrisk} Let $p\in [1,+\infty)$ and consider a polyhedral functional $\rho\colon L^p(\RR)\to [-\infty,+\infty]$ whose parameters are as described in \Cref{defn:poly}. Suppose that the following properties hold:
	\begin{itemize}
		\item \emph{Complete recourse:} $\cb{\bm{{\tilde{b}}}^{\mathsf{T}}\bm{u}\mid \bm{u}\in\mathcal{U}_2}=\RR$.
		\item \emph{Dual feasibility:} There exists $r\in\RR$ such that $\bm{c}+r\bm{b}\in \mathcal{U}_1^+$ and $\bm{{\tilde{c}}}+r\bm{{\tilde{b}}}\in \mathcal{U}_2^+$.
	\end{itemize}
	Then, $\rho$ is finite, convex, and continuous. For each $Y\in L^p(\RR)$, let us define the first-stage attainment set
	\[
	\mathcal{S}_\rho(Y)\coloneqq\cb{\bm{u}\in\mathcal{U}_1\mid \exists \bm{U}\in L^p(\mathcal{U}_2)\colon\ \transpose{\bm{c}}\bm{u}+\EE[\transpose{\bm{{\tilde{c}}}}\bm{U}]=\rho(Y),\ \transpose{\bm{b}}\bm{u}+\transpose{\bm{{\tilde{b}}}}\bm{U}=Y}.
	\]
	In addition to the above properties, if $\mathcal{S}_\rho(0)$ is a nonempty bounded set, then $\mathcal{S}_\rho(Y)$ is a nonempty, convex, compact set for every $Y\in L^p(\RR)$; in particular, the infimum in the definition of $\rho$ is attained.
\end{proposition}

Let $\rho$ be a polyhedral risk measure. In the setting of \Cref{systemic_risk_measures}, let us take $\AAA=\{Y\in L^\infty(\RR)\mid \rho(Y)\leq \rho_0\}$ for some $\rho_0\in\RR$. Then, the corresponding systemic risk measure is given by
\[
R(\bm{X}) = \{\bm{z}\in\ZZZ\mid \rho(\Lambda(\bm{X}+\transpose{\bm{B}}\bm{z}))\leq \rho_0\},\quad \bm{X}\in L^\infty(\RR^n),
\]
where $\Lambda$ is the aggregation function defined through a strictly increasing continuous function $f$ and a set-valued constraint function $\YYY$ via \eqref{aggregation_OPT}. The next theorem is a generalization of \Cref{theorem_OPT}.

\begin{theorem}\label{thm:poly}
	Suppose that \Cref{assumption} holds. Let $p\in [1,+\infty)$ and consider a polyhedral risk measure $\rho\colon L^p(\RR)\to [-\infty,+\infty]$ whose parameters are as described in \Cref{defn:poly}. Suppose further that $\rho$ satisfies complete recourse, dual feasibility and $\mathcal{S}_\rho(0)$ is a nonempty bounded set. Let $\bm{X}\in L^\infty(\RR^n)$ with $R(\bm{X})\neq \emptyset$. Then,
	\begin{align}\label{eq:polyrisk}
		R(\bm{X})=\big\{\bm{z}\in\ZZZ\mid \
		&\transpose{\bm{c}}\bm{u}+\EE[\transpose{\bm{{\tilde{c}}}}\bm{U}]\leq \rho_0,\ \transpose{\bm{b}}\bm{u}+\transpose{\bm{{\tilde{b}}}}\bm{U}\leq f(\bm{P}),\
		(\bm{P},\bm{S})\in L^\infty(\YYY(\bm{X}+\transpose{\bm{B}}\bm{z})),\\
		& \bm{P}\in L^\infty(\RR^n),\ \bm{S}\in L^\infty(\ZZ^d),\  \bm{u}\in\mathcal{U}_1,\ \bm{U}\in L^p(\mathcal{U}_2)\notag 
		\}.
	\end{align}
\end{theorem}

\proof{Proof.}
	Let $\tilde{R}(\bm{X})$ denote the set on the right of \eqref{eq:polyrisk}. Let $\bm{z}\in\tilde{R}(\bm{X})$. Then, there exist $\bm{P}\in L^\infty(\RR^n)$, $\bm{S}\in L^\infty(\ZZ^d)$, $\bm{u}\in\mathcal{U}_1$, $\bm{U}\in L^p(\mathcal{U}_2)$ such that
	\begin{equation}\label{eq:tildeR}
		\transpose{\bm{c}}\bm{u}+\EE[\transpose{\bm{{\tilde{c}}}}\bm{U}]\leq \rho_0,\quad \transpose{\bm{b}}\bm{u}+\transpose{\bm{{\tilde{b}}}}\bm{U}\leq f(\bm{P}),\quad (\bm{P},\bm{S})\in L^\infty(\YYY(\bm{X}+\transpose{\bm{B}}\bm{z})).
	\end{equation}
	In particular, for $\mathbb{P}$-almost every $\omega\in\Omega$, $(\bm{P}(\omega),\bm{S}(\omega))$ is a feasible solution of the problem in \eqref{aggregation_OPT} with $\bm{x}=\bm{X}(\omega)+\transpose{\bm{B}}\bm{z}$ since the definition of $\tilde{R}(\bm{X})$ includes the constraints of the problem in \eqref{aggregation_OPT}. Then, for $\mathbb{P}$-almost every $\omega\in\Omega$, we have $\Lambda(\bm{X}(\omega)+\transpose{\bm{B}}\bm{z})\geq f(\bm{P}(\omega))$. The monotonicity of $\rho$ and \eqref{eq:tildeR} imply that
	\[
	\rho\of{\Lambda(\bm{X}+\transpose{\bm{B}}\bm{z})}\leq \rho\of{f(\bm{P})}\leq \rho\of{\transpose{\bm{b}}\bm{u}+\transpose{\bm{{\tilde{b}}}}\bm{U}}\leq \transpose{\bm{c}}\bm{u}+\EE[\transpose{\bm{{\tilde{c}}}}\bm{U}],
	\]
	where the last inequality follows by the definition of $\rho(\transpose{\bm{b}}\bm{u}+\transpose{\bm{{\tilde{b}}}}\bm{U})$. By \eqref{eq:tildeR} again, we have $\transpose{\bm{c}}\bm{u}+\EE[\transpose{\bm{{\tilde{c}}}}\bm{U}]\leq \rho_0$. Hence, $\rho(\Lambda(\bm{X}+\transpose{\bm{B}}\bm{z}))\leq \rho_0$ so that $\bm{z}\in R(\bm{X})$, which proves $\tilde{R}(\bm{X})\subseteq R(\bm{X})$.
	
	Conversely, let $\bm{z}\in R(\bm{X})$. We have $
	\rho(\Lambda(\bm{X}+\transpose{\bm{B}}\bm{z}))\leq \rho_0$. Note that \Cref{prop:poly} ensures that the infimum in the definition of $\rho(\Lambda(\bm{X}+\transpose{\bm{B}}\bm{z}))$ is indeed a minimum. Hence, there exist $\bm{u}\in\mathcal{U}_1$ and $\bm{U}\in L^p(\mathcal{U}_2)$ such that \begin{equation}\label{eq:tildeR2}
		\transpose{\bm{c}}\bm{u}+\EE[\transpose{\bm{{\tilde{c}}}}\bm{U}] = \rho\of{\Lambda(\bm{X}+\transpose{\bm{B}}\bm{z})}\leq \rho_0,\quad  \transpose{\bm{b}}\bm{u}+\transpose{\bm{{\tilde{b}}}}\bm{U} = \Lambda(\bm{X}+\transpose{\bm{B}}\bm{z}).
	\end{equation}
	In particular, $\Lambda(\bm{X}(\omega)+\transpose{\bm{B}}\bm{z})>-\infty$, i.e., $\YYY(\bm{X}(\omega)+\transpose{\bm{B}}\bm{z})\neq\emptyset$ for $\mathbb{P}$-almost every $\omega\in\Omega$. The assumption that $\YYY$ has compact values implies that
	\[
	\mathcal{D}(\omega)\coloneqq \cb{(\bm{p},\bm{s})\in\YYY(\bm{X}(\omega)+\transpose{\bm{B}}\bm{z})\mid \Lambda(\bm{X}(\omega)+\transpose{\bm{B}}\bm{z})=f(\bm{p})}\neq \emptyset
	\]
	for $\mathbb{P}$-almost every $\omega\in\Omega$. By \Cref{rem:graph}, the set-valued function $\omega \mapsto \YYY(\bm{X}(\omega)+\transpose{\bm{B}}\bm{z})$ is measurable. Thus, by the marginal map theorem \citep[Theorem~8.2.11]{aubin}, $\mathcal{D}$ is a measurable set-valued function and it admits a measurable selection $(\bm{P},\bm{S})$. Then, 
	\[
	\transpose{\bm{b}}\bm{u}+\transpose{\bm{{\tilde{b}}}}\bm{U}(\omega) =\Lambda(\bm{X}(\omega)+\transpose{\bm{B}}\bm{z})=f(\bm{P}(\omega))
	\]
	for $\mathbb{P}$-almost every $\omega\in\Omega$. Moreover, we necessarily have $(\bm{P},\bm{S})\in L^\infty(\RR^n)\times L^\infty(\ZZ^d)$ by \Cref{rem:graph}. Finally, by the first equality in \eqref{eq:tildeR2}, we may conclude that $\bm{z}\in\tilde{R}(\bm{X})$, which proves $R(\bm{X})\subseteq \tilde{R}(\bm{X})$. Therefore, $R(\bm{X})=\tilde{R}(\bm{X})$ is established.
	\Halmos
\endproof

\begin{example}\label{avar}
	Let $\lambda\in (0,1)$ and let $\rho\colon L^1(\RR)\to\RR$ be the \emph{average value-at-risk} at level $\lambda$, i.e.,
	\[
	\rho(Y)=\inf_{r\in \RR}\of{r+\frac{1}{\lambda}\EE[(Y+r)^-]},\quad Y\in L^1(\RR).
	\]
	As discussed in \citet[Example 2.10]{polyrisk}, $\rho$ is a polyhedral risk measure with $n_1=1$, $n_2=2$, $\bm{c}=-1$, $\bm{b}=-1$, $\bm{{\tilde{c}}}=(0,\frac{1}{\lambda})^\mathsf{T}$, $\bm{{\tilde{b}}}=(1,-1)^{\mathsf{T}}$, $\mathcal{U}_1= \RR$, $\mathcal{U}_2=\RR^2_+$. Hence,
	\[
	\rho(Y)=\inf\cb{-u+\frac{1}{\lambda}\EE[U_2]\mid u\in\RR, \bm{U}\in L^1(\RR^2_+),\ -u+U_1-U_2=Y}.
	\]
	It is known that the infimum in the definition of $\rho(Y)$ is attained at $r=-q_\lambda^Y$, where $q_\lambda^Y$ is an arbitrary $\lambda$-quantile of $Y$. Thus, if $\bm{X}\in L^\infty(\RR^n)$ is such that $R(\bm{X})\neq \emptyset$, then \eqref{eq:polyrisk} reduces to
	\begin{align*}
		R(\bm{X})=\big\{\bm{z}\in\ZZZ\mid \
		&-u+\frac{1}{\lambda}\EE[U_2]\leq \rho_0,\ -u+U_1-U_2\leq f(\bm{P}),\
		(\bm{P},\bm{S})\in L^\infty(\YYY(\bm{X}+\transpose{\bm{B}}\bm{z})),\\
		& \bm{P}\in L^\infty(\RR^n),\ \bm{S}\in L^\infty(\ZZ^d),\  u\in\RR,\ \bm{U}\in L^\infty(\RR^2_+)\}.
	\end{align*}
\end{example}

\section{Proofs of the Results in Section~\ref{systemic_risk_measures}}\label{appendixB}

\proof{Proof of \Cref{theorem_OPT}.}

Let us take $\rho(Y)=\EE[-Y]$ for each $Y\in L^1(\RR)$, and $\rho_0=-\gamma$. Note that $\rho$ is a polyhedral risk measure on $L^1(\RR)$ with $n_1=n_2=1$, $\bm{c}=\bm{b}=0$, $\bm{{\tilde{c}}}=-1$, $\bm{{\tilde{b}}}=1$, $\mathcal{U}_1=\{1\}$, $\mathcal{U}_2=\RR$. Hence, complete recourse holds trivially. Moreover, $\mathcal{U}_1^+=\RR_+$, $\mathcal{U}_2^+=\{0\}$, and dual feasibility holds with $r=1$. We also have $\mathcal{S}_\rho(0)=\mathcal{U}_1=\{1\}$. Therefore, by \Cref{thm:poly}, we get
\[
R(\bm{X})=\big\{\bm{z}\in\ZZZ\mid 
\EE[U]\geq \gamma,\ U\leq f(\bm{P}),\
(\bm{P},\bm{S})\in L^\infty(\YYY(\bm{X}+\transpose{\bm{B}}\bm{z})),\ U\in L^1(\RR)
\}.
\]
We claim that
\[
R(\bm{X})= \tilde{R}(\bm{X})\coloneqq \big\{\bm{z}\in\ZZZ\mid 
\EE[f(\bm{P})]\geq \gamma,\ (\bm{P},\bm{S})\in L^\infty(\YYY(\bm{X}+\transpose{\bm{B}}\bm{z}))\}.
\]
The inclusion $R(\bm{X})\supseteq \tilde{R}(\bm{X})$ is trivial since taking $U=f(\bm{P})$ is sufficient. For the reverse inclusion, let $\bm{z}\in R(\bm{X})$. Then, $\EE[U]\geq \gamma$ for some $U\in L^1(\RR)$ with $U\leq f(\bm{P})$ and $(\bm{P},\bm{S})\in L^\infty(\YYY(\bm{X}+\transpose{\bm{B}}\bm{z}))$. Then, $\gamma \leq \EE[U]\leq \EE[f(\bm{P})]$ by the monotonicity of expectation. Therefore, $\bm{z}\in\tilde{R}(\bm{X})$ and $R(\bm{X})\subseteq \tilde{R}(\bm{X})$ follows.
\Halmos
\endproof

\proof{Proof of \Cref{theorem_P1_OPT}.}

Let us denote by $\tilde{\PPP}_1(\bm{w})$ the optimal value of \eqref{z_OPT}. The equality $\PPP_1\of{\bm{w}} = \tilde{\PPP}_1(\bm{w})$ follows immediately by rewriting \eqref{P1_OPT} using \Cref{theorem_OPT}. To prove the second claim, suppose that $\ZZZ\subseteq \bm{z}^{\text{LB}}+\RR^G_+$ for some $\bm{z}^{\text{LB}}\in\RR^G$. Then, $\transpose{\bm{w}}\bm{z}\geq \transpose{\bm{w}}\bm{z}^{\text{LB}}$ for every $\bm{z}\in\ZZZ$. In particular, $\PPP_1(\bm{w})\geq  \transpose{\bm{w}}\bm{z}^{\text{LB}}>-\infty$. Moreover, $\PPP_1(\bm{w})<+\infty$ as we assume $R(\bm{X})\neq\emptyset$. Hence, $\PPP_1(\bm{w})\in\RR$.
\Halmos
\endproof

\proof{Proof of \Cref{theorem_P2_OPT}.}

Let us denote by $\tilde{\PPP}_2^{\bm{c}}(\bm{v})$ the optimal value of \eqref{P2_OPT_lin}. The equality $\PPP^{\bm{c}}_2\of{\bm{v}} = \tilde{\PPP}^{\bm{c}}_2\of{\bm{v}}$ follows immediately by rewriting \eqref{P2_OPT} using \Cref{theorem_OPT}. To prove the second claim, suppose that $\ZZZ\subseteq \bm{z}^{\text{LB}}+\RR^G_+$ for some $\bm{z}^{\text{LB}}\in\RR^G$. In particular, since $\bm{v}\in\ZZZ$, we have $\bm{z}^{\text{LB}}\leq \bm{v}$. Then,
\begin{align*}
	\PPP^{\bm{c}}_2\of{\bm{v}} 	&= \inf \big\{\mu\in\RR \mid\EE[\Lambda(\bm{X}+\transpose{\bm{B}}(\bm{v}+\mu\bm{c}))]\ge\gamma,\ \bm{v}+\mu\bm{c}\in \ZZZ\big\}\\
	&\geq \inf\{\mu\in\RR \mid \bm{v}+\mu\bm{c}\in \ZZZ \}\\
	&\geq \inf\{\mu\in\RR \mid \bm{v}+\mu\bm{c}\in \bm{z}^{\text{LB}}+\RR^G_+\}=\inf\{\mu\in\RR \mid \bm{z}^{\text{LB}}\leq \bm{v}+\mu\bm{c} \}=\max_{\ell\in\GGG\colon c_\ell>0}\frac{z^{\text{LB}}_\ell - v_\ell}{c_\ell}>-\infty.
\end{align*}	
To prove the finiteness claim in (i), let $\bm{z}\in R(\bm{X})$ be such that $z_\ell\leq v_\ell$ for every $\ell\in\GGG$ with $c_\ell=0$. We have $\bm{z}+\RR^G_+\subseteq R(\bm{X})$ by \eqref{upperset}. Hence,
\begin{align*}
	\PPP_2^{\bm{c}}\of{\bm{v}}&=\inf\big\{\mu\in\RR\mid \bm{v}+\mu\bm{c}\in \Rsen\of{\bm{X}}\big\}\\
	&\leq \inf\big\{\mu\in\RR\mid \bm{v}+\mu\bm{c}\in \bm{z}+\RR^G_+\big\}=\inf\{\mu\in\RR\mid \bm{z}\leq \bm{v}+\mu\bm{c}\}=\max_{\ell\in\GGG\colon c_\ell>0}\frac{z_\ell-v_\ell}{c_\ell}<+\infty.
\end{align*}	
The condition in (ii) is a special case of the one in (i) since we assume that $R(\bm{X})\neq\emptyset$. Therefore, $\PPP_2^{\bm{c}}(\bm{v})<+\infty$ in (ii) follows as well.
\Halmos
\endproof 

\section{A Graphical Illustration of \Cref{alg1}}\label{sec:illustration}

In \Cref{figure_algorithmStepByStep}, we provide a graphical illustration of \Cref{alg1} by going through the first three iterations on a hypothetical value of the set $R(\bm{X})$ with $G=2$ groups.

\begin{figure}[!tbp]
	\centering
	\subfloat[\centering
	Construct $\ZZZ=\bm{z}^\text{LB}+\RR^G_+$.]{
		\begin{tikzpicture}
			\draw[step = 0.5, gray!30, thin] (0,0) grid (5,5);
			\draw[name path = A, thick, violet!80] (1,4) arc (180:270:1) -- (2,3) arc (90:0:1) -- (3,2) arc (180:270:1);
			\draw[thin, dashed] (0.5,5) -- (0.5,0.5) -- (5,0.5);
			\draw[name path = B, thin] (1,4) -- (1,5) -- (5,5) -- (5,1) -- (4,1);
			\tikzfillbetween[of= A and B]{violet!40};
			\draw[name path = A, thick, violet!80] (1,4) arc (180:270:1) -- (2,3) arc (90:0:1) -- (3,2) arc (180:270:1);
			\draw[name path = B, thin] (1,4) -- (1,5) -- (5,5) -- (5,1) -- (4,1);
			\draw[black, thin] (1,5) -- (0,5) -- (0,0) -- (5,0) -- (5,1);
			\draw[thick, violet!80] (1,4) -- (1,5);
			\draw[thick, violet!80] (4,1) -- (5,1);
			\node (example-align) [align=center] at (2.1, 4.5) {$\Rsen\of{\bm{X}}$};
			\fill (0.5,0.5) circle[radius=1pt];
			\node at (0.25,0.25) {\footnotesize $\bm{z}^\text{LB}$};
	\end{tikzpicture}}
	\subfloat[\centering 
	Compute $\bm{z}^\text{ideal}$ and $\bm{z}^\text{UB}$.]{
		\begin{tikzpicture}
			\draw[step = 0.5, gray!30, thin] (0,0) grid (5,5);
			\draw[name path = A, thick, violet!80] (1,4) arc (180:270:1) -- (2,3) arc (90:0:1) -- (3,2) arc (180:270:1);
			\draw[thin, dashed] (0.5,5) -- (0.5,0.5) -- (5,0.5);
			\draw[name path = B, thin] (1,4) -- (1,5) -- (5,5) -- (5,1) -- (4,1);
			\tikzfillbetween[of= A and B]{violet!40};
			\draw[name path = A, thick, violet!80] (1,4) arc (180:270:1) -- (2,3) arc (90:0:1) -- (3,2) arc (180:270:1);
			\draw[name path = B, thin] (1,4) -- (1,5) -- (5,5) -- (5,1) -- (4,1);
			\draw[black, thin] (1,5) -- (0,5) -- (0,0) -- (5,0) -- (5,1);
			\draw[thick, violet!80] (1,4) -- (1,5);
			\draw[thick, violet!80] (4,1) -- (5,1);
			\node (example-align) [align=center] at (2.1, 4.5) {$\Rsen\of{\bm{X}}$};
			\fill (0.5,0.5) circle[radius=1pt];
			\node at (0.25,0.25) {\footnotesize $\bm{z}^\text{LB}$};
			\fill (4,4) circle[radius=1pt];
			\node at (3.75,3.75) {\footnotesize $\bm{z}^\text{UB}$};
			\fill (1,1) circle[radius=1pt];
			\node at (1.0,0.77) {\footnotesize $\bm{z}^\text{ideal}$};
	\end{tikzpicture}}
	\subfloat[\centering 
	Initialize $\LLL^0$ and $\UUU^0$.]{
		\begin{tikzpicture}
			\draw[step = 0.5, gray!30, thin] (0,0) grid (5,5);
			\draw[name path = A, thick, violet!80] (1,4) arc (180:270:1) -- (2,3) arc (90:0:1) -- (3,2) arc (180:270:1);
			\draw[thin, dashed] (0.5,5) -- (0.5,0.5) -- (5,0.5);
			\draw[name path = B, thin] (1,4) -- (1,5) -- (5,5) -- (5,1) -- (4,1);
			\tikzfillbetween[of= A and B]{violet!40};
			\draw[name path = A, thick, violet!80] (1,4) arc (180:270:1) -- (2,3) arc (90:0:1) -- (3,2) arc (180:270:1);
			\draw[name path = B, thin] (1,4) -- (1,5) -- (5,5) -- (5,1) -- (4,1);
			\draw[black, thin] (1,5) -- (0,5) -- (0,0) -- (5,0) -- (5,1);
			\draw[thick, violet!80] (1,4) -- (1,5);
			\draw[thick, violet!80] (4,1) -- (5,1);
			\node (example-align) [align=center] at (2.1, 4.5) {$\Rsen\of{\bm{X}}$};
			\fill (0.5,0.5) circle[radius=1pt];
			\node at (0.25,0.25) {\footnotesize $\bm{z}^\text{LB}$};
			\fill (4,4) circle[radius=1pt];
			\node at (3.75,3.75) {\footnotesize $\bm{z}^\text{UB}$};
			\fill (1,1) circle[radius=1pt];
			\node at (1.0,0.77) {\footnotesize $\bm{z}^\text{ideal}$};
			\draw[red, thick, dashed] (1,5) -- (1,1) -- (5,1);
			\draw[blue, thick, dashed] (4,5) -- (4,4) -- (5,4);
	\end{tikzpicture}}
	\newline
	\captionsetup[subfigure]{oneside,margin={0.5cm,0cm}}
	\subfloat[\centering 
	Compute $\PPP_2^{\one_G}(\bm{v}^0)$,
	set $\bm{y}^0 = \bm{v}^0 + \PPP_2^{\one_G}(\bm{v}^0)\one_G$.]{
		\begin{tikzpicture}
			\draw[step = 0.5, gray!30, thin] (0,0) grid (5,5);
			\draw[name path = A, thick, violet!80] (1,4) arc (180:270:1) -- (2,3) arc (90:0:1) -- (3,2) arc (180:270:1);
			\draw[thin, dashed] (0.5,5) -- (0.5,0.5) -- (5,0.5);
			\draw[name path = B, thin] (1,4) -- (1,5) -- (5,5) -- (5,1) -- (4,1);
			\tikzfillbetween[of= A and B]{violet!40};
			\draw[name path = A, thick, violet!80] (1,4) arc (180:270:1) -- (2,3) arc (90:0:1) -- (3,2) arc (180:270:1);
			\draw[name path = B, thin] (1,4) -- (1,5) -- (5,5) -- (5,1) -- (4,1);
			\draw[black, thin] (1,5) -- (0,5) -- (0,0) -- (5,0) -- (5,1);
			\draw[thick, violet!80] (1,4) -- (1,5);
			\draw[thick, violet!80] (4,1) -- (5,1);
			\node (example-align) [align=center] at (2.1, 4.5) {$\Rsen\of{\bm{X}}$};
			\fill (0.5,0.5) circle[radius=1pt];
			\node at (0.25,0.25) {\footnotesize $\bm{z}^\text{LB}$};
			\fill (4,4) circle[radius=1pt];
			\node at (3.75,3.75) {\footnotesize $\bm{z}^\text{UB}$};
			\fill (1,1) circle[radius=1pt];
			\node at (1.3,0.77) {\footnotesize $\bm{z}^\text{ideal} = \bm{v}^0$};
			\draw[red, thick, dashed] (1,5) -- (1,1) -- (5,1);
			\draw[blue, thick, dashed] (4,5) -- (4,4) -- (5,4);
			\fill ({2+cos(45)},{2+sin(45)}) circle[radius=1pt];
			\draw[arrow1] (1,1) -- ({2+cos(45)},{2+sin(45)});
			\node at (2.9,3.0) {\footnotesize $\bm{y}^0$};
	\end{tikzpicture}}
	\subfloat[\centering 
	Update $\LLL^{1} = \LLL^0 \setminus (\bm{y}^0 - \RR^G_{++})$, $\UUU^{1} = \UUU^0 \cup (\bm{y}^0+\RR^G_+)$.]{
			\begin{tikzpicture}
				\draw[step = 0.5, gray!30, thin] (0,0) grid (5,5);
				\draw[name path = A, thick, violet!80] (1,4) arc (180:270:1) -- (2,3) arc (90:0:1) -- (3,2) arc (180:270:1);
				\draw[thin, dashed] (0.5,5) -- (0.5,0.5) -- (5,0.5);
				\draw[name path = B, thin] (1,4) -- (1,5) -- (5,5) -- (5,1) -- (4,1);
				\tikzfillbetween[of= A and B]{violet!40};
				\draw[name path = A, thick, violet!80] (1,4) arc (180:270:1) -- (2,3) arc (90:0:1) -- (3,2) arc (180:270:1);
				\draw[name path = B, thin] (1,4) -- (1,5) -- (5,5) -- (5,1) -- (4,1);
				\draw[black, thin] (1,5) -- (0,5) -- (0,0) -- (5,0) -- (5,1);
				\draw[thick, violet!80] (1,4) -- (1,5);
				\draw[thick, violet!80] (4,1) -- (5,1);
				\node (example-align) [align=center] at (2.1, 4.5) {$\Rsen\of{\bm{X}}$};
				\fill (0.5,0.5) circle[radius=1pt];
				\node at (0.25,0.25) {\footnotesize $\bm{z}^\text{LB}$};
				\fill (1,1) circle[radius=1pt];
				\node at (1.0,0.77) {\footnotesize $\bm{z}^\text{ideal}$};
				\draw[red, thick, dashed] (1,5) -- (1,{2+sin(45)}) -- ({2+cos(45)},{2+sin(45)}) -- ({2+cos(45)},1) -- (5,1);
				\draw[blue, thick, dashed] ({2+cos(45)},5) -- ({2+cos(45)},{2+sin(45)}) -- (5,{2+sin(45)});
				\fill ({2+cos(45)},{2+sin(45)}) circle[radius=1pt];			
				\fill (1,{2+sin(45)}) circle[radius=1pt];
				\fill ({2+cos(45)},1) circle[radius=1pt];
		\end{tikzpicture}}
		\subfloat[\centering
		Compute $\PPP_2^{\one_G}(\bm{v}^1)$, set $\bm{y}^1 = \bm{v}^1 + \PPP_2^{\one_G}(\bm{v}^1)\one_G$.]{
				\begin{tikzpicture}
					\draw[step = 0.5, gray!30, thin] (0,0) grid (5,5);
					\draw[name path = A, thick, violet!80] (1,4) arc (180:270:1) -- (2,3) arc (90:0:1) -- (3,2) arc (180:270:1);
					\draw[thin, dashed] (0.5,5) -- (0.5,0.5) -- (5,0.5);
					\draw[name path = B, thin] (1,4) -- (1,5) -- (5,5) -- (5,1) -- (4,1);
					\tikzfillbetween[of= A and B]{violet!40};
					\draw[name path = A, thick, violet!80] (1,4) arc (180:270:1) -- (2,3) arc (90:0:1) -- (3,2) arc (180:270:1);
					\draw[name path = B, thin] (1,4) -- (1,5) -- (5,5) -- (5,1) -- (4,1);
					\draw[black, thin] (1,5) -- (0,5) -- (0,0) -- (5,0) -- (5,1);
					\draw[thick, violet!80] (1,4) -- (1,5);
					\draw[thick, violet!80] (4,1) -- (5,1);
					\node (example-align) [align=center] at (2.1, 4.5) {$\Rsen\of{\bm{X}}$};
					\fill (0.5,0.5) circle[radius=1pt];
					\node at (0.25,0.25) {\footnotesize $\bm{z}^\text{LB}$};
					\fill (1,1) circle[radius=1pt];
					\node at (1.0,0.77) {\footnotesize $\bm{z}^\text{ideal}$};
					\draw[red, thick, dashed] (1,5) -- (1,{2+sin(45)}) -- ({2+cos(45)},{2+sin(45)}) -- ({2+cos(45)},1) -- (5,1);
					\draw[blue, thick, dashed] ({2+cos(45)},5) -- ({2+cos(45)},{2+sin(45)}) -- (5,{2+sin(45)});
					\fill ({2+cos(45)},{2+sin(45)}) circle[radius=1pt];			
					\fill (1,{2+sin(45)}) circle[radius=1pt];
					\fill ({2+cos(45)},1) circle[radius=1pt];
					\draw[arrow1] (1,{2+sin(45)}) -- ({2-cos(57)},{4-sin(57)});
					\fill ({2-cos(57)},{4-sin(57)}) circle[radius=1pt];
					\node at (1.6,3.4) {\footnotesize $\bm{y}^1$};
					\node at (0.8,2.5) {\footnotesize $\bm{v}^1$};
			\end{tikzpicture}}
			\newline
			\subfloat[\centering
			Update $\LLL^{2} = \LLL^1 \setminus (\bm{y}^1 - \RR^G_{++})$, $\UUU^{2} = \UUU^1 \cup (\bm{y}^1+\RR^G_+)$.]{
					\begin{tikzpicture}
						\draw[step = 0.5, gray!30, thin] (0,0) grid (5,5);
						\draw[name path = A, thick, violet!80] (1,4) arc (180:270:1) -- (2,3) arc (90:0:1) -- (3,2) arc (180:270:1);
						\draw[thin, dashed] (0.5,5) -- (0.5,0.5) -- (5,0.5);
						\draw[name path = B, thin] (1,4) -- (1,5) -- (5,5) -- (5,1) -- (4,1);
						\tikzfillbetween[of= A and B]{violet!40};
						\draw[name path = A, thick, violet!80] (1,4) arc (180:270:1) -- (2,3) arc (90:0:1) -- (3,2) arc (180:270:1);
						\draw[name path = B, thin] (1,4) -- (1,5) -- (5,5) -- (5,1) -- (4,1);
						\draw[black, thin] (1,5) -- (0,5) -- (0,0) -- (5,0) -- (5,1);
						\draw[thick, violet!80] (1,4) -- (1,5);
						\draw[thick, violet!80] (4,1) -- (5,1);
						\node (example-align) [align=center] at (2.1, 4.5) {$\Rsen\of{\bm{X}}$};
						\fill (0.5,0.5) circle[radius=1pt];
						\node at (0.25,0.25) {\footnotesize $\bm{z}^\text{LB}$};
						\fill (1,1) circle[radius=1pt];
						\node at (1.0,0.77) {\footnotesize $\bm{z}^\text{ideal}$};
						\draw[red, thick, dashed] (1,5) -- (1,{4-sin(57)}) -- ({2-cos(57)},{4-sin(57)}) -- ({2-cos(57)},{2+sin(45)}) -- ({2+cos(45)},{2+sin(45)}) -- ({2+cos(45)},1) -- (5,1);
						\draw[blue, thick, dashed] ({2-cos(57)},5) -- ({2-cos(57)},{4-sin(57)}) -- ({2+cos(45)},{4-sin(57)}) -- ({2+cos(45)},{2+sin(45)}) -- (5,{2+sin(45)});
						\fill ({2+cos(45)},{4-sin(57)}) circle[radius=1pt];						
						\fill ({2+cos(45)},{2+sin(45)}) circle[radius=1pt];			
						\fill ({2+cos(45)},1) circle[radius=1pt];
						\fill (1,{4-sin(57)}) circle[radius=1pt];
						\fill ({2-cos(57)},{4-sin(57)}) circle[radius=1pt];
						\fill ({2-cos(57)},{2+sin(45)}) circle[radius=1pt];	
				\end{tikzpicture}}
				\subfloat[\centering 
				Compute $\PPP_2^{\one_G}(\bm{v}^2)$, set $\bm{y}^2 = \bm{v}^2 + \PPP_2^{\one_G}(\bm{v}^2)\one_G$.]{
						\begin{tikzpicture}
							\draw[step = 0.5, gray!30, thin] (0,0) grid (5,5);
							\draw[name path = A, thick, violet!80] (1,4) arc (180:270:1) -- (2,3) arc (90:0:1) -- (3,2) arc (180:270:1);
							\draw[thin, dashed] (0.5,5) -- (0.5,0.5) -- (5,0.5);
							\draw[name path = B, thin] (1,4) -- (1,5) -- (5,5) -- (5,1) -- (4,1);
							\tikzfillbetween[of= A and B]{violet!40};
							\draw[name path = A, thick, violet!80] (1,4) arc (180:270:1) -- (2,3) arc (90:0:1) -- (3,2) arc (180:270:1);
							\draw[name path = B, thin] (1,4) -- (1,5) -- (5,5) -- (5,1) -- (4,1);
							\draw[black, thin] (1,5) -- (0,5) -- (0,0) -- (5,0) -- (5,1);
							\draw[thick, violet!80] (1,4) -- (1,5);
							\draw[thick, violet!80] (4,1) -- (5,1);
							\node (example-align) [align=center] at (2.1, 4.5) {$\Rsen\of{\bm{X}}$};
							\fill (0.5,0.5) circle[radius=1pt];
							\node at (0.25,0.25) {\footnotesize $\bm{z}^\text{LB}$};
							\fill (1,1) circle[radius=1pt];
							\node at (1.0,0.77) {\footnotesize $\bm{z}^\text{ideal}$};
							\draw[red, thick, dashed] (1,5) -- (1,{4-sin(57)}) -- ({2-cos(57)},{4-sin(57)}) -- ({2-cos(57)},{2+sin(45)}) -- ({2+cos(45)},{2+sin(45)}) -- ({2+cos(45)},1) -- (5,1);
							\draw[blue, thick, dashed] ({2-cos(57)},5) -- ({2-cos(57)},{4-sin(57)}) -- ({2+cos(45)},{4-sin(57)}) -- ({2+cos(45)},{2+sin(45)}) -- (5,{2+sin(45)});
							\fill ({2+cos(45)},{4-sin(57)}) circle[radius=1pt];						
							\fill ({2+cos(45)},{2+sin(45)}) circle[radius=1pt];			
							\fill ({2+cos(45)},1) circle[radius=1pt];
							\fill (1,{4-sin(57)}) circle[radius=1pt];
							\fill ({2-cos(57)},{4-sin(57)}) circle[radius=1pt];
							\fill ({2-cos(57)},{2+sin(45)}) circle[radius=1pt];	
							\draw[arrow1] ({2+cos(45)},1) -- ({4-cos(33)},{2-sin(33)});					
							\fill ({4-cos(33)},{2-sin(33)}) circle[radius=1pt];	
							\node at (3.4,1.6) {\footnotesize $\bm{y}^2$};
							\node at (2.5,0.8) {\footnotesize $\bm{v}^2$};
					\end{tikzpicture}}
					\subfloat[\centering 
					Update $\LLL^{3} = \LLL^2 \setminus (\bm{y}^2 - \RR^G_{++})$, $\UUU^{3} = \UUU^2 \cup (\bm{y}^2+\RR^G_+)$.]{
							\begin{tikzpicture}
								\draw[step = 0.5, gray!30, thin] (0,0) grid (5,5);
								\draw[name path = A, thick, violet!80] (1,4) arc (180:270:1) -- (2,3) arc (90:0:1) -- (3,2) arc (180:270:1);
								\draw[thin, dashed] (0.5,5) -- (0.5,0.5) -- (5,0.5);
								\draw[name path = B, thin] (1,4) -- (1,5) -- (5,5) -- (5,1) -- (4,1);
								\tikzfillbetween[of= A and B]{violet!40};
								\draw[name path = A, thick, violet!80] (1,4) arc (180:270:1) -- (2,3) arc (90:0:1) -- (3,2) arc (180:270:1);
								\draw[name path = B, thin] (1,4) -- (1,5) -- (5,5) -- (5,1) -- (4,1);
								\draw[black, thin] (1,5) -- (0,5) -- (0,0) -- (5,0) -- (5,1);
								\draw[thick, violet!80] (1,4) -- (1,5);
								\draw[thick, violet!80] (4,1) -- (5,1);
								\node (example-align) [align=center] at (2.1, 4.5) {$\Rsen\of{\bm{X}}$};
								\fill (0.5,0.5) circle[radius=1pt];
								\node at (0.25,0.25) {\footnotesize $\bm{z}^\text{LB}$};
								\fill (1,1) circle[radius=1pt];
								\node at (1.0,0.77) {\footnotesize $\bm{z}^\text{ideal}$};
								\draw[red, thick, dashed] (1,5) -- (1,{4-sin(57)}) -- ({2-cos(57)},{4-sin(57)}) -- ({2-cos(57)},{2+sin(45)}) -- ({2+cos(45)},{2+sin(45)}) -- ({2+cos(45)},{2-sin(33)}) -- ({4-cos(33)},{2-sin(33)}) -- ({4-cos(33)},1) -- (5,1);
								\draw[blue, thick, dashed] ({2-cos(57)},5) -- ({2-cos(57)},{4-sin(57)}) -- ({2+cos(45)},{4-sin(57)}) -- ({2+cos(45)},{2+sin(45)}) -- ({4-cos(33)},{2+sin(45)})-- ({4-cos(33)},{2-sin(33)}) -- (5,{2-sin(33)});
								\fill ({4-cos(33)},{2+sin(45)}) circle[radius=1pt];						
								\fill ({2+cos(45)},{4-sin(57)}) circle[radius=1pt];						
								\fill ({2+cos(45)},{2+sin(45)}) circle[radius=1pt];			
								\fill (1,{4-sin(57)}) circle[radius=1pt];
								\fill ({2-cos(57)},{4-sin(57)}) circle[radius=1pt];
								\fill ({2-cos(57)},{2+sin(45)}) circle[radius=1pt];	
								\fill ({4-cos(33)},{2-sin(33)}) circle[radius=1pt];	
								\fill ({2+cos(45)},{2-sin(33)}) circle[radius=1pt];					
								\fill ({4-cos(33)},1) circle[radius=1pt];					
						\end{tikzpicture}}
						\caption{
							\centering
							Graphical illustration of the first three iterations of \Cref{alg1} for $G=2$.}
						\label{figure_algorithmStepByStep}
					\end{figure}

\section{Proofs of the Results in \Cref{systemic_risk_models}}\label{appendixC}

\subsection{Proof of \Cref{signed_RV_theorem}}

As a preparation for the proof of \Cref{signed_RV_theorem}, we establish three lemmata below. Throughout this section, we fix an operating cash flow vector $\bm{x}\in\RR^n$.

The first lemma lists some elementary properties of feasible solutions.

\begin{lemma}\label{signed_RV_lemma_1}
	Let $\of{\bm{p}, \bm{s}, \bm{t}}\in \RR^n\times\ZZ^n\times\ZZ^n$ be a feasible solution of \eqref{signed_RV_MILP_explicit}. Let $i\in\NNN$. Then, the following implications hold:
	\begin{enumerate}[(i)]
		\item It holds $s_i+t_i\leq 1$.
		\item If $x_i + (\bm{\pi}\transposeT\bm{p})_i < 0$, then $t_i = 1$. 
		\item If $0 \le x_i + (\bm{\pi}\transposeT\bm{p})_i < \bar{p}_i$, then $s_i = 0$.
		\item If $p_i > 0$, then $t_i = 0$.
		\item If $p_i < \bar{p}_i$, then $s_i = 0$.
		\item If $x_i + (\bm{\pi}\transposeT\bm{p})_i < 0$, then $p_i = 0$.
	\end{enumerate}
\end{lemma}

\proof{Proof.}
	(i) Constraint \eqref{signed_RV_MILP_constraint_2} implies that $\bar{p}_i(s_i+t_i)\leq \bar{p}_i$. Since $\bar{p}_i>0$, we obtain $s_i+t_i\leq 1$.\\
	(ii) Suppose that $x_i + (\bm{\pi}\transposeT\bm{p})_i < 0$. If $t_i = 0$, then, by constraint~\eqref{signed_RV_MILP_constraint_1}, $x_i + (\bm{\pi}\transposeT\bm{p})_i \ge \bar{p}_is_i - m_i t_i = \bar{p}_is_i \ge 0$, which is a contradiction.\\
	(iii) Suppose that $0 \le x_i + (\bm{\pi}\transposeT\bm{p})_i < \bar{p}_i$. If $s_i = 1$, then, by (i), $t_i = 0$ and, by constraint~\eqref{signed_RV_MILP_constraint_1}, $x_i + (\bm{\pi}\transposeT\bm{p})_i \ge \bar{p}_is_i - m_i t_i = \bar{p}_i$, which is a contradiction.\\
	(iv) Suppose that $p_i > 0$. If $t_i = 1$, then, by constraint~\eqref{signed_RV_MILP_constraint_2}, $p_i \le \bar{p}_i\of{1 - t_i} = \bar{p}_i\of{1 - 1} = 0$, which is a contradiction.\\
	(v) Suppose that $p_i < \bar{p}_i$. If $s_i = 1$, then, by constraint~\eqref{signed_RV_MILP_constraint_2}, $p_i \ge \bar{p}_is_i = \bar{p}_i$, which is a contradiction.\\
	(vi) Suppose that $x_i + (\bm{\pi}\transposeT\bm{p})_i < 0$. We have $t_i = 1$ by (ii). Then, by constraint~\eqref{signed_RV_MILP_constraint_2}, $p_i \le \bar{p}_i\of{1 - t_i} = \bar{p}_i\of{1 - 1} = 0$. On the other hand, by constraint~\eqref{signed_RV_MILP_constraint_7}, $p_i \ge 0$. Hence, $p_i = 0$.
\Halmos
\endproof

The next two lemmata are concerned with the properties of optimal solutions. Recall that $\bm{\varphi}^\alpha(\bm{x})=(\alpha\bm{x}) \wedge \bm{x}$ for $\bm{x}\in\RR^n$.

\begin{lemma}\label{signed_RV_lemma_6}
	Let $\of{\bm{p}, \bm{s}, \bm{t}}\in \RR^n\times\ZZ^n\times\ZZ^n$ be an optimal solution of \eqref{signed_RV_MILP_explicit}. Let $i\in\NNN$. If $0 \le x_i + (\bm{\pi}\transposeT\bm{p})_i < \bar{p}_i$, then $p_i = \of{\bm{\varphi}^\alpha(\bm{x}) + \beta\bm{\pi}\transposeT\bm{p}}_i^+$.
\end{lemma}

\proof{Proof.}
	The hypothesis of the lemma implies $s_i = 0$ by \Cref{signed_RV_lemma_1}(iii). We use this observation frequently in the rest of the proof. We consider the following three cases:\\
	\textbf{Case 1:} Suppose that $x_i + \beta(\bm{\pi}\transposeT\bm{p})_i \le 0$. In this case, we must have $x_i \le 0$ since $\beta(\bm{\pi}\transposeT\bm{p})_i \ge 0$. In particular, $\varphi^\alpha_i(\bm{x})= (\alpha x_i)\wedge x_i = x_i$ since $0 < \alpha \le 1$. Then, $\of{\bm{\varphi}^\alpha(\bm{x}) + \beta\bm{\pi}\transposeT\bm{p}}_i^+ = \of{x_i + \beta(\bm{\pi}\transposeT\bm{p})_i}^+ = 0$ thanks to the supposition of this case. Hence, we will prove that $p_i = 0$.
	
	To get a contradiction, assume that $p_i > 0$. We have $t_i=0$ by \Cref{signed_RV_lemma_1}(iv). Together with this, $s_i=0$ and constraint~\eqref{signed_RV_MILP_constraint_4} yield $p_i \le x_i + \beta\of{\bm{\pi}\transposeT\bm{p}}_i \le 0$, contradicting $p_i >0$. Therefore, $p_i = 0$.\\
	\textbf{Case 2:} Suppose that $x_i < 0$ and $x_i + \beta(\bm{\pi}\transposeT\bm{p})_i > 0$. In this case, we have $(\alpha x_i)\wedge x_i= x_i$, which implies $\of{\bm{\varphi}^\alpha(\bm{x}) + \beta\bm{\pi}\transposeT\bm{p}}_i^+ = \of{x_i + \beta(\bm{\pi}\transposeT\bm{p})_i}^+ = x_i + \beta(\bm{\pi}\transposeT\bm{p})_i$. Thus, we will prove that $p_i = x_i + \beta\of{\bm{\pi}\transposeT\bm{p}}_i$.
	
	Suppose otherwise that $p_i \neq x_i + \beta(\bm{\pi}\transposeT\bm{p})_i$. First, let us consider the case $p_i > x_i + \beta(\bm{\pi}\transposeT\bm{p})_i$. In particular $p_i>0$, which implies $t_i=0$ by \Cref{signed_RV_lemma_1}(iv). Using this, $s_i=0$ and constraint~\eqref{signed_RV_MILP_constraint_4}, we get $p_i \le x_i + \beta(\bm{\pi}\transposeT\bm{p})_i$, contradicting $p_i > x_i + \beta(\bm{\pi}\transposeT\bm{p})_i$.
	
	Next, we consider the case $p_i < x_i + \beta(\bm{\pi}\transposeT\bm{p})_i$. Let $\bm{\pnew}\in\RR^n_+$ be equal to $\bm{p}$ in all components except the $i\textsuperscript{th}$ one, and take $\pnew_i = x_i + \beta(\bm{\pi}\transposeT\bm{p})_i > p_i$. Also, let $\bm{\tnew}\in\{0,1\}^n$ be equal to $\bm{t}$ in all components except possibly the $i\textsuperscript{th}$ one, and take $\tnew_i = 0$. Since $\bm{p}\leq \bm{\pnew}$, $p_i<\pnew_i$, and $f$ is strictly increasing, we have $f(\bm{p})< f(\bm{\pnew})$. Hence, we obtain a contradiction to the optimality of $(\bm{p},\bm{s},\bm{t})$ once we show that $(\bm{\pnew}, \bm{s}, \bm{\tnew})$ is a feasible solution of \eqref{signed_RV_MILP_explicit}.
	
	To check the feasibility of $(\bm{\pnew}, \bm{s}, \bm{\tnew})$, let $k\in\NNN\setminus\cb{i}$. We verify the $k\textsuperscript{th}$ constraints of \eqref{signed_RV_MILP_explicit} for $(\bm{\pnew}, \bm{s}, \bm{\tnew})$. Note that
	\begin{align}\label{eq:constraintcheck1}
		(\bm{\pi}\transposeT\bm{\pnew})_k =  \sum_{j\in\NNN\backslash\cb{i}}\pi_{jk}\pnew_j + \pi_{ik}\pnew_i &= \sum_{j\in\NNN\backslash\cb{i}}\pi_{jk}p_j + \pi_{ik}\of{x_i + \beta(\bm{\pi}\transposeT\bm{p})_i} \\ 
		&\geq  \sum_{j\in\NNN\backslash\cb{i}}\pi_{jk}p_j + \pi_{ik}p_i = (\bm{\pi}\transposeT\bm{p})_k, \notag 
	\end{align}
	and hence, $x_k + (\bm{\pi}\transposeT\bm{\pnew})_k \geq  x_k + (\bm{\pi}\transposeT\bm{p})_k \ge \bar{p}_ks_k - m_k t_k = \bar{p}_ks_k - m_k \tnew_k$ by the definition of $(\bm{\pnew}, \bm{s}, \bm{\tnew})$ and the feasibility of $(\bm{p}, \bm{s}, \bm{t})$. Therefore, constraint \eqref{signed_RV_MILP_constraint_1} holds. Moreover, \eqref{eq:constraintcheck1} also implies that
	\[
	\varphi_k^\alpha(\bm{x}) + \beta(\bm{\pi}\transposeT\bm{\pnew})_k + (m_k+\bar{p}_k)(s_k + \tnew_k) \geq
	\varphi_k^\alpha(\bm{x}) + \beta(\bm{\pi}\transposeT\bm{p})_k + (m_k+\bar{p}_k)(s_k + t_k)
	\ge p_k = \pnew_k.
	\]
	Thus, constraint \eqref{signed_RV_MILP_constraint_4} holds. Constraints \eqref{signed_RV_MILP_constraint_2}, \eqref{signed_RV_MILP_constraint_7} hold trivially by the feasibility of $(\bm{p}, \bm{s}, \bm{t})$ because they do not include $\pnew_i$ or $\tnew_i$.
	
	Next, we verify the $i\textsuperscript{th}$ constraints of \eqref{signed_RV_MILP_explicit} for $(\bm{\pnew}, \bm{s}, \bm{\tnew})$. 
	Similar to \eqref{eq:constraintcheck1}, we obtain $(\bm{\pi}\transposeT\bm{\pnew})_i = (\bm{\pi}\transposeT\bm{p})_i$ since $\pi_{ii}=0$. Then, $
	x_i + (\bm{\pi}\transposeT\bm{\pnew})_i = x_i + (\bm{\pi}\transposeT\bm{p})_i \ge x_i + \beta(\bm{\pi}\transposeT\bm{p})_i > 0 = \bar{p}_is_i -m_i \tnew_i$
	since $0 < \beta \le 1$, $x_i + \beta\of{\bm{\pi}\transposeT\bm{p}}_i > 0$ by the assumption of this case, $s_i = 0$ as noted above, and $\tnew_i = 0$ by definition. Hence, constraint \eqref{signed_RV_MILP_constraint_1} holds.
	Similarly, constraint \eqref{signed_RV_MILP_constraint_2} holds as
	\[
	\bar{p}_is_i=0< \pnew_i = x_i + \beta(\bm{\pi}\transposeT\bm{p})_i \le x_i + (\bm{\pi}\transposeT\bm{p})_i < \bar{p}_i = \bar{p}_i\of{1 - \tnew_i}.
	\]
	To see that constraint \eqref{signed_RV_MILP_constraint_4} holds, we note that
	\begin{align*}			
		\pnew_i = x_i + \beta(\bm{\pi}\transposeT\bm{p})_i 
		\le \varphi_i^\alpha(\bm{x}) + \beta(\bm{\pi}\transposeT\bm{p})_i 
		& = \varphi_i^\alpha(\bm{x})+ \beta(\bm{\pi}\transposeT\bm{\pnew})_i \\
		& \le \varphi_i^\alpha(\bm{x})+ \beta(\bm{\pi}\transposeT\bm{\pnew})_i + (m_i +\bar{p}_i)(s_i + \tnew_i).
	\end{align*}
	Since $0 < x_i + \beta(\bm{\pi}\transposeT\bm{p})_i = \pnew_i \le x_i + (\bm{\pi}\transposeT\bm{p})_i < \bar{p}_i$ by the hypothesis of the lemma, the first part of constraint \eqref{signed_RV_MILP_constraint_7} holds as well. Finally,	the second part of constraint \eqref{signed_RV_MILP_constraint_7} holds trivially since $s_i = 0$ and $\tnew_i = 0$. Hence, the feasibility of $(\bm{\pnew}, \bm{s}, \bm{\tnew})$ is established and we conclude that $p_i = x_i + \beta(\bm{\pi}\transposeT\bm{p})_i $.\\
	\textbf{Case 3:} Suppose that $x_i \ge 0$ and $x_i + \beta(\bm{\pi}\transposeT\bm{p})_i > 0$. In this case, we have $(\alpha x_i)\wedge x_i= \alpha x_i \ge 0$, which implies $\of{\bm{\varphi}^\alpha(\bm{x}) + \beta\bm{\pi}\transposeT\bm{p}}_i^+ = \of{\alpha x_i + \beta(\bm{\pi}\transposeT\bm{p})_i}^+ = \alpha x_i + \beta(\bm{\pi}\transposeT\bm{p})_i \ge 0$. Thus, we will prove that $p_i = \alpha x_i + \beta(\bm{\pi}\transposeT\bm{p})_i$.
	
	To get a contradiction, assume that $p_i \neq \alpha x_i + \beta(\bm{\pi}\transposeT\bm{p})_i$. Let us consider the case $p_i > x_i + \beta(\bm{\pi}\transposeT\bm{p})_i$ first. In particular, $p_i>0$ and we have $t_i=0$ by \Cref{signed_RV_lemma_1}(iv). With this, $s_i=0$, and constraint~\eqref{signed_RV_MILP_constraint_4}, we obtain $p_i \le \alpha x_i + \beta(\bm{\pi}\transposeT\bm{p})_i$, contradicting $p_i > \alpha x_i + \beta(\bm{\pi}\transposeT\bm{p})_i$.
	
	Next, we consider the case $p_i < \alpha x_i + \beta(\bm{\pi}\transposeT\bm{p})_i$. Let $\bm{\pnew}\in\RR^n_+$ be equal to $\bm{p}$ in all components except the $i\textsuperscript{th}$ one, and take $\pnew_i = \alpha x_i + \beta(\bm{\pi}\transposeT\bm{p})_i > p_i$. Let $\bm{\tnew}\in\{0,1\}^n$ be equal to $\bm{t}$ in all components except possibly the $i\textsuperscript{th}$ one, and take $\tnew_i = 0$. As in Case 2, we get a contradiction to the optimality of $(\bm{p}, \bm{s}, \bm{t})$ once we show that $(\bm{\pnew}, \bm{s}, \bm{\tnew})$ is feasible for \eqref{signed_RV_MILP_explicit}.
	
	To show the feasibility of $(\bm{\pnew}, \bm{s}, \bm{\tnew})$, let us first verify the $k\textsuperscript{th}$ constraints in \eqref{signed_RV_MILP_explicit}, where $k\in\NNN\backslash\cb{i}$. By the assumption of this case, we have
	\begin{align}			
		(\bm{\pi}\transposeT\bm{\pnew})_k  =  \sum_{j\in\NNN\backslash\cb{i}}\pi_{jk}\pnew_j + 		\pi_{ik}\pnew_i =& \sum_{j\in\NNN\backslash\cb{i}}\pi_{jk}p_j + \pi_{ik}\of{\alpha x_i + \beta(\bm{\pi}\transposeT\bm{p})_i}  \label{eq:constraintcheck2}\\ 
		\geq & \sum_{j\in\NNN\backslash\cb{i}}\pi_{jk}p_j + \pi_{ik}p_i = (\bm{\pi}\transposeT\bm{p})_k.\notag 
	\end{align}
	Hence, $x_k + (\bm{\pi}\transposeT\bm{\pnew})_k \geq  x_k + (\bm{\pi}\transposeT\bm{p})_k \ge \bar{p}_ks_k - m_k t_k = 	\bar{p}_ks_k - m_k \tnew_k$ by the feasibility of $(\bm{p}, \bm{s}, \bm{t})$ so that constraint \eqref{signed_RV_MILP_constraint_1} holds. Similarly, constraint \eqref{signed_RV_MILP_constraint_4}
	holds as
	\[			
	\varphi_k^\alpha(\bm{x}) + \beta(\bm{\pi}\transposeT\bm{\pnew})_k + (m_k +\bar{p}_k)(s_k + 	\tnew_k) \\
	\geq \alpha x_k + \beta(\bm{\pi}\transposeT\bm{p})_k + (m_k +\bar{p}_k)(s_k + t_k) 
	\ge p_k = \pnew_k.
	\]
	Constraints \eqref{signed_RV_MILP_constraint_2}, \eqref{signed_RV_MILP_constraint_7} hold trivially by the feasibility of $(\bm{p}, \bm{s}, \bm{t})$ as they do not include $\pnew_i$ or $\tnew_i$.
	
	Next, we verify the $i\textsuperscript{th}$ constraints in \eqref{signed_RV_MILP_explicit} for $(\bm{\pnew}, \bm{s}, \bm{\tnew})$. Similar to \eqref{eq:constraintcheck2}, we have $(\bm{\pi}\transposeT\bm{\pnew})_i = (\bm{\pi}\transposeT\bm{p})_i$. Thus, $x_i + (\bm{\pi}\transposeT\bm{\pnew})_i = x_i + (\bm{\pi}\transposeT\bm{p})_i \ge x_i + \beta(\bm{\pi}\transposeT\bm{p})_i > 0 	= \bar{p}_is_i - m_i \tnew_i$ since $0 < \beta \le 1$, $x_i + \beta(\bm{\pi}\transposeT\bm{p})_i > 0$ by the assumption in this case, $s_i = 0$ by \Cref{signed_RV_lemma_1}(iii), and $\tnew_i = 0$ by definition. Therefore, constraint \eqref{signed_RV_MILP_constraint_1} holds. Similarly, constraint \eqref{signed_RV_MILP_constraint_2} holds as
	\begin{align*}			
		\bar{p}_is_i= 0 < \pnew_i = \alpha x_i + \beta(\bm{\pi}\transposeT\bm{p})_i \le x_i + (\bm{\pi}\transposeT\bm{p})_i < \bar{p}_i = \bar{p}_i(1 - 	\tnew_i),
	\end{align*}
	which follows by the hypothesis of the lemma and the assumptions of this case. Note that
	\[
	\pnew_i = \alpha x_i+ \beta(\bm{\pi}\transposeT\bm{p})_i 
	= \varphi_i^\alpha(\bm{x})+ \beta(\bm{\pi}\transposeT\bm{\pnew})_i 
	\le \varphi_i^\alpha(\bm{x})+ \beta(\bm{\pi}\transposeT\bm{\pnew})_i + (m_i +\bar{p}_i)(s_i + \tnew_i).
	\]
	Thus, constraint \eqref{signed_RV_MILP_constraint_4} holds. By the hypothesis of this lemma, we have
	\begin{align*}			
		0 \le \alpha x_i + \beta(\bm{\pi}\transposeT\bm{p})_i = \pnew_i \le x_i + (\bm{\pi}\transposeT\bm{p})_i < \bar{p}_i.
	\end{align*}
	Hence, constraint \eqref{signed_RV_MILP_constraint_7} holds. Constraint \eqref{signed_RV_MILP_constraint_7} holds trivially since $s_i = 0$ and $\tnew_i = 0$. Therefore, $(\bm{\pnew}, \bm{s}, \bm{\tnew})$ is feasible for \eqref{signed_RV_MILP_explicit} and $p_i = \alpha x_i + \beta(\bm{\pi}\transposeT\bm{p})_i$ follows.
	
	In all three cases, we obtain $p_i = \of{\bm{\varphi}^\alpha(\bm{x}) + \beta\bm{\pi}\transposeT\bm{p}}_i^+$ for every $i\in\NNN$.
\Halmos
\endproof

\begin{lemma}\label{signed_RV_lemma_7}
	Let $\of{\bm{p}, \bm{s}, \bm{t}}\in \RR^n\times\ZZ^n\times\ZZ^n$ be an optimal solution of \eqref{signed_RV_MILP_explicit}. Let $i\in\NNN$. If $x_i + (\bm{\pi}\transposeT\bm{p})_i \ge \bar{p}_i$, then $p_i = \bar{p}_i$.
\end{lemma}

\proof{Proof.}
	To get a contradiction, assume that $p_i \neq \bar{p}_i$. Then $p_i < \bar{p}_i$ since we already have $p_i \le \bar{p}_i$ by constraint~\eqref{signed_RV_MILP_constraint_7}. Let $\bm{\pnew}\in\RR^n_+$ be equal to $\bm{p}$ in all components except the $i\textsuperscript{th}$ one, and take $\pnew_i = \bar{p}_i > p_i$. Let $\bm{\snew}\in\{0,1\}^n$ be equal to $\bm{s}$ in all components except the $i\textsuperscript{th}$ one, and take $\snew_i = 1$. Finally, let $\bm{\tnew}\in\{0,1\}^n$ be equal to $\bm{t}$ in all components except possibly the $i\textsuperscript{th}$ one, and take $\tnew_i = 0$. We claim that $\of{\bm{\pnew}, \bm{\snew}, \bm{\tnew}}$ is a feasible solution of \eqref{signed_RV_MILP_explicit}. Once the claim is shown, we get a contradiction to the optimality of $(\bm{p}, \bm{s}, \bm{t})$ because $f(\bm{\pnew})>f(\bm{p})$ by the strict monotonicity of $f$.
	
	To show the feasibility of $(\bm{\pnew}, \bm{\snew}, \bm{\tnew})$, we first verify the $k\textsuperscript{th}$ constraints in \eqref{signed_RV_MILP_explicit} for fixed $k\in\NNN\backslash\cb{i}$. Note that
	\begin{align}\label{eq:constraintcheck3}
		(\bm{\pi}\transposeT\bm{p})_k & = \sum_{j\in\NNN\backslash\cb{i}}\pi_{jk}p_j + \pi_{ik}p_i \\ 
		& \le \sum_{j\in\NNN\backslash\cb{i}}\pi_{jk}p_j + \pi_{ik}\bar{p}_i = \sum_{j\in\NNN\backslash\cb{i}}\pi_{jk}\pnew_j + \pi_{ik}\pnew_i = (\bm{\pi}\transposeT\bm{\pnew})_k.\notag 
	\end{align}
	Then, $\bar{p}_k\snew_k - m_k \tnew_k = \bar{p}_ks_k - m_k t_k \le x_k + (\bm{\pi}\transposeT\bm{p})_k \le x_k + (\bm{\pi}\transposeT\bm{\pnew})_k$ by the feasibility of $(\bm{p}, \bm{s}, \bm{t})$. Hence, constraint \eqref{signed_RV_MILP_constraint_1} holds. Note that constraint \eqref{signed_RV_MILP_constraint_4} holds since
	\begin{align*}
		\pnew_k = p_k & \le \varphi_k^\alpha(\bm{x}) + \beta(\bm{\pi}\transposeT\bm{p})_k + (m_k +\bar{p}_k)(s_k + t_k) \\
		& \le \varphi_k^\alpha(\bm{x}) + \beta(\bm{\pi}\transposeT\bm{\pnew})_k + (m_k +\bar{p}_k)(\snew_k + \tnew_k),
	\end{align*}
	which follows by the feasibility of $(\bm{p}, \bm{s}, \bm{t})$. Constraints \eqref{signed_RV_MILP_constraint_2}, \eqref{signed_RV_MILP_constraint_7} hold trivially by the feasibility of $(\bm{p}, \bm{s}, \bm{t})$ because they do not include $\pnew_i$, $\snew_i$ or $\tnew_i$.
	
	Next, we verify the $i\textsuperscript{th}$ constraints in \eqref{signed_RV_MILP_explicit}. Similar to \eqref{eq:constraintcheck3}, we have $(\bm{\pi}\transposeT\bm{p})_i=(\bm{\pi}\transposeT\bm{\pnew})_i$ since $\pi_{ii}=0$. Then, constraint \eqref{signed_RV_MILP_constraint_1} holds since
	\[
	\bar{p}_i\snew_i - m_i \tnew_i 
	= \bar{p}_i \le x_i + (\bm{\pi}\transposeT\bm{p})_i  =x_i + (\bm{\pi}\transposeT\bm{\pnew})_i,
	\]
	thanks to the hypothesis of the lemma. Since $\snew_i = 1$ and $\tnew_i = 0$ by definition, we have
	$\bar{p}_i\snew_i =\bar{p}_i =  \pnew_i  =\bar{p}_i\of{1 - \tnew_i}$, which verifies constraint \eqref{signed_RV_MILP_constraint_2}. Note that $\beta(\bm{\pi}\transposeT\bm{p})_i \ge 0$, $\alpha x_i + m_i \geq \alpha x_i + x_i^- \geq 0$, and $x_i+m_i\geq x_i + x_i^- \ge 0$. Then, constraint \eqref{signed_RV_MILP_constraint_4} holds since
	\begin{align*}
		\pnew_i = \bar{p}_i 
		&\le\bar{p}_i + \varphi_i^\alpha(\bm{x}) + m_i + \beta(\bm{\pi}\transposeT\bm{p})_i \\ 
		& = \varphi_i^\alpha(\bm{x}) + \beta(\bm{\pi}\transposeT\bm{p})_i + (m_i +\bar{p}_i)
		= \varphi_i^\alpha(\bm{x}) + \beta(\bm{\pi}\transposeT\bm{p})_i + (m_i +\bar{p}_i)(\snew_i + \tnew_i).
	\end{align*}
	Constraint \eqref{signed_RV_MILP_constraint_7} holds trivially. Hence,
	$(\bm{\pnew}, \bm{\snew}, \bm{\tnew})$ is a feasible solution of \eqref{signed_RV_MILP_explicit}.
\Halmos 
\endproof

We combine the results of the above lemmata to complete the proof of \Cref{signed_RV_theorem}.

\proof{Proof of \Cref{signed_RV_theorem}.}
	We first argue that the feasible region of \eqref{signed_RV_MILP_explicit} is free of the choice of $(m_i)_{i\in\NNN}$. Let $(\bm{p},\bm{s},\bm{t})$ be a feasible solution of \eqref{signed_RV_MILP_explicit} with the given choice of $(m_i)_{i\in\NNN}$. Let us fix $i\in\NNN$. By \Cref{signed_RV_lemma_1}(i), we have $t_i=0$ or $s_i=0$ (or both). If $t_i=0$, then $\bar{p}_is_i-m_it_i=\bar{p}_is_i-x_i^-t_i$, trivially. If $s_i=0$, then $\bar{p}_is_i-x_i^- t_i= -x_i^-\leq x_i\leq x_i+(\bar{\bm{\pi}}^{\mathsf{T}}\bm{p})_i$. Hence, constraint \eqref{signed_RV_MILP_constraint_1} still holds when $m_i$ is replaced with $x_i^-$. If $s_i+t_i=0$, then $(m_i+\bar{p}_i)(s_i+t_i)=0=(x_i^-+\bar{p}_i)(s_i+t_i)$. If $s_i+t_i=1$, then
	\[
	\varphi^\alpha_i(\bm{x})+\beta(\bm{\pi}^{\mathsf{T}}\bm{p})_i+(x_i^-+\bar{p}_i)(s_i+t_i)\geq \varphi^\alpha_i(\bm{x})+x_i^-+\bar{p}_i\geq \bar{p}_i\geq p_i
	\] 
	since $\varphi_i^\alpha(\bm{x})+x_i^-=x_i+x_i^-=x_i^+\geq 0$ when $x_i<0$ and $\varphi_i^\alpha(\bm{x})+x_i^-=\alpha x_i\geq 0$ when $x_i\geq 0$. Hence, constraint \eqref{signed_RV_MILP_constraint_4} still holds when $m_i$ is replaced with $x_i^-$. It follows that $(\bm{p},\bm{s},\bm{t})$ is a feasible solution of \eqref{signed_RV_MILP_explicit} with $(m_i)_{i\in\NNN}$ replaced by $(x_i^-)_{i\in\NNN}$. The reverse observation can be seen easily by inspection since increasing these constants potentially relaxes constraints \eqref{signed_RV_MILP_constraint_1}, \eqref{signed_RV_MILP_constraint_4}, and does not affect constraints \eqref{signed_RV_MILP_constraint_2}, \eqref{signed_RV_MILP_constraint_7}. Therefore, the feasible region remains the same for different choices of $(m_i)_{i\in\NNN}$.

	It is easy to check that $(\bm{p},\bm{s}, \bm{t}) = (\zero_n,\zero_n, \bm{1}_n)\in\RR^n\times\ZZ^n\times\ZZ^n$ is a feasible solution of \eqref{signed_RV_MILP_explicit}. Given $(\bm{s},\bm{t})\in\{0,1\}^n\times\{0,1\}^n$, it is clear that the set of all $\bm{p}\in \RR^n_+$ for which $(\bm{p},\bm{s},\bm{t})$ is feasible for \eqref{signed_RV_MILP_explicit} is a closed set. Thus, the feasible region of \eqref{signed_RV_MILP_explicit} is a closed set as a finite union of closed sets. Moreover, the feasible region is a subset of $[\zero_n,\bm{{\bar{p}}}]\times\{0,1\}^n\times\{0,1\}^n$; therefore, it is also compact. Since $f$ is a continuous function, it follows that \eqref{signed_RV_MILP_explicit} has at least one optimal solution and $\Lambda^{\text{RV}}(\bm{x})\in\RR$.
	
	Let $\of{\bm{p}, \bm{s}, \bm{t}}$ be an optimal solution of \eqref{signed_RV_MILP_explicit}. To prove that $\bm{p}$ is a clearing vector, we show that $\bm{\Phi}^{\text{RV}}(\bm{p}) = \bm{p}$. Let $i\in\NNN$. Recalling \eqref{fixed_point_signed_RV}, we consider three cases:
	\begin{enumerate}[(i)]
		\item Assume that $(\bm{\pi}\transposeT\bm{p} + \bm{x})_i < 0$. Then, $\Phi^{\text{RV}}_i\of{\bm{p}} = 0$ by \eqref{fixed_point_signed_RV}, and $p_i = 0$ by \Cref{signed_RV_lemma_1}(vi).
		\item Assume that $0 \le (\bm{\pi}\transposeT\bm{p} + \bm{x})_i < \bar{p}_i$. Then, $\Phi^{\text{RV}}_i(\bm{p}) = \of{\bm{\varphi}^\alpha(\bm{x}) + \beta\bm{\pi}\transposeT\bm{p}}_i^+$ by \eqref{fixed_point_signed_RV}, and $p_i = \of{\bm{\varphi}^\alpha(\bm{x}) + \beta\bm{\pi}\transposeT\bm{p}}_i^+$ by \Cref{signed_RV_lemma_6}.
		\item Assume that $(\bm{\pi}\transposeT\bm{p} + \bm{x})_i \ge \bar{p}_i$. Then, $\Phi^{\text{RV}}_i(\bm{p}) = \bar{p}_i$ by \eqref{fixed_point_signed_RV}, and $p_i = \bar{p}_i$ by \Cref{signed_RV_lemma_7}.
	\end{enumerate}
	In each case, we have $p_i =  \Phi^{\text{RV}}_i\of{\bm{p}}$.	Therefore, $\bm{p}$ is a clearing vector.
\Halmos
\endproof

\subsection{Proofs of the Corollaries of \Cref{signed_RV_theorem}}

As a preparation for the proof of \Cref{pos_RV_cor}, we prove a lemma that is analogous to \Cref{signed_RV_lemma_7}.

\begin{lemma}\label{lem:RV+}
	Let $(\bm{p},\bm{s})\in\RR^n\times\ZZ^n$ be an optimal solution of \eqref{pos_RV_MILP_explicit}. Let $i\in\NNN$. If $x_i+(\bm{\pi}^{\mathsf{T}}\bm{p})_i\geq \bar{p}_i$, then $p_i=\bar{p}_i$.
	\end{lemma}

\proof{Proof.}
The proof follows a similar argument as in the proof of \Cref{signed_RV_lemma_7}. Hence, we skip some details for brevity. To get a contradiction, let us assume that $p_i<\bar{p}_i$. Let $\bm{\pnew}\in\RR^n_+$ be equal to $\bm{p}$ in all components except the $i\textsuperscript{th}$ one, and take $\pnew_i = \bar{p}_i > p_i$. Let $\bm{\snew}\in\{0,1\}^n$ be equal to $\bm{s}$ in all components except possibly the $i\textsuperscript{th}$ one, and take $\snew_i = 1$. It is easy to check that $(\bm{\pnew},\bm{\snew})$ is feasible for \eqref{pos_RV_MILP_explicit} with $f(\bm{\pnew})>f(\bm{p})$, contradicting the optimality of $(\bm{p},\bm{s})$.
\Halmos 
\endproof

\proof{Proof of \Cref{pos_RV_cor}.}
First, note that $(\bm{0}_n,\bm{0}_n)$ is a feasible solution of \eqref{pos_RV_MILP_explicit}. By the same arguments as in the proof of \Cref{signed_RV_theorem}, it follows that \eqref{pos_RV_MILP_explicit} has an optimal solution and $\Lambda^{\text{RV}_+}(\bm{x})\in\RR$.

For the rest of the proof, we assume that $m_i=x_i^-=0$ for each $i\in\NNN$ in the formulation of \eqref{signed_RV_MILP_explicit}. This causes no loss of generality thanks to \Cref{signed_RV_theorem}. Let $(\bm{p},\bm{s},\bm{t})$ be a feasible solution of \eqref{signed_RV_MILP_explicit}. Let $i\in\NNN$ be such that $t_i=1$. Then, \eqref{signed_RV_MILP_constraint_2} implies that $p_i=0$ and $s_i=0$. With this observation, the following inequalities hold trivially:
\[
\bar{p}_is_i \leq x_i+(\transpose{\bm{\pi}}\bm{p})_i,\quad  p_i\leq \alpha x_i +\beta (\transpose{\bm{\pi}}\bm{p})_i+\bar{p}_i s_i,\quad 0\leq p_i \leq \bar{p}_i.
\]
Moreover, these inequalities hold trivially when $i\in\NNN$ is such that $t_i=0$. Therefore, $(\bm{p},\bm{s})$ is feasible for \eqref{pos_RV_MILP_explicit}.

Finally, let $(\bm{p},\bm{s})$ be an optimal solution of \eqref{pos_RV_MILP_explicit}. We first show that $(\bm{p},\bm{s},\bm{0}_n)$ is a feasible solution of \eqref{signed_RV_MILP_explicit}. Let $i\in\NNN$. The only constraint that needs to be checked is $\bar{p}_is_i\leq p_i$. This holds trivially if $s_i=0$. Let us suppose that $s_i=1$. Then, constraint \eqref{pos_RV_MILP_constraint_1} yields $\bar{p}_i\leq x_i+(\bm{\pi}^{\mathsf{T}}\bm{p})_i$. Hence, $p_i=\bar{p}_i$ by \Cref{lem:RV+} and the constraint $\bar{p}_is_i\leq p_i$ is verified. This establishes the feasibility of $(\bm{p},\bm{s},\bm{0}_n)$. We claim that $(\bm{p},\bm{s},\bm{0}_n)$ is indeed optimal for \eqref{signed_RV_MILP_explicit}. Suppose that this is not the case. Then, by \Cref{signed_RV_theorem}, there exists an optimal solution $(\bm{{p}^\prime}, \bm{{s}^\prime},\bm{{t}^\prime})$. In particular, $f(\bm{{p}^\prime})>f(\bm{p})$. By the previous paragraph, $(\bm{{p}^\prime},\bm{{s}^\prime})$ is feasible for \eqref{pos_RV_MILP_explicit}, which contradicts the optimality of $(\bm{p},\bm{s})$. Thus, the claim follows and we also have $\Lambda^{\text{RV}_+}(\bm{x})=\Lambda^{\text{RV}}(\bm{x})$. By \Cref{signed_RV_theorem} again, $\bm{p}$ is a clearing vector for $\systemRV$.
\Halmos
\endproof

\proof{Proof of \Cref{signed_EN_cor}.}
The proofs of the existence of an optimal solution and the finiteness of $\Lambda^{\text{EN}}(\bm{x})$ follow as in the proof of \Cref{pos_RV_cor}. Let $(\bm{p},\bm{t})$ be a feasible solution of \eqref{signed_EN_MILP_explicit}. We claim that $(\bm{p},\bm{0}_n,\bm{t})$ is a feasible solution of \eqref{signed_RV_MILP_explicit}. Indeed, constraints \eqref{signed_RV_MILP_constraint_2}, \eqref{signed_RV_MILP_constraint_4}, \eqref{signed_RV_MILP_constraint_7} hold trivially. Let $i\in\NNN$. If $t_i=1$, then \eqref{signed_RV_MILP_constraint_1} holds since $-m_i\leq -x_i^- \leq x_i + (\bm{\pi}\transposeT\bm{p})_i$. If $t_i=0$, then \eqref{signed_EN_MILP_constraint_4} gives $p_i\leq x_i+(\transpose{\bm{\pi}}\bm{p})_i$ so that \eqref{signed_RV_MILP_constraint_1} holds. This completes the proof of the claim. Conversely, let $(\bm{p},\bm{s},\bm{t})$ be a feasible solution of \eqref{signed_RV_MILP_explicit}. We claim that $(\bm{p},\bm{t})$ is feasible for \eqref{signed_EN_MILP_explicit}. It is clear that constraints \eqref{signed_EN_MILP_constraint_2}, \eqref{signed_EN_MILP_constraint_7} are satisfied. Let $i\in\NNN$. If $s_i=0$, then \eqref{signed_EN_MILP_constraint_4} is satisfied obviously. If $s_i=1$, then \Cref{signed_RV_lemma_1}(v) yields $p_i=\bar{p}_i$, and \eqref{signed_RV_MILP_constraint_2} yields $t_i=0$. With these, rewriting \eqref{signed_RV_MILP_constraint_1} gives $p_i=\bar{p}_i\leq x_i+(\transpose{\bm{\pi}}\bm{p})_i$. Therefore, \eqref{signed_EN_MILP_constraint_4} is satisfied and the claim follows. From these claims and \Cref{signed_RV_theorem}, it follows that the feasible region of \eqref{signed_EN_MILP_explicit} is free of the choice of $(m_i)_{i\in\NNN}$.

Let $(\bm{p},\bm{t})$ be an optimal solution of \eqref{signed_EN_MILP_explicit}. Using the above claims and following the same arguments as in the proof of \Cref{pos_RV_cor}, it can be checked that $(\bm{p},\bm{0}_n,\bm{t})$ is optimal for \eqref{signed_RV_MILP_explicit} and $\Lambda^{\text{EN}}(\bm{x})=\Lambda^{\text{RV}}(\bm{x})$. By \Cref{signed_RV_theorem}, $\bm{p}$ is a clearing vector for $\systemRV$.
\Halmos 
\endproof

\proof{Proof of \Cref{pos_EN_cor}.}
Clearly, if $\bm{p}$ is a feasible solution of \eqref{pos_EN_LP_explicit}, then $(\bm{p},\bm{0}_n)$ is a feasible solution of \eqref{signed_EN_MILP_explicit}. Conversely, let $(\bm{p},\bm{t})$ be a feasible solution of \eqref{signed_EN_MILP_explicit} and fix $i\in\NNN$. If $t_i=0$, then constraint \eqref{signed_EN_MILP_constraint_4} yields $p_i\leq x_i+(\transpose{\bm{\pi}}\bm{p})_i$. If $t_i=1$, then constraint \eqref{signed_EN_MILP_constraint_2} implies $p_i=0$ so that $p_i=0\leq x_i+(\transpose{\bm{\pi}}\bm{p})_i$ holds. Hence, $\bm{p}$ is feasible for \eqref{pos_EN_LP_explicit}. Similar to the proof of \Cref{pos_RV_cor}, these observations and \Cref{signed_EN_cor} imply that if $\bm{p}$ is an optimal solution of \eqref{pos_EN_LP_explicit}, then $(\bm{p},\bm{0}_n)$ is optimal for \eqref{signed_EN_MILP_explicit} so that $\bm{p}$ is a clearing vector and $\Lambda^{\text{EN}_+}(\bm{x})=\Lambda^{\text{RV}}(\bm{x})\in\RR$.
\Halmos 
\endproof 

\section{The Signed Eisenberg-Noe Model: Justifying the Seniority-Based Approach}\label{sec:naiveapproach}

The aim of this section is to analyze the sink node approach proposed in \citet[Section 2.2]{eisenberg-noe} and compare it with our seniority-based approach in \Cref{systemic_risk_models} for treating negative operating cash flows. We argue that following the sink node approach and constructing an extended network with $n+1$ nodes whose parameters depend on the signed operating cash flow vector have some major drawbacks both from the network models and the systemic risk measures perspectives.

\subsection{The Extended Network in the Sink Node Approach}

Let $\systemRV$ be a signed Eisenberg-Noe network; in particular, $\bm{x}\in\RR^n$ and $\alpha=\beta=1$. According to the conjecture in \cite{eisenberg-noe} (see \Cref{rem:naiveapproach}), the network can be extended by a sink node, to which each node with a negative operating cash flow owes the absolute value of that amount. The resulting extended network will become a standard Eisenberg-Noe network; as a result, operating costs and interbank liabilities will have equal priority.

To construct the extended network, we first define the total obligations vector $\bm{{\bar{p}}}^{\bm{x}}\in\RR^{n+1}$ by
\[
\bar{p}^{\bm{x}}_i \coloneqq \bar{p}_i + x_i^- ,\quad i\in\NNN;\quad\quad  \bar{p}^{\bm{x}}_{n+1}\coloneqq 0.
\]
Then, for every $i, j\in\{1,\ldots,n+1\}$, the relative liability of node $i$ to node $j$ is given by
\[
	\pi^{\bm{x}}_{ij} \coloneqq \begin{cases}
		\frac{\pi_{ij}\bar{p}_i}{\bar{p}_i + x_i^-} &\quad\text{if }i, j \in \NNN, \\
		\frac{x_i^-}{\bar{p}_i + x_i^-} &\quad\text{if } i\in \NNN, j=n+1, \\
		\frac{1}{n+1}	&\quad\text{if }i=n+1.
	\end{cases}
\]
Here, since the sink node does not have any obligations to the other nodes, we select $\pi^{\bm{x}}_{n+1,j} = \frac{1}{n+1}$ without loss of generality. Clearly, ${\bm{\pi}}^{\bm{x}}\in\RR^{(n+1)\times(n+1)}$ is a right stochastic matrix. Finally, we define the operating cash flow vector $\bm{x}^{\bm{x}}\in\RR^{n+1}_+$ by
\[
x_i^{\bm{x}} = x_i^+,\quad i\in \NNN; \quad \quad x^{\bm{x}}_{n+1}=0.
\]
This extended network satisfies the model assumptions of \citet[Section 2.2]{eisenberg-noe} since $\bm{x}^{\bm{x}}\in\RR^{n+1}_+$. Observe that, if $\bm{x}\in\RR^n_+$, then the extended network reduces to a network of $n$ (possibly) connected nodes and one isolated node.

According to the fixed point characterization in \citet[Section 2.3]{eisenberg-noe}, a clearing vector for the extended network is characterized as a fixed point of the mapping $\bm{{\tilde{\Phi}}}^{\text{EN}_+}\colon[\bm{0}_{n+1},{\bm{{\bar{p}}}}^{\bm{x}}]\to[\bm{0}_{n+1},{\bm{{\bar{p}}}}^{\bm{x}}]$ defined by
\[
	\bm{{\tilde{\Phi}}}^{\text{EN}_+}(\bm{{\tilde{p}}}) \coloneqq \of{\transpose{(\bm{\pi}^{\bm{x}})}\bm{{\tilde{p}}} +\bm{x}^{\bm{x}}}\wedge {\bm{{\bar{p}}}}^{\bm{x}},
\]
and the existence of a clearing vector is guaranteed by \citet[Theorem 1]{eisenberg-noe}.

\subsection{Comparing the Sink Node and the Seniority-Based Approaches: An Example}

We provide an example of a signed Eisenberg-Noe network with its extended version and discuss some counter-intuitive implications. Let us take
\[
\bm{x} = \begin{bmatrix} 1 \\ -1 \end{bmatrix},\quad  \bm{{\bar{p}}} = \begin{bmatrix} 1 \\ 1 \end{bmatrix},\quad \bm{\pi} = \begin{bmatrix}0 & 1 \\ 1 & 0\end{bmatrix}.
\]
By the sink node approach, the extended network has
\[
\bm{x}^{\bm{x}} = \begin{bmatrix} 1 \\ 0 \\ 0 \end{bmatrix},\quad  \bm{{\bar{p}}}^{\bm{x}}= \begin{bmatrix} 1 \\ 2 \\ 0 \end{bmatrix},\quad \bm{\pi}^{\bm{x}}= \begin{bmatrix} 0 & 1 & 0 \\ \frac12 & 0 & \frac12 \\ \frac{1}{3} & \frac{1}{3} & \frac{1}{3}\end{bmatrix}.
\]

Let us consider the payment vectors $\bm{p}=\transpose{(1,0)}$ for the original network and $\bm{{\tilde{p}}}=\transpose{(1,1,0)}$ for the extended network. It is easy to check that $\bm{\Phi}^{\text{RV}}(\bm{p})= \bm{p}$ (see \eqref{fixed_point_signed_RV}) and $\bm{{\tilde{\Phi}}}^{\text{EN}_+}(\bm{{\tilde{p}}})= \bm{{\tilde{p}}}$. Hence, $\bm{p}$ is a clearing vector for the original network in the sense of \Cref{clearing_vector_defn_signed_RV}, and $\bm{{\tilde{p}}}$ is a clearing vector for the extended network in the sense of our \Cref{clearing_vector_defn_signed_RV} and \citet[Definition~1]{eisenberg-noe}. While $\bm{p}$ suggests that node $2$ pays nothing, $\bm{{\tilde{p}}}$ suggests that node $2$ pays one unit in total: it pays $0.5$ to node $1$ and $0.5$ to the sink node. However, in the scope of the original network, the use of $\bm{{\tilde{p}}}$ implies the violation of the limited liability property in the following sense (see \Cref{rem:lim_absRV}). The net amount of the assets of node $2$ is $x_2+\tilde{p}_1\pi_{12}=-1+1=0$. While, according to limited liability, node $2$ should not make any payments to node $1$, the extended network makes node $2$ pay $\tilde{p}_2\pi^{\bm{x}}_{21}=\frac12$ to node $1$. Moreover, due to lack of seniority between external and internal liabilities, node $2$ ignores half of its external liabilities. These suggest that a clearing vector produced by the sink node approach might not yield a payment vector for the original network with a clear interpretation. On the other hand, making external liabilities senior over internal ones as in our approach ensures that limited liability is respected, and the interpretation of a clearing vector is quite natural as we already stay within the original network.

Furthermore, from a modeling perspective, it is possible that operating costs are very short-termed liabilities (as suggested by \citet{eisenberg-noe}, they can be liabilities to ``\ldots workers, suppliers, and so forth\ldots"), which usually should be satisfied first. A negative operating cash flow can be considered as a highly liquid asset for the external counterparty. In a situation where the external counterparty belongs to a different Eisenberg-Noe network (with a disjoint set of nodes), this liquid asset will be the source of payment for the internal liabilities of the counterparty. Then, the equal seniority assumption of the sink node approach would suggest that both networks should already be combined into a larger network where the operating cash flows become internal liabilities. Thus, assuming that external liabilities have seniority over the internal ones provides a more reasonable approach in these cases.

\subsection{Aggregation Function in the Sink Node Approach}

Following the sink node approach, we may use the mathematical programming formulation in \Cref{pos_EN_cor} for the extended network. Writing this formulation explicitly in terms of the original network yields
\begin{align*}
	\text{maximize} \quad 			& f(\bm{{\tilde{p}}}) && \\ 
	\text{subject to}\quad & \tilde{p}_i \le  x^+_i + \sum_{j=1}^{n}\frac{\pi_{ji}\bar{p}_j}{\bar{p}_j + x_j^-}\tilde{p}_j, &&i\in\NNN,   \\ 
	& \tilde{p}_{n+1} \le  \sum_{j=1}^{n}\frac{x_j^-}{\bar{p}_j + x_j^-}\tilde{p}_j, &&\text{(redundant due to last constraint)}  \\ 
	& 0 \le \tilde{p}_i \le \bar{p}_i + x_i^-,&& i\in\NNN, \\
	& 0 \le \tilde{p}_{n+1} \le 0,&& \text{(hence $\tilde{p}_{n+1}=0$)}
\end{align*}
where $f\colon [\bm{0}_{n+1},\bm{{\bar{p}}}^{\bm{x}}]\to\RR$ is a strictly increasing continuous function. An optimal solution $\tilde{\bm{p}}$ of this problem is a clearing vector for the extended network. Let us denote by $\tilde{\Lambda}^{\text{EN}_+}(\bm{x})$ the optimal value of this problem. We can observe that the constraints are highly nonlinear in $\bm{x}$, which, in the scope of systemic risk measures, leads to intractable formulations for weighted-sum and Pascoletti-Serafini scalarizations. Indeed, for these scalarizations, $\bm{x}$ is to be replaced with $\bm{X}(\omega)+\transpose{\bm{B}}\bm{z}$, where $\omega$ is a scenario and $\bm{z}$ is a capital allocation vector of interest. Thus, we change the structure of the network both \emph{randomly} and depending on the \emph{decision variable} of the problem! For this reason, we find it quite impractical to follow the sink node approach for computing systemic risk measures.

\section{Proofs of the Results in \Cref{RVsystrisk}}\label{app:RVsystrisk}

\proof{Proof of \Cref{lem:RV_RM}.}

Clearly, (a) implies (b).

Suppose that (b) holds and let $\bm{z}\in R^{\text{RV}}(\bm{X})$. Then, the definitions of $\Lambda^{\text{RV}}$ and $f$ imply that $\transpose{\one}_n\bm{{\bar{p}}}=f(\bm{{\bar{p}}})\geq \EE[\Lambda^{\text{RV}}(\bm{X}+\transpose{\bm{B}}\bm{z})]\geq \gamma$. Hence, (c) holds.

Finally, suppose that (c) holds. Let us define $\bm{P} \equiv  \bm{{\bar{p}}}$, $\bm{S} = \bm{T} \equiv \zero_n$. Let $\omega\in\Omega$. We show that $(\bm{P}(\omega),\bm{S}(\omega),\bm{T}(\omega))$ is a feasible solution of $\text{RV}(\bm{X}(\omega)+\transpose{\bm{B}}\bm{{\hat{z}}})$, see \Cref{signed_RV_theorem}. Let $i\in\NNN_\ell$ for some $\ell\in\GGG$. We have $(\transpose{\bm{B}}\bm{{\hat{z}}})_i=\hat{z}_\ell= \|\bm{X}^-\|_\infty + \frac{1}{\alpha}\norminfpbar$. Then, constraint \eqref{signed_RV_MILP_constraint_1} holds since
\[
	X_i(\omega) + (\transpose{\bm{B}}\bm{{\hat{z}}})_i + (\bm{\pi}\transposeT\bm{P}(\omega))_i
	=X_i(\omega) + \|\bm{X}^-\|_\infty + \frac{1}{\alpha}\norminfpbar+ (\bm{\pi}\transposeT\bm{{\bar{p}}})_i
	\geq 0=\bar{p}_i S_i(\omega)-M_iT_i(\omega).
\]
Next, we have
\begin{align*}
	&\alpha \of{X_i(\omega)+(\transpose{\bm{B}}\bm{{\hat{z}}})_i}+\beta (\transpose{\bm{\pi}}\bm{P}(\omega))_i+(M_i+\bar{p}_i)(S_i(\omega)+T_i(\omega))\\
	&=\alpha X_i(\omega) + \alpha\|\bm{X}^-\|_\infty + \norminfpbar+\beta (\transpose{\bm{\pi}}\bm{{\bar{p}}})_i  \geq  \bar{p}_i=P_i(\omega).
\end{align*}
Similarly, we also have 
\begin{align*}
	&X_i(\omega)+(\transpose{\bm{B}}\bm{{\hat{z}}})_i+\beta (\transpose{\bm{\pi}}\bm{P}(\omega))_i+(M_i+\bar{p}_i)(S_i(\omega)+T_i(\omega))\\
	&= X_i(\omega) + \|\bm{X}^-\|_\infty + \frac{1}{\alpha}\norminfpbar+\beta (\transpose{\bm{\pi}}\bm{{\bar{p}}})_i 
	\geq  \bar{p}_i=P_i(\omega).
\end{align*}
Hence, constraint \eqref{signed_RV_MILP_constraint_4} holds. Constraints \eqref{signed_RV_MILP_constraint_2} and \eqref{signed_RV_MILP_constraint_7} hold trivially and the feasibility of $(\bm{P}(\omega),\bm{S}(\omega),\bm{T}(\omega))$  follows. Then, by the definition of $\Lambda^{\text{RV}}$, we have $\EE[\Lambda^{\text{RV}}(\bm{X}+\transpose{\bm{B}}\bm{{\hat{z}}})]\geq \EE[\transpose{\one}_n\bm{P}]= \transpose{\one}_n\bm{{\bar{p}}} \ge \gamma$. Therefore, $\bm{{\hat{z}}}\in R^{\text{RV}}(\bm{X})$, i.e., (a) holds.
\Halmos 
\endproof

\proof{Proof of \Cref{P1_signed_RV_proposition}.}

By \Cref{lem:RV_RM}, $\bm{{\hat{z}}}\in R^{\text{RV}}(\bm{X})\neq\emptyset$. Thus, the inequality in \eqref{eq:P1RV} follows. Since $\bm{{\hat{z}}}=(\|\bm{X}^-\|_\infty + \frac{1}{\alpha}\norminfpbar)\one_G$, the inequality in \eqref{eq:P1RV} follows as well. Let us define a set-valued function $\YYY^\text{RV}\colon \RR^n\to 2^{\RR^n\times\ZZ^n\times\ZZ^n}$ by
\begin{align*}
	\YYY^\text{RV}(\bm{x}) \coloneqq \Big\{ \of{\bm{p}, \bm{s}, \bm{t}} \in\RR^n\times\ZZ^n\times\ZZ^n \mid\ 
	& \bm{{\bar{p}}}\odot\bm{s} - \bm{M}\odot \bm{t} \le \bm{x} + \transpose{\bm{\pi}}\bm{p}; \ \bm{{\bar{p}}}\odot\bm{s}\leq \bm{p} \le \bm{{\bar{p}}}\odot\of{\one_n- \bm{t}};\\
	& \bm{p} \le\bm{\varphi}^\alpha (\bm{x})+ \beta \transpose{\bm{\pi}}\bm{p} + (\bm{M}+\bm{{\bar{p}}})\odot (\bm{s} + \bm{t}); \\
	& \bm{p} \in \sqb{\zero_n, \bm{{\bar{p}}}}; \bm{s},\bm{t} \in \cb{0,1}^n
	\Big\},
\end{align*}
where $\bm{M}\coloneqq \transpose{(M_1,\ldots,M_n)}$, and $\bm{a}\odot\bm{b} \coloneqq \transpose{(a_1 b_1,\ldots,a_n b_n)}$ denotes the Hadamard product of $\bm{a},\bm{b}\in\RR^n$.  For each $\bm{x}\in\RR^n$, the set $\YYY^{\text{RV}}(\bm{x})$ is compact as a closed subset of the compact set $[\bm{0}_n,\bm{{\bar{p}}}]\times\{0,1\}^n\times\{0,1\}^n$. Then, by \Cref{theorem_P1_OPT} with $\YYY = \YYY^\text{RV}$, the optimal value of \eqref{P1_MILP_signed_RV_explicit} coincides with $\PPP_1^\text{RV}(\bm{w})$. Finally, since \eqref{P1_MILP_signed_RV_explicit} is an MILP problem, having finite optimal value implies the existence of an optimal solution.
\Halmos
\endproof

\proof{Proof of \Cref{corollary_P2_signed_RV}.}

Let $\bm{z}\in R^{\text{RV}}(\bm{X})$ be such that $z_\ell\leq v_\ell$ for every $\ell\in\GGG$ with $c_\ell=0$. In particular, $\bm{z}+\RR^G_+\subseteq R^\text{RV}(\bm{X})$. Then,
\begin{align*}
	\PPP_2^{\text{RV},\bm{c}}(\bm{v})&=\inf\{\mu\in\RR\mid \bm{v}+\mu\bm{c}\in R^{\text{RV}}(\bm{X})\}\\
	&\leq \inf\{\mu\in\RR\mid \bm{v}+\mu\bm{c}\in \bm{z}+\RR^G_+\}=\inf\{\mu\in\RR\mid \bm{z}\leq \bm{v}+\mu\bm{c}\}=\max_{\ell\in\GGG\colon c_\ell>0}\frac{z_\ell -v_\ell}{c_\ell}.
\end{align*}
Hence, \eqref{eq:P2RV} follows. Taking $\YYY=\YYY^\text{RV}$, where $\YYY^\text{RV}$ is the set-valued function defined in the proof of \Cref{P1_signed_RV_proposition}, and applying \Cref{theorem_P2_OPT} yield that the optimal value of \eqref{P2_MILP_signed_RV_explicit} coincides with $\PPP_2^{\text{RV},\bm{c}}(\bm{v})$. Since \eqref{P2_MILP_signed_RV_explicit} is an MILP problem, with constraint \eqref{P2_MILP_signed_RV_constraint_5} treated as two linear constraints, the finiteness of the optimal value implies the existence of an optimal solution.
\Halmos 
\endproof

\section{Remarks on the Nonconcavity of Aggregation Functions}\label{sec:counter}

In this section, by providing counterexamples, we show that the aggregation functions of the Rogers-Veraart and the signed Eisenberg-Noe models are not concave in general (see \Cref{noconcave_signed_EN}). Accordingly, the corresponding systemic risk measures may fail to have convex values. This is in contrast to what is usually assumed in the literature on systemic risk measures. At the end of this section, we discuss the lack of convexity/concavity in comparison with some recent approaches in similar contexts where one deviates from convexity assumptions.

\subsection{A Counterexample for the Rogers-Veraart Model}

In the setting of the Rogers-Veraart model (\Cref{pos_RV_cor}), let us take $f(\bm{p}) \coloneqq \transpose{\one}_n\bm{p}$ for $\bm{p}\in [\bm{0}_n,\bm{{\bar{p}}}]$; see \Cref{rem:f} for the interpretation of the corresponding aggregation function as the total debt paid at clearing. Consider a two-node network with
\[
\bm{{\bar{p}}} = \begin{bmatrix} 1 \\ 3 \end{bmatrix},\quad \bm{\pi} = \begin{bmatrix}0 & 1 \\ 1 & 0\end{bmatrix}, \quad \alpha = \beta = 0.1.
\]
Then, for a given operating cash flow vector $\bm{x}\in\RR^2_+$, the problem \eqref{pos_RV_MILP_explicit} of calculating the aggregation function reads as
\begin{align*}
	\text{maximize} \quad 			&p_1+p_2 && \\ 
	\text{subject to} \quad 	& s_1 \le x_1 + p_2, \quad 3s_2 \le x_2 + p_1, \\
	& p_1 \le 0.1 x_1 + 0.1 p_2 + s_1 , \quad p_2 \le 0.1 x_2 + 0.1 p_1 + 3s_2 , \\ 
	& 0 \le p_1 \le 1, \quad 0 \le p_2 \le 3, \quad s_1, s_2 \in \cb{0,1}.
\end{align*}
This is an MILP problem with only two binary variables; as an alternative to using MILP solvers, one can even solve it by enumerating through four linear programming problems each of which can be solved by the simple graphical method. We consider the following choices for $\bm{x}$: $\bm{x}^1 = \bm{0}_2$, $\bm{x}^2 =\bm{1}_2$, $\bm{x}^3 = \transpose{(3,3)}$. For $\bm{x}^1$, $(\bm{p}^1,\bm{s}^1)=(\bm{0}_2,\bm{0}_2)$ is an optimal solution and we have $\Lambda^{\text{RV}_+}(\bm{x}^1) = 0$. For $\bm{x}^2$, 
$(\bm{p}^2, \bm{x}^2) = (\transpose{(1, \frac15)},\transpose{(1,0)})$ is an optimal solution with optimal value $\Lambda^{\text{RV}_+}(\bm{x}^2) = 1.2$. Finally, for $\bm{x}^3$, $(\bm{p}^3, \bm{x}^3) =(\transpose{(1, 3)}, \one_2)$ is an optimal solution with optimal value $\Lambda^{\text{RV}_+}(\bm{x}^3) = 4$. If we let $\lambda=\frac{2}{3}$, then $\lambda\bm{x}^1 + (1-\lambda)\bm{x}^3 =\bm{x}^2$ so that $\Lambda^{\text{RV}_+}(\lambda\bm{x}^1 + (1-\lambda)\bm{x}^3) = \Lambda^{\text{RV}_+}(\bm{x}^2) = 1.2$. On the other hand, $\lambda\Lambda^{\text{RV}_+}(\bm{x}^1) + (1-\lambda)\Lambda^{\text{RV}_+}(\bm{x}^3) = 
\frac{2}{3} \times 0 + \frac{1}{3} \times 4 = \frac{4}{3}$,
which yields $\Lambda^{\text{RV}_+}(\lambda\bm{x}^1 + (1-\lambda)\bm{x}^3) < \lambda\Lambda^{\text{RV}_+}(\bm{x}^1) + (1-\lambda)\Lambda^{\text{RV}_+}(\bm{x}^3)$. Hence, $\Lambda^{\text{RV}_+}$ is not concave.

\subsection{A Counterexample for the Signed Eisenberg-Noe Model}

In the setting of the signed Eisenberg-Noe model (\Cref{signed_EN_cor}), let us take $f(\bm{p}) \coloneqq \transpose{\one}_n\bm{p}$ for $\bm{p}\in[\bm{0}_n,\bm{{\bar{p}}}]$. Consider a two-node network with
$\bm{{\bar{p}}} = (1, 4)^{\mathsf{T}}$, $\pi_{11}=\pi_{22}=0$, $\pi_{12}=\pi_{21}=1$. Then, the problem \eqref{signed_EN_MILP_explicit} for calculating the aggregation function becomes
\begin{align*}
	\text{maximize} \quad 			& p_1+p_2&& \\ 
	\text{subject to} \quad 	& p_1 \le 1 - t_1, \quad p_2 \le 4 - 4 t_2, \\
	& p_1 \le x_1 + p_2 + (m_1+1)t_1 , \quad p_2 \le x_2 + p_1 + (m_2+4)t_2 , \\ 
	& 0 \le p_1 \le 1, \quad 0 \le p_2 \le 4, \quad t_1, t_2 \in \cb{0,1}.
\end{align*}
We consider the following choices of $\bm{x}$: $\bm{x}^1 =\transpose{(-3, -3)}$, $\bm{x}^2 =\bm{0}_2$, $\bm{x}^3 = \transpose{(3,3)}$. For $\bm{x}^1$, $(\bm{p}^1,\bm{t}^1)=(\bm{0}_2,\bm{1}_n)$ is an optimal solution with $\Lambda^{\text{EN}}(\bm{x}^1) = 0$. For $\bm{x}^2$, $(\bm{p}^2, \bm{t}^2) =(\bm{1}_2,\bm{0}_2)$ is an optimal solution with $\Lambda^{\text{EN}}(\bm{x}^2) = 2$. Finally, for $\bm{x}^3$, $(\bm{p}^3,\bm{t}^3) = (\transpose{(1, 4)},\bm{0}_2)$ is an optimal solution with $\Lambda^{\text{EN}}(\bm{x}^3) = 5$. Taking $\lambda=\frac{1}{2}$, we have $\lambda\bm{x}^1 + (1-\lambda)\bm{x}^3 = \bm{x}^2$ so that $\Lambda^{\text{EN}}(\lambda\bm{x}^1 + (1-\lambda)\bm{x}^3) = \Lambda^{\text{EN}}(\bm{x}^2) = 2$. However, $\lambda\Lambda^{\text{EN}}(\bm{x}^1) + (1-\lambda)\Lambda^{\text{EN}}(\bm{x}^3) = \frac{1}{2} \times 0 + \frac{1}{2} \times 5 = 2.5$, which yields $\Lambda^{\text{EN}}(\lambda\bm{x}^1 + (1-\lambda)\bm{x}^3) < \lambda\Lambda^{\text{EN}}(\bm{x}^1) + (1-\lambda)\Lambda^{\text{EN}}(\bm{x}^3)$. Hence, $\Lambda^{\text{EN}}$ is not concave.

\subsection{Nonconcavity and Connection to Star-Shaped Risk Measures}

In the literature on systemic risk measures, it is usually assumed that the aggregation function is concave (and monotone), which is motivated by economic considerations. For instance, in \citet[Section 2]{ararat}, it is argued that the aggregation function is a quantification of the impact of wealth, i.e., operating cash flow, on society and concavity encodes the principle that a diversified wealth should have better consequences for society.

An aggregation function can be considered as a multivariate extension of a utility function for the setting of financial networks. The motivations for assuming monotonicity and concavity are similar for both types of functions. In \citet[Section 3]{ararat}, using a concave aggregation function, the so-called multivariate $g$-divergences are defined as extensions of the $g$-divergence in information theory, and an economic interpretation is provided as a probabilistic measure of distance between the network and the societal node.

In economic theory, criticisms on the concavity assumption for utility functions are not new: we refer the reader to \cite{savage,prospect} for earlier works, and to \cite{miklos,starshaped} for more recent uses of nonconcave utility functions in financial mathematics. In our setting, we obtain a nonconcave aggregation function in the standard Rogers-Veraart model as a consequence of the discontinuity caused by the default costs. Hence, the lack of concavity can be attributed to the underlying clearing mechanism. On the other hand, in the signed Eisenberg-Noe model, the aggregation function is nonconcave due to possible immediate defaults. The next result shows that the aggregation function satisfies a weaker diversification-related property in this model.

\begin{lemma}\label{prop:Lambdastar}
	In the setting of \Cref{signed_EN_cor}, suppose that $f(\bm{p})=\transpose{\one}_n\bm{p}$ for $\bm{p}\in [\zero_n,\bm{{\bar{p}}}]$. Then, the aggregation function $\Lambda^{\textnormal{EN}}$ is star-shaped, i.e., $\Lambda^{\textnormal{EN}}(\lambda \bm{x})\leq \lambda \Lambda^{\textnormal{EN}}(\bm{x})$ for every $\lambda> 1$ and $\bm{x}\in\RR^n$.
	\end{lemma}

\proof{Proof.}
Let $\lambda> 1$ and $\bm{x}\in\RR^n$. By \Cref{signed_EN_cor}, $\Lambda^{\text{EN}}(\lambda \bm{x})$ is the optimal value of the problem
	\begin{align*}
	\text{maximize} \quad 			& \transpose{\one}_n\bm{p} && \\ 
	\text{subject to}\quad & p_i \le \bar{p}_i(1 - t_i), &&i\in\NNN, \\
	& p_i \le  \lambda x_i + (\bm{\pi}\transposeT\bm{p})_i + (m_i+\bar{p}_i)t_i, &&i\in\NNN,  \\ 
	& 0 \le p_i \le \bar{p}_i,\quad t_i \in \cb{0,1}, && i\in\NNN,
\end{align*}
and $\lambda \Lambda^{\text{EN}}(\bm{x})$ is the optimal value of the problem
	\begin{align*}
	\text{maximize} \quad 			& \transpose{\one}_n\bm{p} && \\ 
	\text{subject to}\quad & p_i \le \lambda \bar{p}_i(1 - t_i), &&i\in\NNN, \\
	& p_i \le  \lambda x_i + (\bm{\pi}\transposeT\bm{p})_i + \lambda (m_i+\bar{p}_i)t_i, &&i\in\NNN,  \\ 
	& 0 \le p_i \le \lambda\bar{p}_i,\quad t_i \in \cb{0,1}, && i\in\NNN.
\end{align*}
Since $\lambda\geq 1$, it is clear that the feasible region of the first problem is a subset of that of the second problem. Moreover, both problems are maximizing the same objective function. It follows that $\Lambda^{\text{EN}}(\lambda \bm{x})\leq \lambda \Lambda^{\text{EN}}(\bm{x})$.
\Halmos
\endproof

Within the broader literature on risk measures, it has been an active discussion over the last decade whether convexity is the right mathematical property for expressing the impact of diversification. An immediate and satisfactory generalization can be made by removing translativity and replacing convexity with quasiconvexity, yielding \emph{quasiconvex risk measures}; see \cite{simone,kupper,biagini} in the univariate case and \cite{biagini} in the setting of systemic risk measures. While this family is quite rich and includes, e.g., certainty equivalents and indices of riskiness in addition to convex risk measures, it does not cover \emph{value-at-risk}, which is a positively homogeneous monetary risk measure.

Very recently, \emph{star-shaped risk measures} have been proposed in \cite{starshaped}, where a normalized (i.e., $\rho(0)=0$) monetary risk measure $\rho\colon L^\infty(\RR)\to\RR$ is called star-shaped if 
\[
\rho(\lambda Y)\geq \lambda \rho(Y)
\]
for every $Y\in L^\infty(\RR)$ and $\lambda >1$. Star-shapedness discourages concentration of a portfolio; e.g., the risk of doubling an investment is at least twice as much as the risk of the original position. Clearly, if a monetary risk measure $\rho$ is positively homogeneous, i.e., $\rho(\lambda Y)=\lambda \rho(Y)$ for every $Y\in L^\infty(\RR)$ and $\lambda \geq 0$, then it is star-shaped; in particular, value-at-risk is star-shaped.

We will show that a systemic risk measure with a star-shaped aggregation function, e.g., the one for the signed Eisenberg-Noe model, is star-shaped in a set-valued sense. Our discussion will not rely on the particular structure of the acceptance set. Thus, we consider a slightly more general setting than the ones in Sections \ref{systemic_risk_measures} and \ref{sec:polyrisk}. 

\begin{proposition}\label{prop:starshaped}
	Consider the systemic risk measure $R$ defined in \eqref{senSystemicRiskMeasureOPT}, where $\ZZZ\subseteq\RR^n$ is a nonempty closed convex set such that $\ZZZ+\RR^n_+\subseteq\ZZZ$ and $\ZZZ\subseteq \lambda\ZZZ$ for every $\lambda>1$, $\Lambda$ is an increasing aggregation function, and $\AAA$ is the acceptance set of a monetary risk measure $\rho\colon L^\infty(\RR)\to\RR$ (see \Cref{rem:extension} and \Cref{sec:polyrisk}). 
		\begin{enumerate}[(i)]
	\item If $\Lambda$ is a star-shaped function and $\rho$ is a star-shaped risk measure, then $R$ is star-shaped, i.e., $R(\lambda \bm{X})\subseteq \lambda R(\bm{X})$ for every $\lambda >1$ and $\bm{X}\in L^\infty(\RR^n)$.
	\item If $\Lambda=\Lambda^{\textnormal{EN}}$ and $\ZZZ=\bm{z}^{\textnormal{LB}}+\RR^G_+$ for some $\bm{z}^{\textnormal{LB}}\in -\RR^G_+$, then $R=R^{\textnormal{EN}}$ is star-shaped.
	\end{enumerate}
\end{proposition}

\proof{Proof.}
(i) Let $\lambda>1$ and $\bm{X}\in L^\infty(\RR^n)$. Then, for every $\bm{z}\in\RR^n$, we have 
$\Lambda(\lambda\bm{X}+\transpose{\bm{B}}\bm{z})\leq \lambda \Lambda\of{\bm{X}+\frac{1}{\lambda}\transpose{\bm{B}}\bm{z}}$, which implies that
\[
\rho(\Lambda(\lambda\bm{X}+\transpose{\bm{B}}\bm{z}))\geq \rho\of{\lambda \Lambda\of{\bm{X}+\frac{1}{\lambda}\transpose{\bm{B}}\bm{z}}}\geq \lambda \rho\of{\Lambda\of{\bm{X}+\frac{1}{\lambda}\transpose{\bm{B}}\bm{z}}}
\] 
by the monotonicity and star-shapedness of $\rho$. Thus,
\begin{align*}
	R(\lambda\bm{X})&=\{\bm{z}\in\ZZZ\mid \Lambda(\lambda \bm{X}+\transpose{\bm{B}}\bm{z})\in\AAA\}=\{\bm{z}\in\ZZZ\mid \rho(\Lambda(\lambda \bm{X}+\transpose{\bm{B}}\bm{z}))\leq 0\}\\
	&\subseteq \cb{\bm{z}\in\ZZZ\mid \lambda \rho\of{\Lambda\of{\bm{X}+\frac{1}{\lambda}\transpose{\bm{B}}\bm{z}}}\leq 0}= \lambda\cb{\frac{1}{\lambda}\bm{z}\mid \rho\of{\Lambda\of{\bm{X}+\frac{1}{\lambda}\transpose{\bm{B}}\bm{z}}}\leq 0,\ \bm{z}\in\ZZZ}\\
	&\subseteq\lambda \cb{\bm{{\tilde{z}}}\in\ZZZ\mid  \rho\of{\Lambda\of{\bm{X}+\transpose{\bm{B}}\bm{{\tilde{z}}}}}\leq 0}=\lambda R(\bm{X}).
	\end{align*}
(ii) By \Cref{prop:Lambdastar}, $\Lambda^{\text{EN}}$ is star-shaped. Since $\bm{z}^{\text{LB}}\leq \bm{0}_G$, we have $\ZZZ\subseteq \lambda \ZZZ$ for every $\lambda>1$. Hence, $R^\text{EN}$ is star-shaped by (i).
\Halmos
\endproof

While \cite{starshaped} establishes some theoretical features of star-shaped risk measures in the univariate framework, \Cref{prop:starshaped} suggests that their set-valued extensions have immediate applications in systemic risk measures. We leave it for future research to study the theoretical properties of set-valued star-shaped risk measures in a general multivariate framework.

\section{Additional Computational Results}\label{app:comp}

In this section, we perform additional computational experiments as a continuation of \Cref{computational_results}. We consider the signed Eisenberg-Noe ($\alpha=\beta=1$) and the signed Rogers-Veraart ($\alpha=0.7$, $\beta=0.9$) models. For the random operating cash flow vector in both models, we generate $K$ instances of a Gaussian random vector whose mean vector and covariance matrix are as described in \Cref{sec:datagen}.

For both models, we consider a two-group network with $n = 45$ banks that are decomposed as $n_1 = 15$ big banks and $n_2 = 30$ small banks. As a base case, we take $K = 50$, $\sigma = 100$, $\varrho = 0.05$, $\gamma^p = 0.3$, $\bm{\nu}=\transpose{(-50, -100)}$, $\epsilon=1$, $\bm{z}^{\text{LB}}=(-500,-500)^{\mathsf{T}}$, and
\[
\bm{q}^\text{con} = 
\begin{bmatrix}
	0.5 & 0.1 \\
	0.3 & 0.5
\end{bmatrix},\quad
\bm{l}^\text{gr} = 
\begin{bmatrix}
	200 & 100 \\
	50 & 50
\end{bmatrix}.
\]
In the rest of this section, we perform sensitivity analyses on this network with respect to the connectivity probabilities and the number of scenarios.

\subsection{Connectivity Probability}

\begin{figure}[!hbp]
	\centering
		\subfloat[\centering
				Signed Eisenberg-Noe model]{\includegraphics[width=0.52\textwidth]{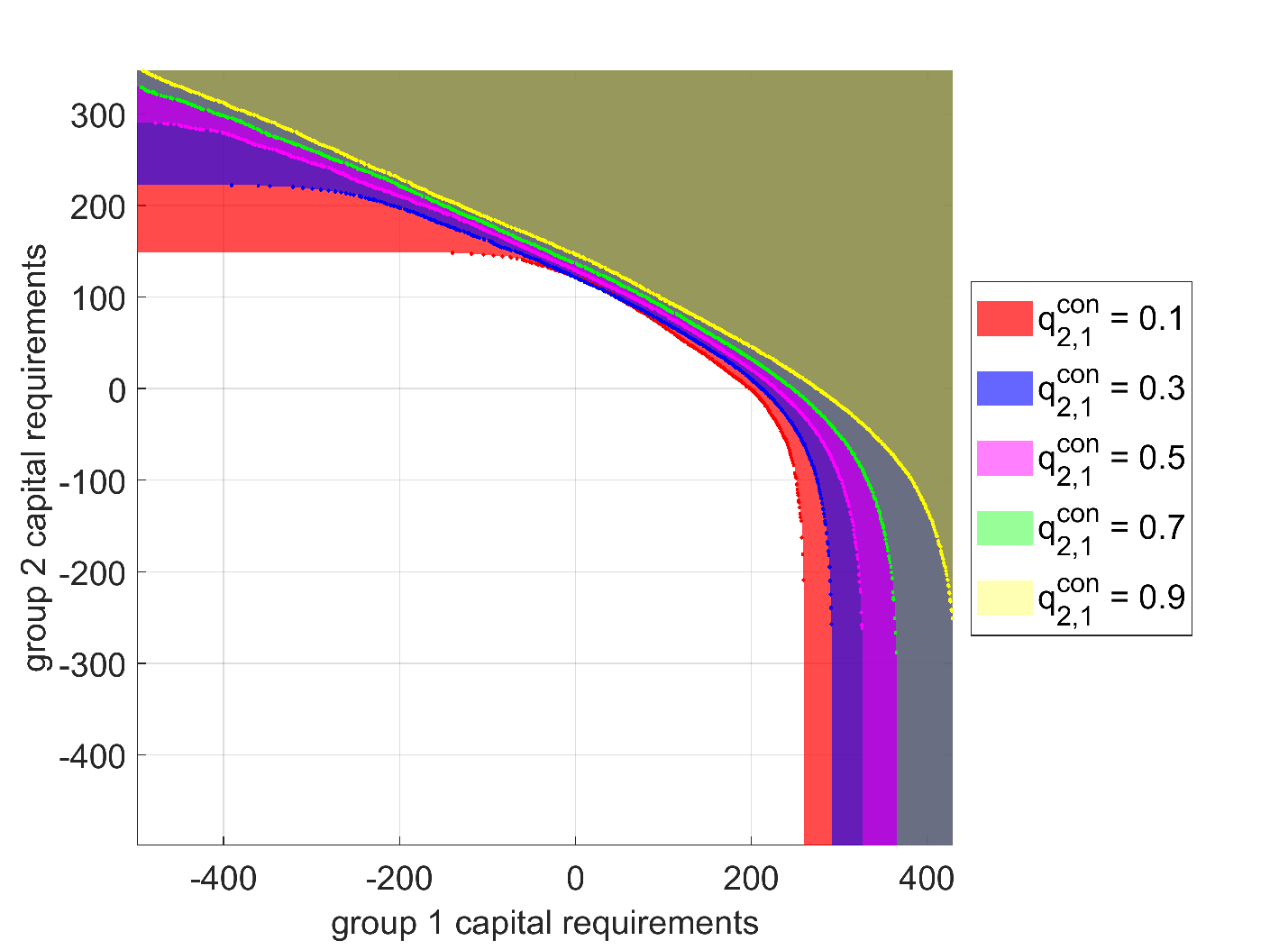}}
			\negthinspace\negthinspace\negthinspace 
		\subfloat[\centering
		Signed Rogers-Veraart model
		($\alpha=0.7, \beta=0.9$)]{\includegraphics[width=0.52\textwidth]{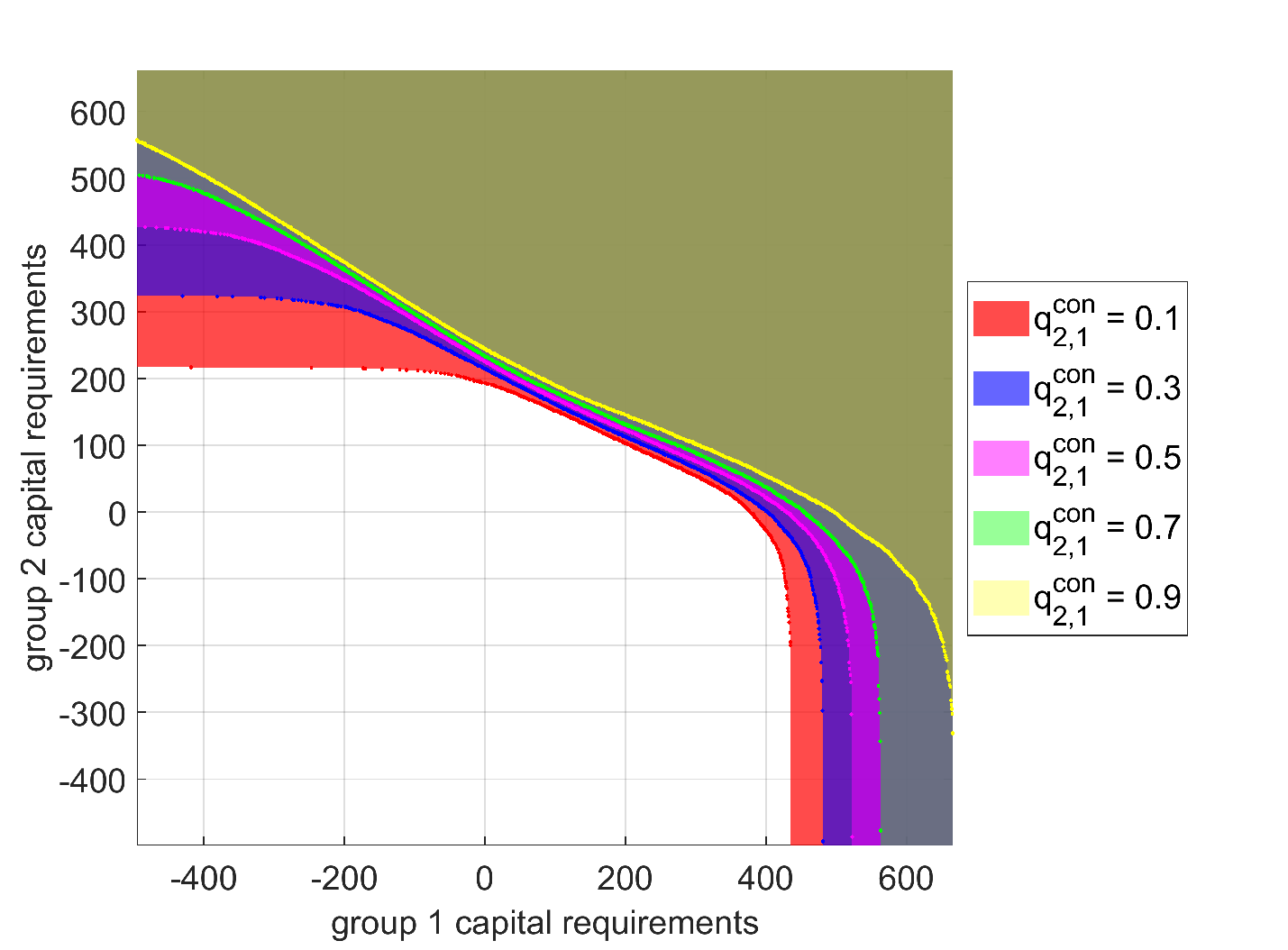}}
	\caption{Inner approximations of the systemic risk measure for different $q^{\text{con}}_{2,1}$ values.}
	\label{figure_pbigsmall_2D_inner}
\end{figure}

Connectivity probabilities play a major role in determining the topology of the network because they define the existence of liabilities between the banks. We study the sensitivity of the systemic risk measure with respect to the connectivity probability $q^\text{con}_{2,1}$ corresponding to the liabilities of small banks to big banks. We present in \Cref{table_pbigsmall_2D} the computational performance of the algorithm for $q^\text{con}_{2,1} \in \cb{0.1, 0.3, 0.5, 0.7, 0.9}$ and \Cref{figure_pbigsmall_2D_inner} consists of the corresponding inner approximations. Note that $q^\text{con}_{2,1} = 0.3$ in the base case.

\begin{table}[!tbp]
	\centering
\subfloat[\centering
Signed Eisenberg-Noe model]{
	\resizebox{0.4\textwidth}{!}{
	\begin{tabular}{|c|c|c|c|c|}
		\hline
		$q^\text{con}_{2,1}$ &
		$T$ & 
		$\#(\PPP_2)$& 
		$\overline{\text{time}}(\PPP_2)$ (sec.)& 
		$\text{time}(\text{total})$ (sec.)\\ \hline\hline
		0.1	&  782	&  392	& 5.54 	&    2171	\\ \hline
		0.3 &  1168		&  585	&  6.92	&   4047	\\ \hline
		0.5 &  1310	&  655	&  8.81	&    5773	\\ \hline
		0.7 &  1512	&  757	&  9.05	&   6848	\\ \hline
		0.9 &  1548		& 774	&  8.95	&  6926\\ \hline
	\end{tabular}
}
}
\quad
\subfloat[\centering
Signed Rogers-Veraart model
($\alpha=0.7, \beta=0.9$)]{\resizebox{0.4\textwidth}{!}{
	\begin{tabular}{|c|c|c|c|c|}
		\hline
		$q^\text{con}_{2,1}$ &
		$T$ & 
		$\#(\PPP_2)$& 
		$\overline{\text{time}}(\PPP_2)$ (sec.)& 
		$\text{time}(\text{total})$ (sec.)\\ \hline\hline
		0.1	&  964	&  481	& 5.38 	&    2588	\\ \hline
		0.3 &  1556		&  777	&  5.81	&   4513	\\ \hline
		0.5 &  1760	&  879	&  7.74	&    6804	\\ \hline
		0.7 &  1934	&  966	&  10.21 &   9859	\\ \hline
		0.9 &  2182	& 1090	&  9.27	&  10105\\ \hline
	\end{tabular}
}
}
\vspace{5pt}
	\caption{Computational performance of \Cref{alg1} relative to $q^\text{con}_{2,1}$.}
	\label{table_pbigsmall_2D}
\end{table}

In \Cref{table_pbigsmall_2D}, we note that the average time spent per Pascoletti-Serafini scalarization first increases with $q^\text{con}_{2,1}$ and decreases after $q^\text{con}_{2,1}=0.7$ in both models. The increase might be attributed to the fact that, as $q^\text{con}_{2,1}$ increases up to the break-even value $q^\text{con}_{2,1}=0.7$, big and small banks in the network become more connected and the corresponding MILP formulations of the clearing vectors tend to require more computation time. On the other hand, \Cref{table_pbigsmall_2D} suggests that it gets easier to solve the MILP problems as we approach to the fully connected case after $q^\text{con}_{2,1}=0.7$.

As $q^\text{con}_{2,1}$ increases, we note that the corresponding systemic risk measure in \Cref{figure_pbigsmall_2D_inner} gets smaller for both models. This may be interpreted as follows: a network with increased connectivity is more vulnerable to random shocks, hence systemic risk increases and the set of admissible capital allocation vectors becomes smaller.

\subsection{Number of Scenarios}\label{scsen1}

\begin{figure}[!htbp]
	\centering
\subfloat{\includegraphics[width=0.52\textwidth]{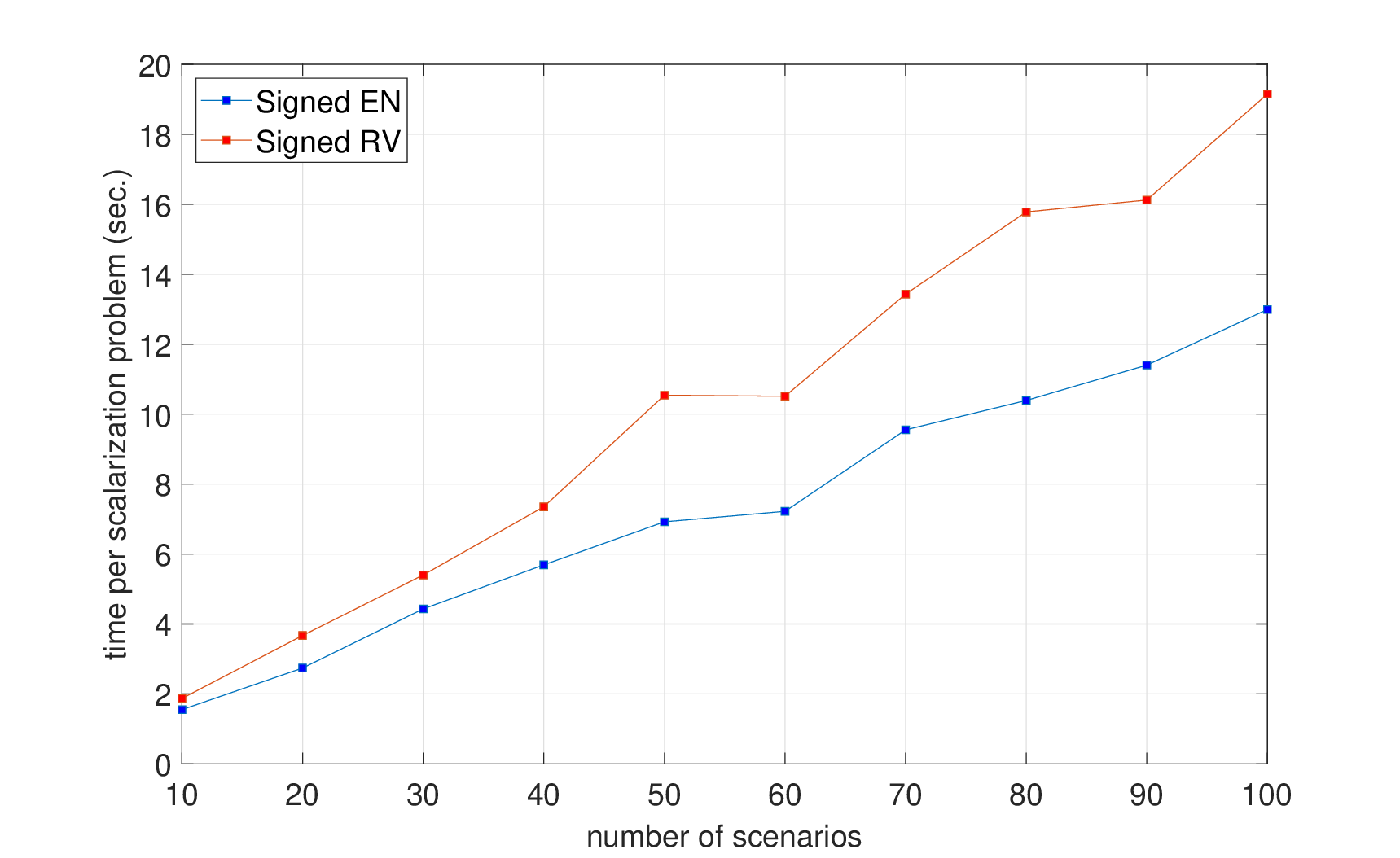}}
		\negthinspace\negthinspace\negthinspace 
			\subfloat{\includegraphics[width=0.52\textwidth]{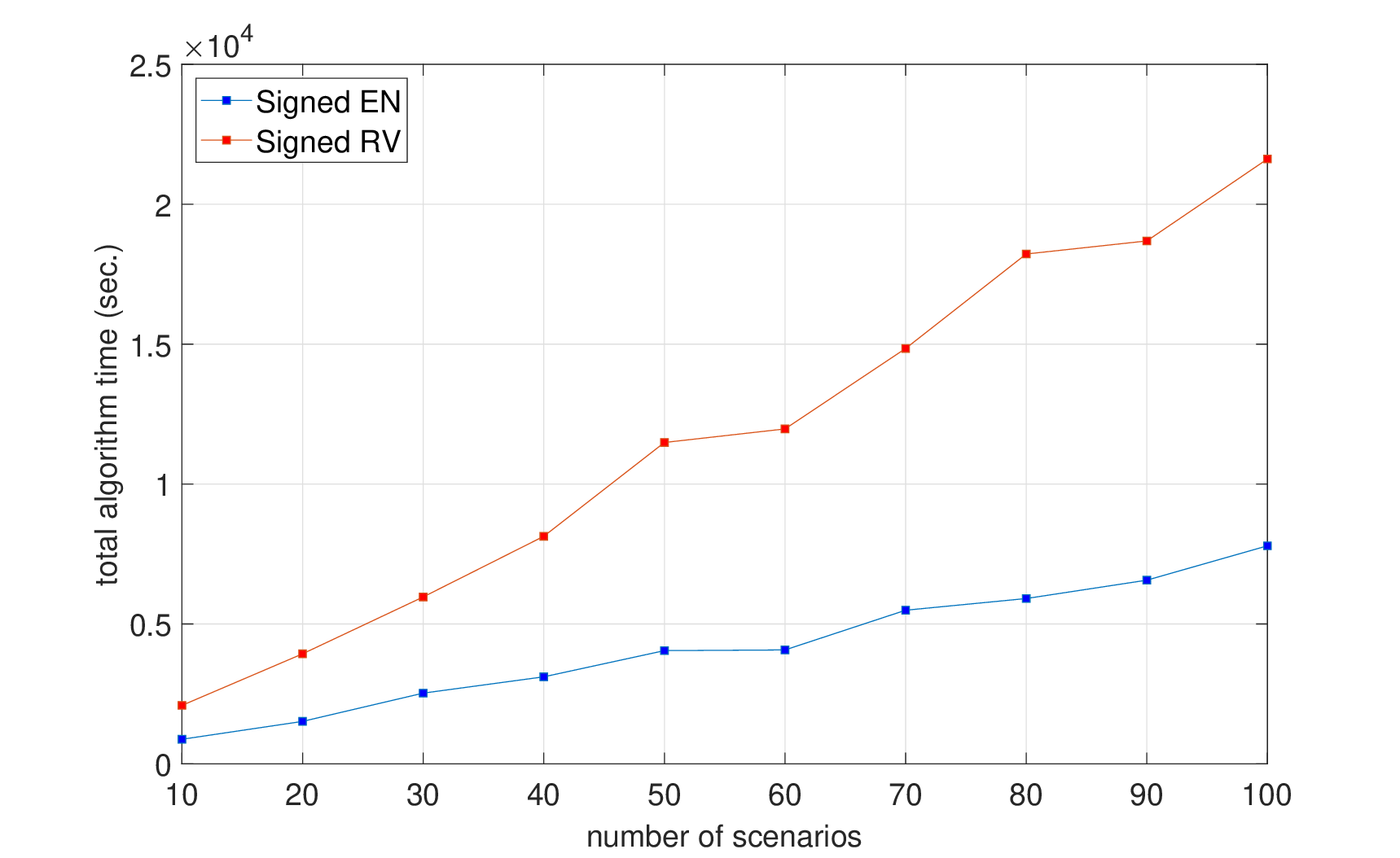}}
	\caption{Number of scenarios vs. average time per $\PPP_2$ problem (left) and number of scenarios vs. total algorithm time (right) plots.}
	\label{figure_scenarios_KvsAvg_KvsTotal}
\end{figure}

Next, we analyze the sensitivity of computation times with respect to $K$, the number of scenarios. Since the network structure remains the same for all choices of $K$, we anticipate that there will be no major changes in the shape of the systemic risk measures. This is also verified by our computations; we omit the corresponding figures for brevity. On the other hand, since each scenario adds one more MILP problem to each iteration of the bisection search for the Pascoletti-Serafini scalarization, one would expect major changes in computation times. This is indeed the case for both models as shown in \Cref{table_scenarios_2D}. The plots in \Cref{figure_scenarios_KvsAvg_KvsTotal} suggest that the average time per Pascoletti-Serafini scalarization and the total algorithm time increase linearly with $K$. The linear increase is a major benefit of the bisection search; though not reported here, using the deterministic equivalent formulations for Pascoletti-Serafini scalarization consumes much more time and the computation times increase faster than linearly with the number of scenarios. We also see that the rate of increase in both plots is higher for the signed Rogers-Veraart model. This increased difficulty may be explained by the fact that the signed Rogers-Veraart model has an extra binary variable (for each node and scenario) in the MILP formulations compared to the signed Eisenberg-Noe model.

We conclude this section with a remark on the computational performance of \Cref{alg1} for the standard and the signed Rogers-Veraart models.

\begin{remark}\label{rem:comparetwomodels}
	For the two-group examples of the standard (\Cref{sec:RVsc}) and signed Rogers-Veraart models, we use the same network size and structure, and comparable parameter configurations. The tables on the computational performance of \Cref{alg1} suggest that computing the systemic risk measure is a more demanding task with our method in the case of the signed Rogers-Veraart model. This may be attributed to the existence of big-$M$ constants in the mixed-integer programming formulation of clearing vectors. The computational difficulties related to such constants are known in the stochastic programming literature, see \citet[Section 4]{tutorial} for a discussion in the setting of chance-constrained programming and also for some techniques to avoid these constants. We leave it for future research to develop similar techniques in our setting which may speed up computations for signed models.
	\end{remark}

\begin{table}[!tbp]
	\centering
\subfloat[\centering
Signed Eisenberg-Noe model]{\resizebox{0.4\textwidth}{!}{
	\begin{tabular}{|c|c|c|c|c|}
		\hline
		$K$ &
		$T$ & 
		$\#(\PPP_2)$& 
		$\overline{\text{time}}(\PPP_2)$ (sec.)& 
		$\text{time}(\text{total})$ (sec.)\\ \hline\hline
			10 &  1130		&  566	&  	1.55	&  879	\\ \hline
			20 &  1106		&  554	&   2.74	&  1518	\\ \hline
			30 &  1138		&  570	&  	4.43	&  2528	\\ \hline
			40 &  1092		&  547	& 5.69 	&  3111 	\\ \hline	
			50 &  1168		&  585	&  6.92	&   4047	\\ \hline
			60 &  1126		&  564	&  7.22	&  4072	\\ \hline
			70 &  1148		&  575	&  	9.55	&  5491	\\ \hline
			80 &  1136		&  569	&  	10.39	&  5909	\\ \hline
			90 &  1150		&  576	&  	11.40	&  6565	\\ \hline
			100 &  1198		&  600	&  	12.99	&  7793	\\ \hline
		\end{tabular}
	}
}
\quad
\subfloat[\centering
Signed Rogers-Veraart model
($\alpha=0.7, \beta=0.9$)]{\resizebox{0.4\textwidth}{!}{
	\begin{tabular}{|c|c|c|c|c|}
		\hline
		$K$ &
		$T$ & 
		$\#(\PPP_2)$& 
		$\overline{\text{time}}(\PPP_2)$ (sec.)& 
		$\text{time}(\text{total})$ (sec.)\\ \hline\hline
		10 &  2234		&  1116	&  	1.87	&  2090	\\ \hline
		20 &  2144		&  1071	&   3.67	&  3933	\\ \hline
		30 &  2210		&  1104	&  	5.40	&  5965	\\ \hline
		40 &  2216		&  1107	& 7.35	&  8133 	\\ \hline	
		50 &  2182		&  1090	&  10.54	&   11485	\\ \hline
		60 &  2280		&  1139	&  10.51	&  11970	\\ \hline
		70 &  2212		&  	1105 &  13.43	&  14845	\\ \hline
		80 &  2312		&  	1155 &  15.78	&  18221	\\ \hline
		90 &  2320		&  1159 &  	16.12	&  18688	\\ \hline
		100 & 2260		&  1129&  	19.15&  21617	\\ \hline
	\end{tabular}
}
}
\vspace{5pt}
	\caption{Computational performance of \Cref{alg1} relative to $K$. 
	}
	\label{table_scenarios_2D}
\end{table}

\end{document}